\documentclass[]{aa}
\usepackage{natbib,twoopt}
\usepackage[breaklinks=true]{hyperref}
\usepackage{multirow,threeparttable}
\usepackage{txfonts}
\usepackage{color}
\usepackage{xcolor}
\usepackage{mathtools}
\usepackage{url}

\begin{document}

\title{X-ray study of the merging galaxy cluster Abell 3411-3412 with \emph{XMM-Newton} and \emph{Suzaku}}
\author{X. Zhang\inst{1,2} 
\and A. Simionescu\inst{2,1,3} 
\and H. Akamatsu\inst{2} 
\and J. S. Kaastra\inst{2,1}
\and J. de Plaa\inst{2} 
\and R. J. van Weeren\inst{1}
}
\institute{Leiden Observatory, Leiden University, PO Box 9513, 2300 RA Leiden, The Netherlands \\\email{xyzhang@strw.leidenuniv.nl}\label{inst1}
\and SRON Netherlands Institute for Space Research, Sorbonnelaan 2, 3584 CA Utrecht, The Netherlands\label{inst2}
\and Kavli Institute for the Physics and Mathematics of the Universe (WPI), The University of Tokyo, Kashiwa, Chiba 277-8583, Japan\label{inst3}
}
\abstract
{
%Abell 3411-3412 is a merging galaxy cluster system. Previous \emph{Chandra} observations have found an outbound bullet-like sub-cluster in the northern part and many surface brightness edges at the southern periphery, where radio observations have also discovered multiple diffuse sources. Among them, one radio relic is associate with an X-ray edge and therefore shows a direct evidence of shock re-acceleration. So far, the properties of the reported shock front and other edges need to be constrained from a thermodynamic view.  
\emph{Chandra} observations of the \object{Abell 3411}-3412 merging galaxy cluster system have previously revealed an outbound bullet-like sub-cluster in the northern part and many surface brightness edges at the southern periphery, where multiple diffuse sources are also reported from radio observations. Notably, a south-eastern radio relic associated with fossil plasma from a radio galaxy and with a detected X-ray edge provides direct evidence of shock re-acceleration. The properties of the reported surface brightness features have yet to be constrained from a thermodynamic view.
}
{We use the \emph{XMM-Newton} and \emph{Suzaku} observations of Abell 3411-3412 to reveal the thermodynamical nature of the previously reported re-acceleration site and other X-ray surface brightness edges. Meanwhile, we aim to investigate the temperature profile in the low-density outskirts with \emph{Suzaku} data. 
}
{ We perform both imaging and spectral analysis to measure the density jump and the temperature jump across multiple known X-ray surface brightness discontinuities. We present a new method to calibrate the vignetting function and spectral model of the \emph{XMM-Newton} soft proton background. Archival \emph{Chandra}, \emph{Suzaku}, and \emph{ROSAT} data are used to estimate the cosmic X-ray background and Galactic foreground levels with improved accuracy compared to standard blank sky spectra.}
{
At the south-eastern edge, both \emph{XMM-Newton} and \emph{Suzaku}'s temperature jumps point to a $\mathcal{M}\sim1.2$ shock, which agrees with the previous result from surface brightness fits with \emph{Chandra}. The low Mach number supports the re-acceleration scenario at this shock front. The southern edge shows a more complex scenario, where a shock and the presence of stripped cold material may coincide. There is no evidence for a bow shock in front of the north-western ``bullet'' sub-cluster. The \emph{Suzaku} temperature profiles in the southern low density regions are marginally higher than the typical relaxed cluster temperature profile. The measured value $kT_{500}=4.84\pm0.04\pm0.19$ keV with \emph{XMM-Newton} and $kT_{500}=5.17\pm0.07\pm0.13$ keV with \emph{Suzaku} are significantly lower than previously inferred from \emph{Chandra}.
}
{}
\titlerunning{\emph{XMM-Newton} and \emph{Suzaku} study on A3411}
\authorrunning{X. Zhang et al.}

\maketitle

\section{Introduction}
Galaxy clusters are the largest gravitationally bound objects in the Universe. They grow hierarchically by merging with sub-clusters and accreting matter from the intergalactic medium. During mergers, the gravitational energy is converted to thermal energy of the intracluster medium (ICM) via merging induced shocks and turbulence. Shocks compress and heat the ICM, which exhibits surface brightness, temperature, and pressure jumps. As a consequence, the pressure is discontinuous across a shock front. In galaxy clusters, there is another type of surface brightness discontinuity namely ``cold fronts'', which are produced by the motion of relatively cold gas clouds in the ambient high-entropy gas \citep{2007PhR...443....1M}.  In merger systems, cold fronts indicate sub-cluster cores under disruption. It is hard to determine whether a surface brightness discontinuity is a shock or a cold front based only on imaging analysis, especially when when the merging scenario is complicated or still unclear. On the other hand, the temperature and pressure profiles across shocks and cold fronts shows different trends. For cold front, the denser side of the discontinuity has a lower temperature such that the pressure profile remains continuous. Hence, temperature measurements from spectroscopic analysis are necessary to distinguish shocks and cold fronts. 

Besides heating and compressing the ICM, shocks can accelerate a small proportion of particles into the relativistic regime as cosmic ray protons (CRp) and electrons (CRe). The interaction of CRe with the magnetic field in the ICM leads to synchrotron radiation that is observable at radio wavelengths as radio relics. Radio relics are often observed in galaxy cluster peripheries with elongated (0.5 to 2 Mpc) arched morphologies and high polarisation ($\gtrsim20\%$, \citealt{1998A&A...332..395E}). The basic idea of the shock acceleration mechanism is diffusive shock acceleration (DSA, \citealt{1987PhR...154....1B,1991SSRv...58..259J}). According to DSA theory, the acceleration efficiency depends on the shock Mach number $\mathcal{M}$. The Mach number can be derived either from X-ray observations using the Rankine-Hugoniot jump condition \citep{1959flme.book.....L} or from the radio injection spectral index $\alpha_\mathrm{inj}$ on the assumption of DSA. Since the first clear detection of an X-ray shock co-located with the north western radio relic in Abell 3667 \citep{2010ApJ...715.1143F}, around 20 X-ray-radio coupled shocks have been found (see \citealt{2019SSRv..215...16V} for a review). However, there are still some remaining questions from the observational results so far. First, both X-ray and radio observations suggest low Mach numbers for cluster merging shocks ($\mathcal{M}<4$). In weak shocks, particles from the thermal pool are less efficiently accelerated due to the steep injection spectrum \citep{2002JKAS...35..159K} and less effective thermal-leakage-injection \citep{2002ApJ...579..337K}. The re-acceleration scenario has been proposed to alleviate this problem \citep{2005ApJ...627..733M}. With the presence of pre-existing fossil plasma, the acceleration efficiency would be highly increased \citep{2005ApJ...620...44K,2011ApJ...734...18K}. Second, the Mach numbers derived from X-ray observations are not always identical to those from radio observations. This could be explained from both sides. The X-ray estimations from surface brightness or temperature jumps may suffer from projection effects \citep{2017A&A...600A.100A}. In radio, when using the integrated spectral index $\alpha_\mathrm{int}$ to calculate Mach numbers, the simple approximation that $\alpha_\mathrm{int}=\alpha_\mathrm{inj}+0.5$ \citep{1962SvA.....6..317K,1998A&A...332..395E} would be incorrect when the underlying assumptions fail \citep{2015JKAS...48....9K,2016MNRAS.455.2402S}. The systematics of both methods need to be studied well before we can ascribe the discrepancy to problems in the DSA theory.

 Abell 3411-3412 is a major merger system where the first direct evidence of the re-acceleration scenario was observed \citep{2017NatAs...1E...5V}. From the dynamic analysis with optical samples, it is a probable binary merger at redshift $z=0.162$, about 1 Gyr after the first passage. The two sub-clusters have comparable masses of $\sim1\times10^{15} M_{\sun}$. Later, \citet{2019ApJ...882...69G} increased the optical sample from 174 to 242 galaxies and confirmed the redshift. From the same dataset, recently, \citet{2019ApJ...887...31A} use the $Y_X-M$ scaling relation to find $r_{500}\sim1.3$ Mpc, $kT=6.5\pm0.1$ keV, and $M_{500}=(7.1\pm0.7)\times10^{14}M_\sun$, which is much lower than the result from the previous dynamic analysis. Based on the \emph{Chandra} X-ray flux map, the core of one sub-cluster is moving towards the northeast and shows bullet-like morphology while another sub-cluster core has been entirely stripped during the previous passage. From radio images, at least four ``relics'' are located at the southern periphery of the system \citep{2013ApJ...769..101V,2013MNRAS.435..518G}. The most north-western of these four is associated with a radio galaxy, where the spectral index decreases along the radio jet and starts to increase at a certain location of the relic. The flattening edge is co-located with an X-ray surface brightness jump. All the evidence points to a scenario in which CRes lose energy via synchrotron and inverse Compton radiation in the jet, and then are re-accelerated when crossing the shock. \citet{2017NatAs...1E...5V}'s analysis shows the Mach number from radio observation is $\mathcal{M}_\mathrm{radio}=1.9$, and the compression factor of the shock from the X-ray surface brightness profile fitting is $C=1.3\pm0.1$, corresponding to $\mathcal{M}_\mathrm{SB}=1.2$. Later \citet{2019ApJ...887...31A} report that the compression factor at this discontinuity based on \emph{Chandra} data is $C=1.19\substack{ +0.21\\-0.13}$. Additionally, they provide the temperature measurements of both pre-shock and post-shock regions. However, they use large radii sector (annulus) regions to extract spectra, which makes the temperature ratio biased by the ICM far away from the shock location. \citet{2019ApJ...882...69G} suggest this shock could be produced by an optically poor group. Besides the south-west shock, \citet{2019ApJ...887...31A} report a south surface brightness discontinuity as a cold front from the sub-cluster Abell 3412's core debris; a potential surface brightness discontinuity in front of the south-east shock; and a bow shock in front of the ``bullet'' with $\mathcal{M}_\mathrm{SB}=1.15\substack{+0.14\\-0.09}$.
 
In this paper, we analyse archival \emph{XMM-Newton} and \emph{Suzaku} data to constrain the thermodynamical property of the reported shock as well as to characterise the other X-ray surface brightness discontinuities. This paper is organised as follows. In Sect. \ref{sec:data-reduction}, we describe the data reduction processes. In Sects. \ref{sec:imaging} and \ref{sec:spectrum}, we describe imaging and spectral analysis methods, selection regions, model components and systematics. We present results in Sect. \ref{sec:results}. We discuss and interpret our results in Sect. \ref{sec:discussion}. We summarise our results in Sect. \ref{sec:conclusion}. We assume $H_0=70$ km s$^{-1}$ Mpc$^{-1}$, $\Omega_\mathrm{m}=0.3$, and $\Omega_\Lambda=0.7$. At the redshift $z=0.162$, $1\arcmin$ corresponds to 167.2 kpc.

\section{Data Reduction}
\label{sec:data-reduction}

\begin{figure*}[t!]
\begin{tabular}{ccc}
\resizebox{0.3\hsize}{!}{\includegraphics{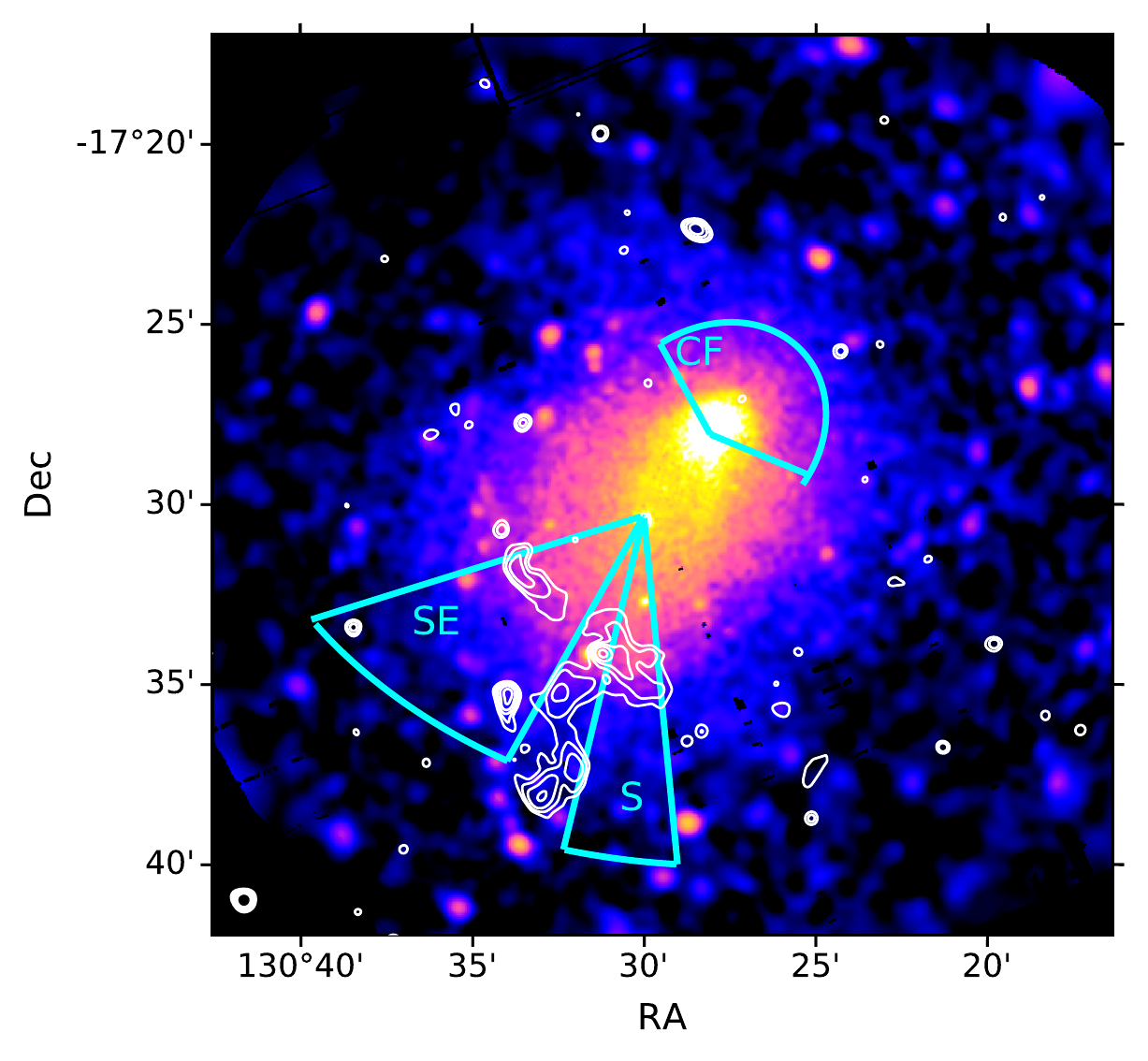}} &
\resizebox{0.3\hsize}{!}{\includegraphics{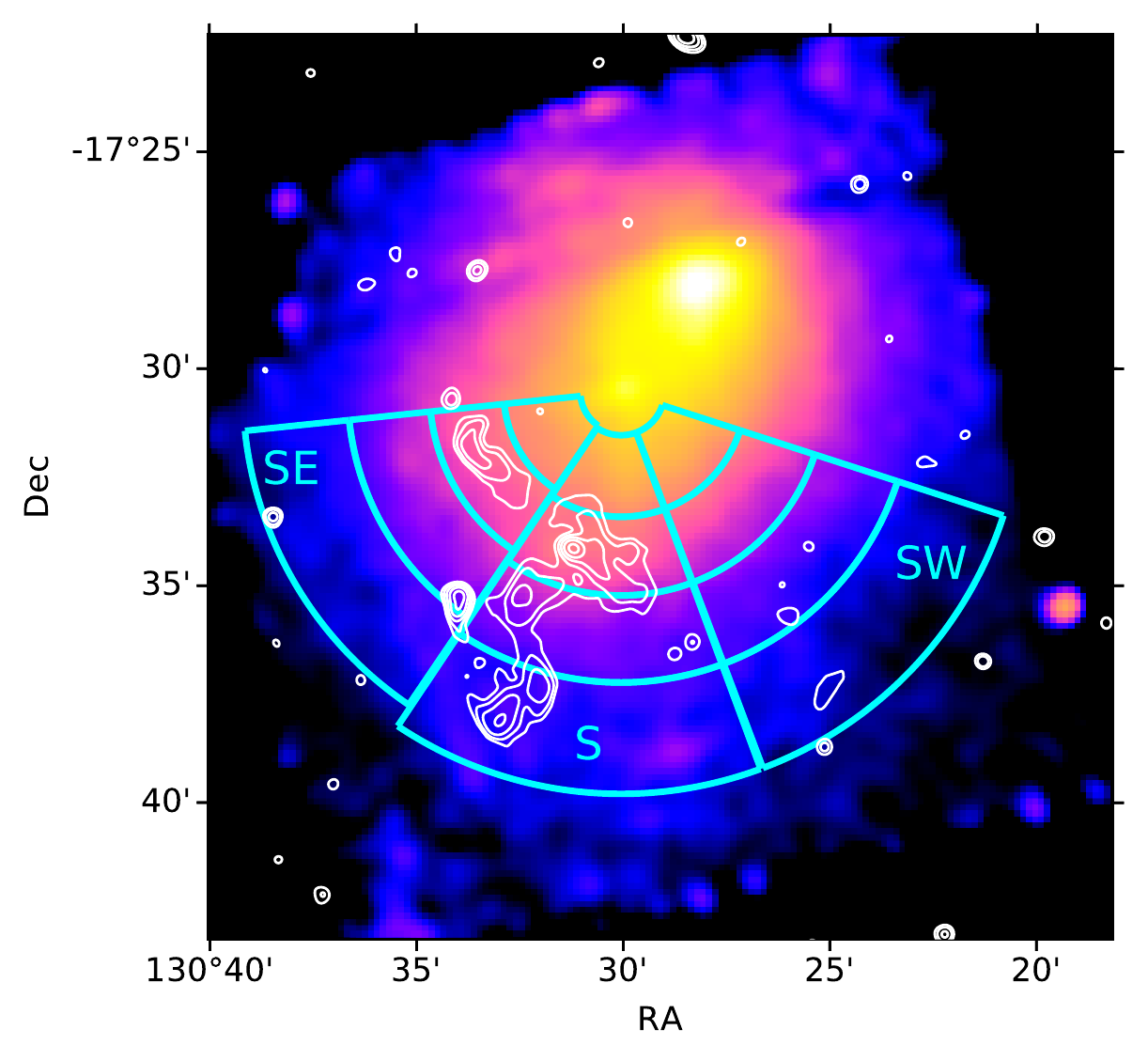}} &
\resizebox{0.3\hsize}{!}{\includegraphics{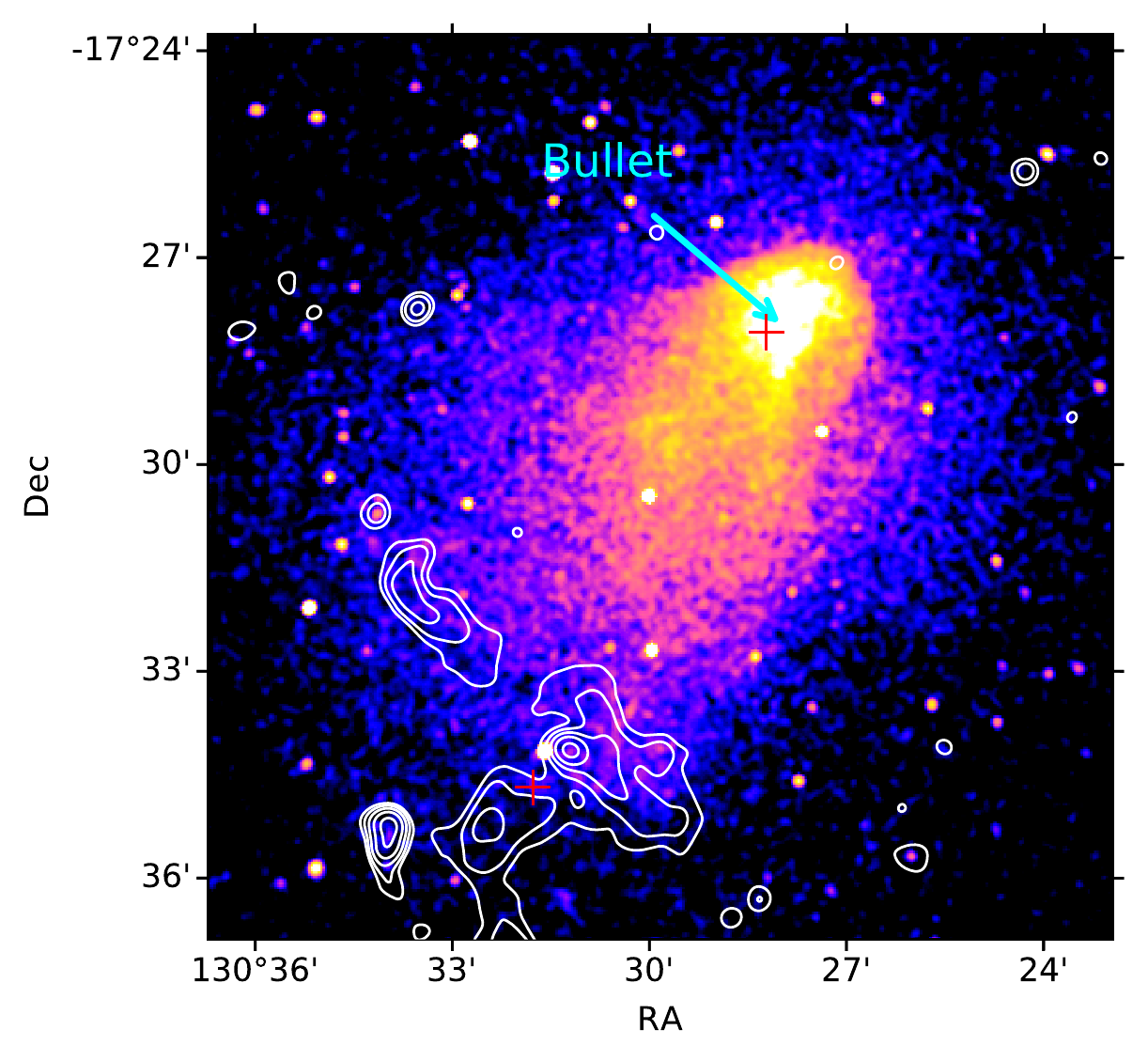}}
\end{tabular}

\caption{Smoothed flux image of Abell 3411 combined from 1.2 -- 4.0 keV \emph{XMM-Newton} EPIC CCDs (top left), 0.7 -- 7.0 keV \emph{Suzaku} XIS CCDs (top right), and 1.2 -- 4.0 keV \emph{Chandra} ACIS-I (bottom). Particle background and vignetting effect have been corrected. White contours are GMRT 610 MHz radio intensity. \emph{XMM-Newton} and \emph{Suzaku} analysis regions are plotted with cyan sectors. The locations of two brightest cluster galaxies (BCGs) are plotted with red crosses in the \emph{Chandra} image. The BCGs' coordinates are obtained from \citet{2019ApJS..240...39G}.}
\label{fig:abell3411}
\end{figure*}

\subsection{XMM-Newton}
We have analysed 137 ks \emph{XMM-Newton} EPIC archival data (ObsID: 0745120101) for this target. The \emph{XMM-Newton} Science Analysis System (SAS) v17.0.0 is used for data reduction. MOS and pn event files are obtained from the observation data files with the tasks \texttt{emproc} and \texttt{epproc}. The out-of-time event file of pn is produced by \texttt{epproc} as well.

This observation suffers from strong soft proton contamination. To minimise the contamination of soft proton flares, we adopt strict good time intervals (GTI) filtering criteria. For each detector, we first bin the 10 -- 12 keV light curve in 100 s intervals. We take the median value of the histogram as the mean flux of the source. All bins with count rate more than $\mu+1\sigma$ are rejected, where the $\sigma$ is derived from a Poissonian distribution. To exclude the contamination of some extremely fast flares, we then bin the residual 10 -- 12 keV light curve in 20 s intervals and reject bins with a count rate more than $\mu+2\sigma$. After GTI filtering, the clean exposure time of MOS1, MOS2, and pn are 89 ks, 97 ks, and 74 ks, respectively. For both imaging and spectral analysis, we select single to quadruple MOS events (\texttt{PATTERN<=12}) and single to double pn events (\texttt{PATTERN<=4}). 

Particle backgrounds are generated from integrated Filter Wheel Closed (FWC) data \footnote{\url{https://www.cosmos.esa.int/web/xmm-newton/filter-closed}} 2017v1. The FWC spectra of MOS are normalised using the unexposed area as described by \citet{2008A&A...478..575K}. The normalisation factors of MOS1 and MOS2 FWC spectra are 0.97 and 0.98, respectively. For pn, there is no "clean" out-of-FOV area (see Appendix \ref{appendix:oofov}). Therefore we normalise the integrated FWC spectrum using the FWC observation in revolution 2830, which is performed six months after our observation and is the closest FWC observation time. The normalisation factor of the integrated pn FWC spectrum is 0.82.

\subsection{Suzaku}
Abell 3411 was observed by Suzaku for 127 ks (ObsID: 809082010). A $21\arcmin$ offset area was observed for 39 ks (ObsID: 809083010). We use standard screened X-ray Imaging Spectrometer (XIS) event files for analysis. Two clocking mode ($5\times5$ and $3\times3$) events lists are combined. Additionally, geomagnetic cut-off rigidity (COR) $>8$ selection is applied to filter the event files and generate the non-X-ray background (NXB). The latest recommended recipe for removing flickering pixels \footnote{\url{https://heasarc.gsfc.nasa.gov/docs/suzaku/analysis/xisnxbnew.html}} is applied to both observation and NXB event files. After COR screening, the valid source exposure time is 105 ks for XIS0, and 108 ks for XIS1 and XIS3. The NXB spectra are generated by using the task \texttt{xisnxbgen} \citep{2008PASJ...60S..11T} and are subtracted directly. The normalisation of NXB spectra is scaled by the 10 -- 14 keV count rates. To estimate the systematics contributed by the NXB in the spectral analysis, we assume a fluctuation of $3\%$ around the nominal value \citep{2008PASJ...60S..11T}.

The \emph{Suzaku} XIS astrometry shift could be as large as $50\arcsec$ \citep{2007PASJ...59S...9S}. To measure the offset of our observation, we first make a combined 0.7 -- 7.0 keV XIS flux map to detect point sources and then compare the XIS coordinates with EPIC coordinates from the 3XMM-DR8 catalogue \citep{2016A&A...590A...1R}. We follow the instruction\footnote{\url{https://heasarc.gsfc.nasa.gov/docs/suzaku/analysis/expomap.html}} to correct the vignetting effect. Only four point sources are detected by \texttt{wavdetect} in the CIAO package. The mean XIS RA offset is $25.0\pm0.3\arcsec$ to the east, and the mean Declination offset is $6.8\pm0.3\arcsec$ to the south.

\subsection{Chandra}

We use the same \emph{Chandra} dataset as \citet{2017NatAs...1E...5V}. Event files, as well as auxiliary files, are reproduced by task \texttt{chandra\_repro} in the Chandra Interactive Analysis of Observations (CIAO) package v4.10 with \texttt{CALDB} 4.8.0. We use \texttt{merge\_obs} to merge all observations and create a 1.2 -- 4.0 keV flux map. Stowed event files are used as particle backgrounds. The normalisations are scaled by the 10 -- 12 keV band count rate of each observation. 

The observation IDs, instruments, pointing coordinates, and clean exposure times of the observations taken with all three satellites are listed in Table \ref{table:obs}.

\begin{table*}[t!]
\caption{Observation information.}
\label{table:obs}
\centering
\begin{tabular}{ccccc}
\hline\hline
Telescope & ObsID & Instrument & Pointing Coordinate (RA, Dec) & Valid Exp. (ks)\\
\hline
\multirow{3}{*}{\emph{XMM-Newton}} & \multirow{3}{*}{0745120101} &EPIC-MOS1 &\multirow{3}{*}{08:41:55, -17:28:43} & 89\\
 & &EPIC-MOS2 & & 97\\
 & &EPIC-pn & & 74\\
\hline
\multirow{6}{*}{\emph{Suzaku}} & \multirow{3}{*}{809082010} &XIS 0& \multirow{3}{*}{08:42:03, -17:34:12} &105\\
 & &XIS 1& &108\\
 & &XIS 3& &108\\
 \cline{2-5}
 &\multirow{3}{*}{809083010 (Offset)} &XIS 0& \multirow{3}{*}{08:43:06, -17:19:34}&32\\
 & &XIS 1& &32\\
 & &XIS 3& &32\\
\hline
\multirow{8}{*}{\emph{Chandra}} &13378 & \multirow{8}{*}{ACIS-I} & 08:42:05, -17:32:16& 10 \\
&15316& & 08:42:03, -17:29:53 &39\\
&17193& & 08:42:01, -17:29:56 &22\\
&17496& & 08:42:04, -17:29:02 &32\\
&17497& & 08:42:01, -17:29:19 &22\\
&17583& & 08:42:01, -17:29:56 &32\\
&17584& & 08:42:02, -17:29:29&33\\
&17585& & 08:42:01, -17:29:19 &24\\
\hline
\end{tabular}
\end{table*}

\section{Imaging analysis}
\label{sec:imaging}
We use the \emph{XMM-Newton} 1.2 -- 4.0 keV band for surface brightness analysis. The vignetting-corrected exposure maps are generated by the task \texttt{eexpmap}. Pixels with less than 0.3 of the maximum exposure value are masked by \texttt{emask} and then excluded. Because half of the photons from mirrors 1 and 2 are deflected by the RGS system, and the quantum efficiency of MOS is different from that of pn, we need to scale the MOS exposure maps to make the MOS fluxes to match the pn flux. We first derive the radial surface brightness profiles of the three detectors with unscaled exposure maps. The selection region is a circle centred at the pn focal point. We fit $0\arcmin<r<6\arcmin$ MOS-to-pn surface brightness ratios with a constant model. The ratios are 0.37 and 0.38 for MOS1 and MOS2, respectively. We combine the net count maps and scaled exposure maps from three detectors to produce a flux map. The particle background subtracted, vignetting corrected, smoothed image is shown in Fig. \ref{fig:abell3411}. We exclude point sources before we extract the surface brightness profiles. Point sources' coordinates are obtained from the 3XMM-DR8 catalogue \citep{2016A&A...590A...1R} and checked by visual inspection. The exclusion shape of each point source is generated by the \texttt{psfgen} in SAS with the PSF model \texttt{ELLBETA}.

We extract surface brightness profiles along four regions, which are marked on the \emph{XMM-Newton} flux map in Fig. \ref{fig:abell3411}. The first selection region (the south-west region) is the previously reported shock \citep{2017NatAs...1E...5V}. From the \emph{XMM-Newton} flux map, this discontinuity is unlikely to be seen by the naked eye. With the help of the \emph{Chandra} flux map, we are able to define an elliptical sector whose side is parallel to the discontinuity. The second region (south) is crossing the south discontinuity seen in the \emph{Chandra} flux map \citep{2019ApJ...887...31A} as well as a diffuse radio emission. The third one (cold front) stretches along the direction of the "bullet" and probably hosts a bow shock. The last one is the "bullet" itself. We set the region boundary carefully to be parallel to the surface brightness edge.

We extract surface brightness profiles from both \emph{XMM-Newton} and \emph{Chandra} datasets. We use a projected double power law density model to fit discontinuities, whose unprojected density profile is 
\begin{equation}
n(r)=\begin{dcases*}
C n_\mathrm{edge} \left(\frac{r}{r_\mathrm{edge}}\right)^{-\alpha_1}& When $r\le r_\mathrm{edge}$,\\
n_\mathrm{edge} \left(\frac{r}{r_\mathrm{edge}}\right)^{-\alpha_2}& When $r>r_\mathrm{edge}$.
\end{dcases*}
\end{equation}
$C$ is the compression factor at the shock or cold front. $r_\mathrm{edge}$ and $n_\mathrm{edge}$ are the radius and the density at the edge respectively. We assume the curvature radius along the line of sight is equal to the average radius of the surface brightness discontinuity (i.e. the ellipticity along the third axis is zero). The projected surface brightness profile is 
\begin{equation}
S(r) = \int_{-\infty}^{\infty}n^2 \left(\sqrt{z^2+r^2}\right) \mathrm{d}z + S_\mathrm{bg},
\end{equation}
where $z$ is the coordinate along the line of sight, $S_\mathrm{bg}$ is the surface brightness contributed by the X-ray background. For \emph{Chandra}, we measure $S_\mathrm{bg}=7\times10^{-7}$ count s$^{-1}$ cm$^{-2}$ arcmin$^{-2}$ from the front-illuminated ACIS-S chips. For XMM-Newton, this value is more difficult to be properly estimated. Because soft protons are less vignetted than photons, we can see an artificial surface brightness increases beyond $10\arcmin$. We therefore avoid regions located beyond $10\arcmin$ from the focal point. C-statistics \citep{1979ApJ...228..939C} is adopted to calculate the likelihood function for fitting.

\section{Spectral analysis}

\begin{table*}
\caption{Spectral fitting components and models.}
\label{table:components}
\centering
\begin{tabular}{ccccc}
\hline\hline
Component & Model\tablefootmark{a} & RMF & ARF &Coupling\\
\hline
\multicolumn{5}{c}{\emph{XMM-Newton} EPIC}\\
\hline
ICM & $cie*reds*hot\tablefootmark{b}$ & Yes & Yes & -\\
LHB & $cie$ & Yes & Yes & RASS\\
GH & $cie*hot$ & Yes & Yes & RASS\\
CXB & $pow*hot$ & Yes & Yes & -\\
\hline
FWC continuum & $pow$ & Yes & No & -\\
FWC lines & $delt$s & Yes & No & -\\
SP & $pow$& Dummy \tablefootmark{c} & No & -\\
\hline
\multicolumn{5}{c}{\emph{Suzaku} XIS}\\
\hline
ICM & $cie*reds*hot$ & Yes & Yes & -\\
LHB & $cie$ & Yes & Yes & RASS\\
GH & $cie*hot$ & Yes & Yes & RASS\\
CXB & $pow*hot$ & Yes & Yes & -\\
\hline
\multicolumn{5}{c}{\emph{Suzaku} XIS offset observation}\\
\hline
LHB & $cie$ & Yes & Yes & RASS\\
GH & $cie*hot$ & Yes & Yes & -\\
CXB & $pow*hot$ & Yes & Yes & -\\
SWCX & $delt$ & Yes & Yes & -\\
\hline
\multicolumn{5}{c}{\emph{RASS}}\\
\hline
LHB & $cie$ & Yes & Yes & -\\
GH & $cie*hot$ & Yes & Yes & -\\
CXB & $pow*hot$ & Yes & Yes & -\\
\hline
\end{tabular}
\tablefoot{\\
\tablefoottext{a}{For details of different models, please see the SPEX Manual (\url{https://spex-xray.github.io/spex-help/index.html).}}\\
\tablefoottext{b}{We set the temperature of the \emph{hot} model to $5\times10^{-4}$ keV to mimic the absorption of a neutral plasma.}\\
\tablefoottext{c}{The dummy RMF has a uniform photon redistribution function, see Appendix \ref{appendix:sp} for details.}
}
\end{table*}

\label{sec:spectrum}
To study the thermodynamic structure of the cluster, in particular, across known surface brightness discontinuities, we perform spectroscopic analysis and obtain the temperature from different selection regions.
For spectral analysis, the spectral fitting package SPEX v3.05 \citep{1996uxsa.conf..411K, kaastra2018.2419563} is used. The reference proto-solar element abundance table is from \citet{LandoltBornstein2009:sm_lbs_978-3-540-88055-4_34}. OGIP format spectra and response matrices are converted to SPEX format by the \texttt{trafo} task. All spectra are optimally binned \citep{2016A&A...587A.151K} and fitted with C-statistics \citep{1979ApJ...228..939C}. The Galactic hydrogen column density is calculated using the method of \citet{2013MNRAS.431..394W}\footnote{\url{https://www.swift.ac.uk/analysis/nhtot/}}, which takes both atomic and molecular hydrogen into account. The weighted effective column density is $n_\mathrm{H}=5.92\times10^{20}$ cm$^{-2}$. We use the ROSAT All-Sky Survey (RASS) spectra generated by the X-Ray Background Tool\footnote{\url{https://heasarc.gsfc.nasa.gov/cgi-bin/Tools/xraybg/xraybg.pl}} \citep{2019ascl.soft04001S} to help us constrain two foreground thermal components: the local hot bubble (LHB) and Galactic halo (GH). The RASS spectrum is selected from a $1\degr-2\degr$ annulus centred at our galaxy cluster. The two foreground components are modelled using single temperature collisional ionisation equilibrium (CIE) models in SPEX. The GH is absorbed by the Galactic hydrogen while the LHB is unabsorbed. We fix the abundance to the proto-solar abundance for those two components. The best fit foreground parameters are shown in Table \ref{table:xbg}. These temperatures are consistent with previous studies (e.g. \citealt{2009PASJ...61..805Y}).

\begin{table}[!]
\caption{X-ray foreground components constrained by the RASS spectrum. The normalisations are scaled to a 1 arcmin$^2$ area.}
\label{table:xbg}
\centering
\begin{tabular}{ccc}
\hline\hline
& Flux ( 0.1 -- 2.4 keV) & $kT$ \\
& $10^{-2}$ ph s$^{-1}$ m$^{-2}$ & keV \\
\hline
LHB & $3.61\pm0.07$ & $0.11\pm0.01$ \\
GH & $1.49\pm0.28$ & $0.20\pm0.02$ \\
\hline
\end{tabular}
\end{table}

\begin{figure*}[t!]
\begin{tabular}{cc}
\resizebox{0.45\hsize}{!}{\includegraphics{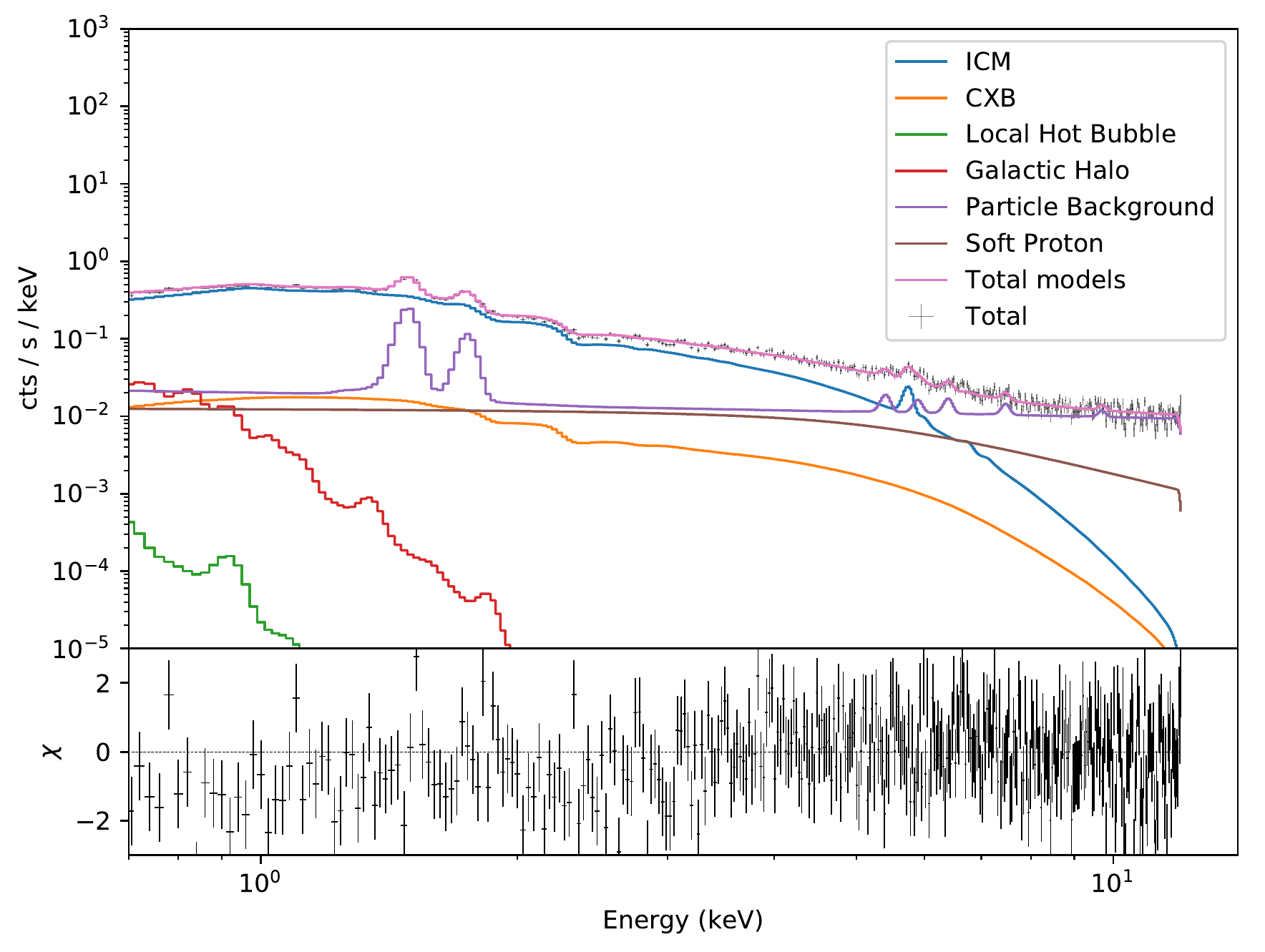}}&
\resizebox{0.45\hsize}{!}{\includegraphics{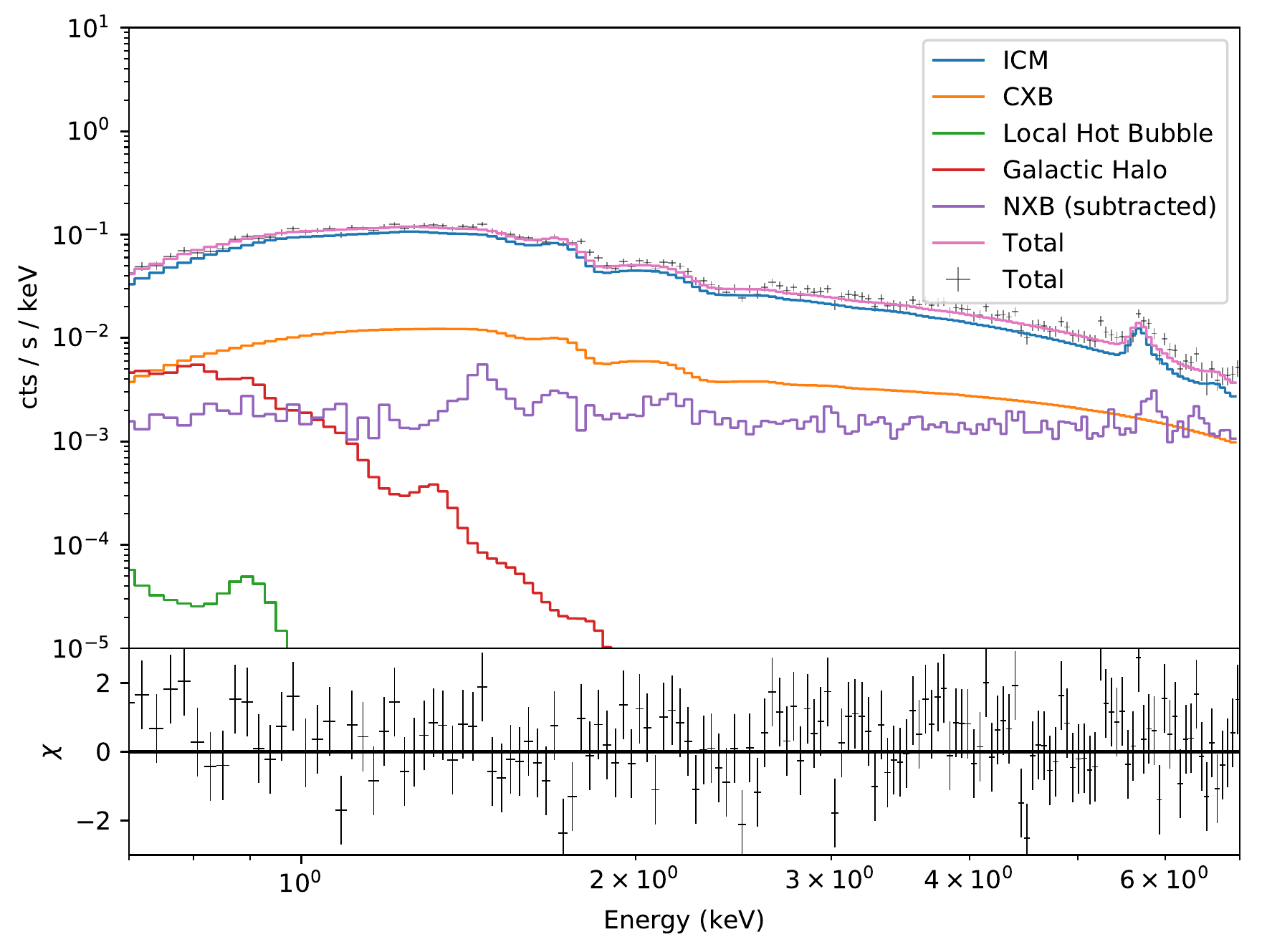}}\\
\resizebox{0.45\hsize}{!}{\includegraphics{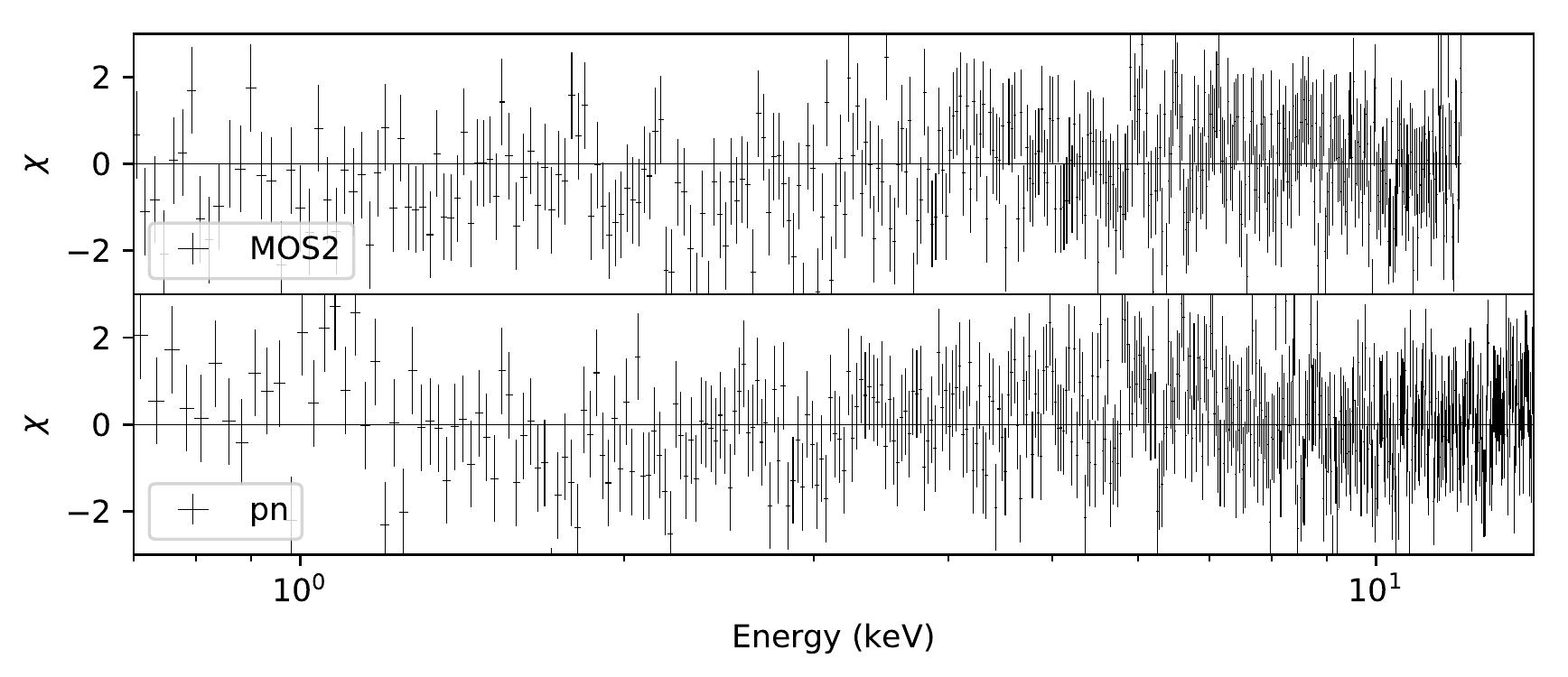}}&
\resizebox{0.45\hsize}{!}{\includegraphics{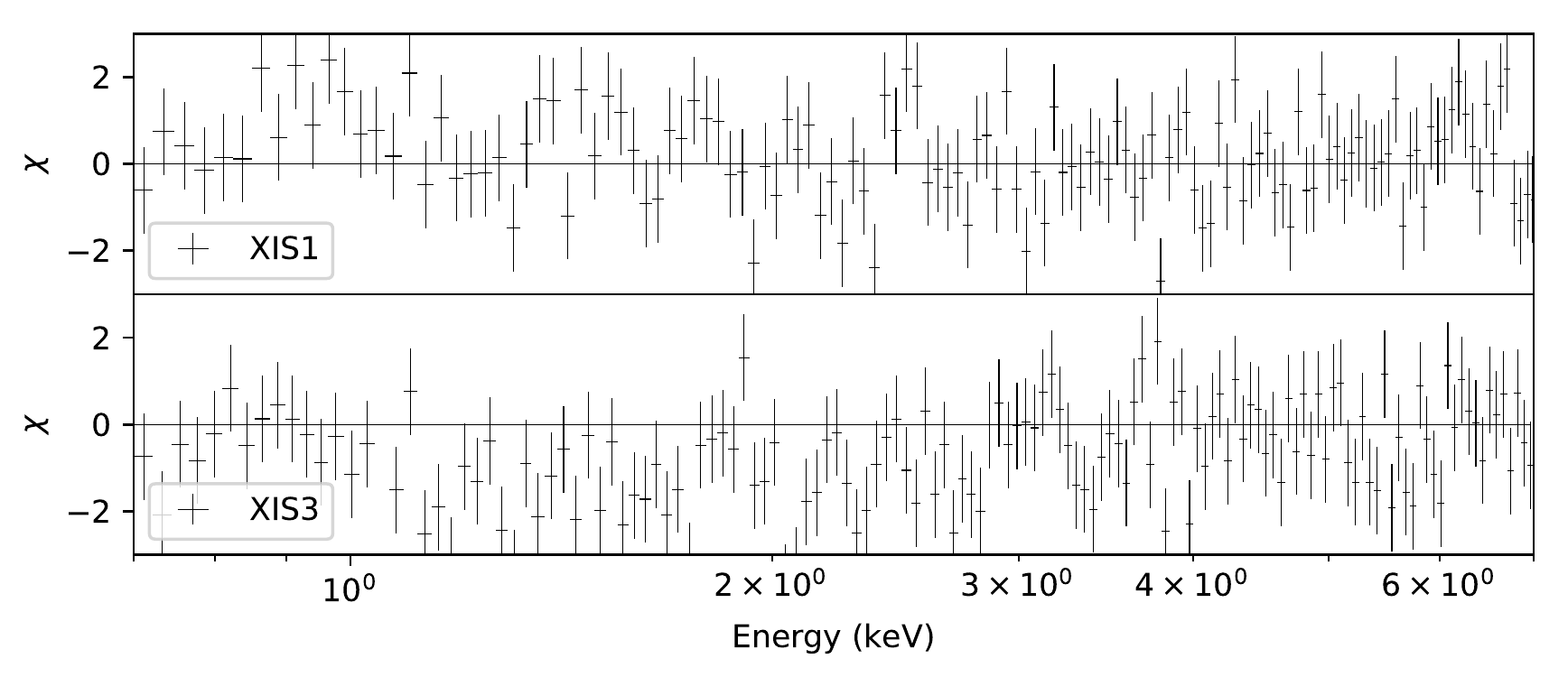}}\\
\end{tabular}
\caption{The $r_{500}$ \emph{XMM-Newton}-EPIC MOS1 (top left) and \emph{Suzaku} XIS0 (top right) spectra as well as individual spectral components. We also plot residuals from the other two EPIC detectors (bottom left) and XIS detectors (bottom right). The fit statistics are $\mathrm{C-stat / d.o.f}=1992/1554$ for \emph{XMM-Newton} EPIC spectra and $\mathrm{C-stat / d.o.f} =563/423$ for \emph{Suzaku} XIS spectra.}
\label{fig:spec_example}
\end{figure*}

\subsection{XMM-Newton}\label{section:spectrum-xmm}
In the \emph{XMM-Newton} spectral analysis, the effective extraction region areas of spectra from different detectors are calculated using the SAS task \texttt{backscale}. To ensure that the extracted spectra from different detectors cover the same sky area, we exclude the union set of the bad pixels of all three detectors from each spectrum. This method will lead to lower photon statistics but can reduce the spectral discrepancies due to different selection regions when we perform the parallel fitting. With the calculated \texttt{backscale} parameter, we determine the sky area of each spectrum with respect to 1 arcmin$^2$ and set the region normalisation to that value. The spectral components and models are listed in Table \ref{table:components}. We fit all spectra from different detectors simultaneously. We plot the MOS1 spectrum within $r_{500}$ in Fig. \ref{fig:spec_example} as an example to show all spectral components. The components of the MOS2 and pn spectra are similar, so we only additionally plot the fit residuals of these two detectors in Fig. \ref{fig:spec_example}.

The ICM is modelled with a single temperature CIE. The abundances of metal elements are coupled with the Fe abundance. With the FWC data, we find the particle background continuum can be fit by a broken power law with break energy at 2.5 and 2.9 keV for MOS and pn, respectively. Because the instrumental lines in particle backgrounds are spatially variable, we fit instrumental lines as delta functions with free normalisations. Instrumental lines' energies are taken from \citet{2015A&A...575A..37M}. If the selection region includes MOS1 CCD4 or MOS2 CCD5 pixels, channels below 1.0 keV are ignored because of the low energy noise plateau \footnote{\url{https://xmm-tools.cosmos.esa.int/external/xmm\_user\_support/documentation/uhb/epicdetbkgd.html}}. The two foreground components are coupled with the components of the RASS spectrum. To determine the point source detection limit and calculate the cosmic X-ray background (CXB) flux, we first use the CIAO package tool \texttt{wavdetect} to detect point sources in a 1 -- 8 keV combined \emph{XMM-Newton} EPIC flux image. We set \texttt{wavdetect} parameters \texttt{scale=``1.0 2.0 4.0''}, \texttt{ellsigma=4}, and \texttt{sigthresh=1e-5}, which is roughly the reciprocal of the image size in our case. Other parameters' values are left as the default. In the detected source list, we select the four lowest detection significance sources to extract and combine their MOS and pn spectra. The source extraction regions are directly obtained from the output of the \emph{wavdetect} task. Local backgrounds are extracted and subtracted from the total source spectra with elliptical annuli, whose inner radii are the radii of the source regions, and the width is $15\arcsec$. We fit the point source spectra using $abs*pow$ models with free power law normalisation and photon index. The best-fit flux in 2 -- 8 keV range is $(6.0\pm0.5)\times10^{-15}$ erg s$^{-1}$ cm$^{-2}$. Two out of four sources are in our \emph{Chandra} point source catalogue (see Appendix \ref{appendix:cxb}). Their \emph{Chandra} fluxes are $(6.8\pm1.1)\times10^{-15}$ and $(9.2\pm1.2)\times10^{-15}$ erg s$^{-1}$ cm$^{-2}$, respectively. Therefore, we use $6.0\times10^{-15}$ erg s$^{-1}$ cm$^{-2}$ from 2 -- 8 keV as a detection limit to calculate the CXB surface brightness. The corresponding CXB surface brightness is $3.5\times10^{-14}$ erg s$^{-1}$ cm$^{-2}$ arcmin$^{-2}$ with a fixed photon index $\Gamma=1.41$, see Appendix \ref{appendix:cxb} for details. The CXB deviation is calculated by Eq.\ref{equation:cxberror}, we fit spectra with $\pm1\sigma_\mathrm{sys}$ CXB luminosity to obtain the systematics contributed by CXB uncertainty. We also include the GH systematics for \emph{XMM-Newton} spectral analysis with the uncertainty measured from the \emph{Suzaku} offset observation (see Sect. \ref{section:offset}). We calibrate the soft proton background in terms of spectral models and vignetting functions with an observation of the Lockman Hole (see Appendix \ref{appendix:sp}). The best-fit parameters and the systematic uncertainties of each soft proton component are listed in Table \ref{table:sp_calibrated}. When studying the systematics from the soft proton components, we fit spectra with $\pm1\sigma_\mathrm{sys}$ of the MOS1, MOS2, and pn soft proton luminosity individually. The envelope of the highest and the lowest fitted temperatures are taken as the systematics from the soft proton model.

\subsection{Suzaku}
\label{section:spectrum-suzaku}
In the \emph{Suzaku} spectral analysis, the energy range 0.7 -- 7.0 keV is used for spectral fitting. ARFs are generated by the task \texttt{xissimarfgen} \citep{2007PASJ...59S.113I} with the parameter \texttt{source\_mode=UNIFORM}. X-ray spectral components are the same as those of the EPIC spectra. We exclude sources with 2 -- 8 keV flux $S_\mathrm{2-8 keV}>2\times10^{-14}$ erg s$^{-1}$ cm$^{-2}$ in our catalog (see Appendix \ref{appendix:cxb}) using $1\arcmin$ radius circles. The unresolved CXB flux, as well as its uncertainty for each selected region, are calculated by Eq. \ref{equation:cxberror}. All spectra from different detectors are fitted simultaneously as well. Because the \emph{Suzaku} ARFs are normalised to $400\pi$ arcmin$^2$, we set region normalisations in SPEX to $400\pi$. In that case, the fitted luminosity value corresponds to 1 arcmin$^2$. An example of the XIS0 $r_{500}$ spectrum is shown in Fig. \ref{fig:spec_example} to illustrate all spectral components. Same as the EPIC spectra, we additionally plot the residuals of XIS1 and XIS3.

\subsubsection{Offset observation}\label{section:offset}
We use the offset observation to study systematics from the foreground X-ray components. We extract spectra from the full field of view but exclude the XIS0 bad region and point sources by visual inspection. We fit the spectrum from 0.4 to 7.0 keV with LHB, GH, and CXB components. Additionally, we add a delta line component at 0.525 keV to fit an extremely strong O I K$\alpha$ line, which is generated by the fluorescence of solar X-rays with neutral oxygen in the Earth's atmosphere \citep{2014PASJ...66L...3S}. Because the LHB flux is prominent at energies much lower than 0.4 keV, we still couple the normalisation and temperature with the RASS LHB component. From 0.4 to 1 keV, the spectrum is dominated by the GH. We free the normalisation of the GH but still couple the temperature with the RASS GH component. The CXB power law index is set as $\Gamma=1.41$, and the normalisation is thawed. Best-fit parameters are listed in Table \ref{table:offset}. The best-fit GH normalisation is $40\%$ lower than the best-fit value from RASS. We include the $40\%$ GH normalisation to study the systematics. 

\begin{table*}[t!]
\caption{Best fit parameters of \emph{Suzaku} offset spectra. The distance of model components is set to $z=0.162$ to calculate the emissivity. Normalisations are scaled to a 1 arcmin$^{2}$ area.}
\label{table:offset}
\centering
\begin{tabular}{ccccc}
\hline\hline
Component & Parameter & Unit & Value & Status\\
\hline
\multirow{2}{*}{LHB} & $norm$ & $10^{64}$ m$^{-3}$& $4.7\times10^{5}$&Fixed\\
& $kT$ & keV & 0.11 &Fixed\\
\hline
\multirow{2}{*}{GH} & $norm$ & $10^{64}$ m$^{-3}$ & $(5.6\pm0.7)\times10^{5}$ &Free\\
& $kT$ & keV & 0.20 &Fixed\\
\hline
\multirow{2}{*}{CXB} & $lum$ & $10^{30}$ W& $(2.15\pm0.07)\times10^{4}$ &Free\\
& $\Gamma$ & - & 1.41 &Fixed\\
\hline
\end{tabular}
\end{table*}

\subsubsection{Selection regions}
Because of the large radius of the point spread function (PSF) of \emph{Suzaku}, structures on small scales are not resolved. We use sector regions centred at the centre of the cluster and extending towards the south-east (SE), south (S), and south-west (SW) directions (see Fig. \ref{fig:abell3411}) to extract spectra and measure the temperature profiles.

\subsection{XMM-Newton-Suzaku cross-calibration}
\begin{figure}[t!]
\resizebox{\hsize}{!}{\includegraphics{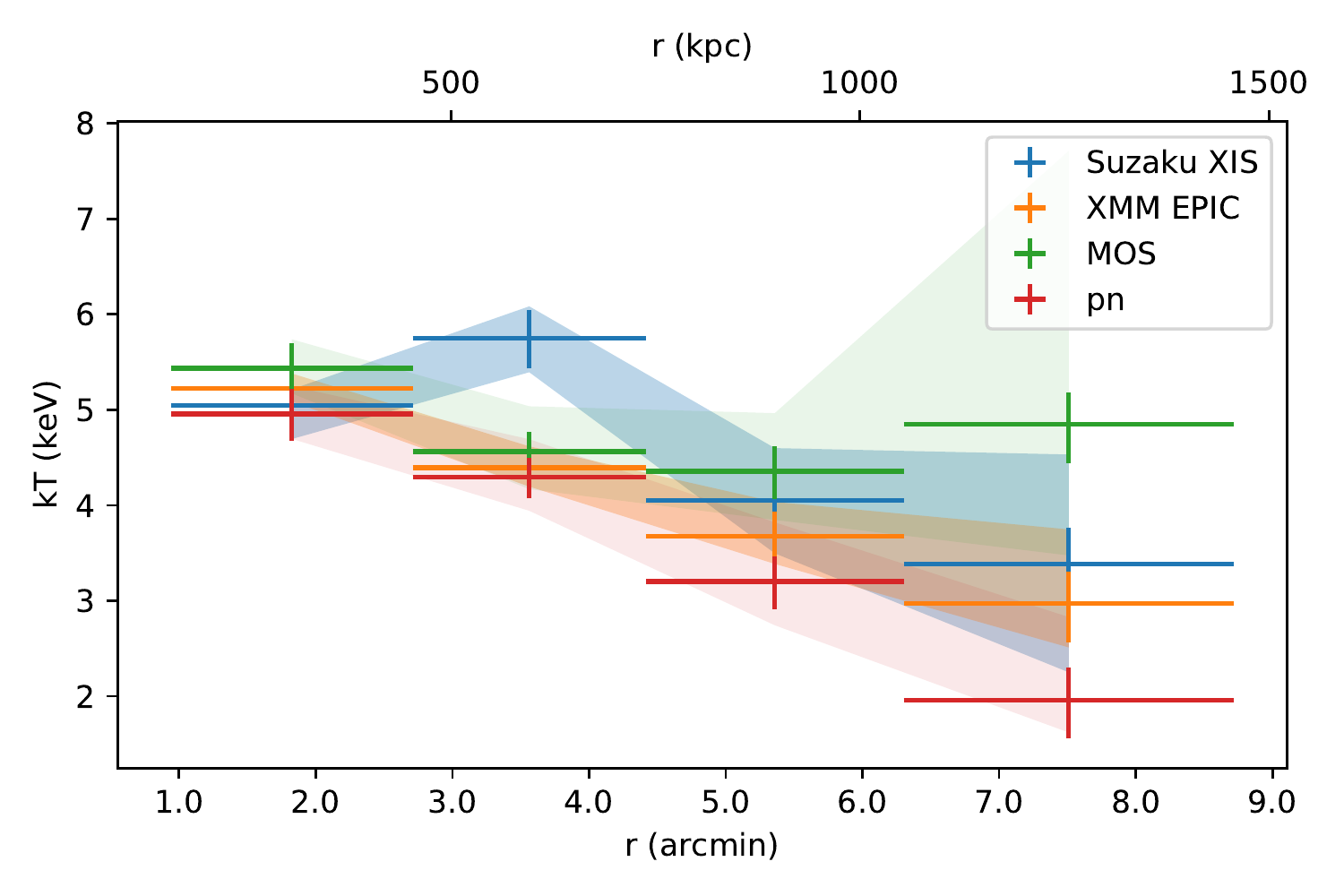}}
\caption{Temperature profiles of both \emph{Suzaku} and \emph{XMM-Newton} in the \emph{Suzaku} SE selection region. Filled bands of each profile represent the major systematics, i.e. SP for \emph{XMM-Newton} and CXB$+$GH for \emph{Suzaku}.}
\label{fig:se_crosscheck}
\end{figure}

Because \emph{Suzaku} XIS has a lower instrumental background and doesn't suffer from soft proton contamination due to its low orbit, its temperature measurements in faint cluster outskirts can be considered more reliable than those from \emph{XMM-Newton} EPIC. Thereby, we use the \emph{Suzaku} temperature profiles to cross-check the validity of the \emph{XMM-Newton} temperature profile and verify our soft proton modelling approach. We use the \emph{Suzaku} SE selection region for the cross-check because the S and SW regions cover the missing MOS1 CCD. We extract EPIC spectra from the exact same regions as the XIS spectra except for the point source exclusion regions. All the spectra are fitted by the method described in Sects. \ref{section:spectrum-xmm} and \ref{section:spectrum-suzaku}. We plot \emph{Suzaku} and \emph{XMM-Newton} temperature profiles as well as profiles from only MOS and pn in Fig. \ref{fig:se_crosscheck}. Except for the second subregion from the cluster centre, the MOS temperatures are globally higher than \emph{Suzaku} XIS temperatures, which are themselves higher than pn temperatures. The total EPIC temperatures are in agreement with XIS temperatures within the systematics. 

\section{Results}
\label{sec:results}

\subsection{Properties of surface brightness discontinuities}
We calculate surface brightness profiles from each selection region shown in Fig. \ref{fig:abell3411}. We use the double power law model introduced in Sect. \ref{sec:imaging} to fit surface brightness profiles. Because \emph{Chandra} has a narrow PSF, we first fit \emph{Chandra} profiles to obtain precise $r_\mathrm{edge}$s. For \emph{XMM-Newton} profiles, we convolve a $\sigma=0.1\arcmin$ gaussian kernel to the model to mimic the PSF effect. We fix $r_\mathrm{edge}$ for the \emph{XMM-Newton} profile fitting based on the value determined with \emph{Chandra}. We compare the C-statistic value when fixing C to the \emph{Chandra} result or allowing it to be free in the fit.

\begin{figure*}[ht]
\begin{tabular}{ccc}
\resizebox{0.32\hsize}{!}{\includegraphics{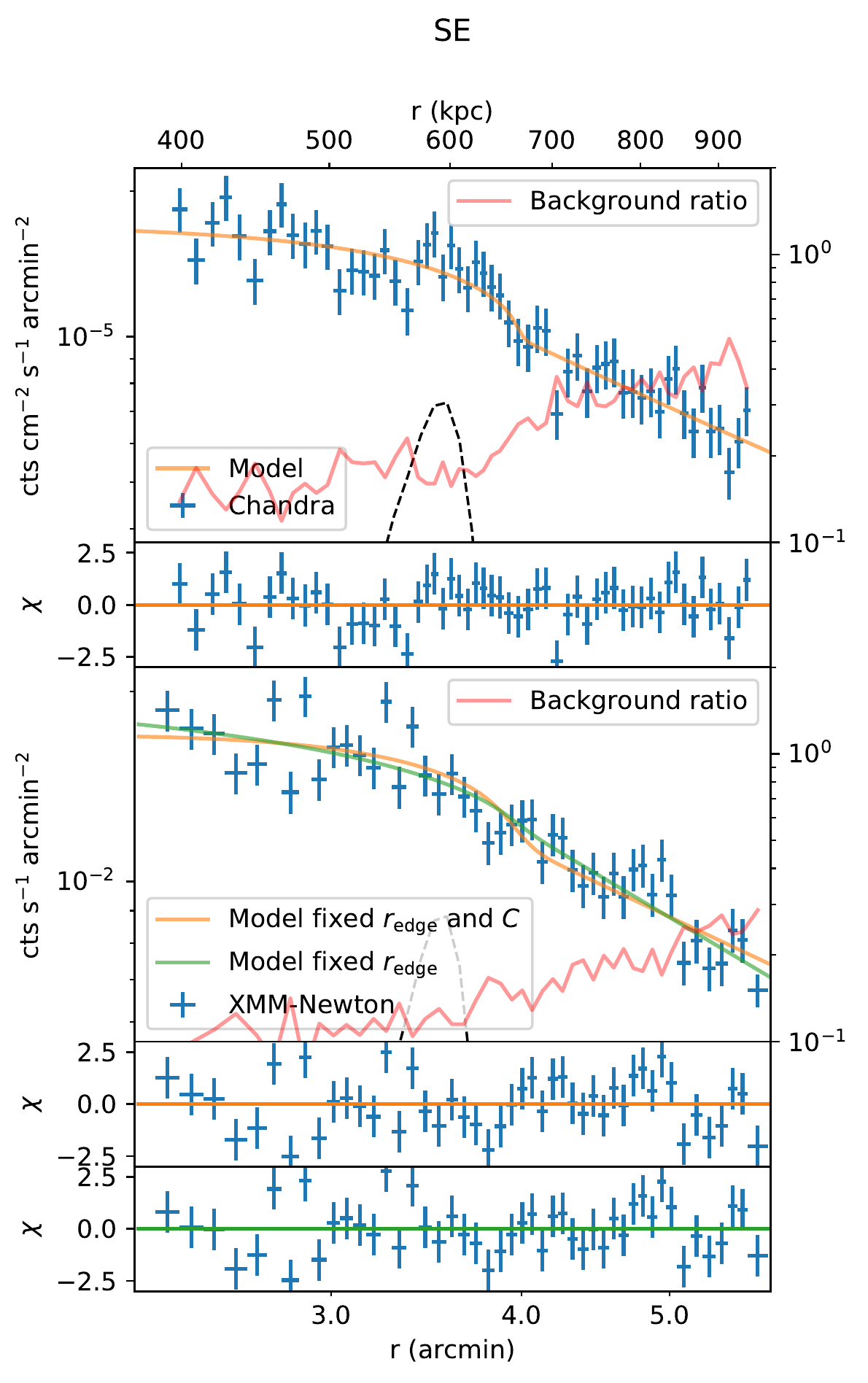}}&
\resizebox{0.32\hsize}{!}{\includegraphics{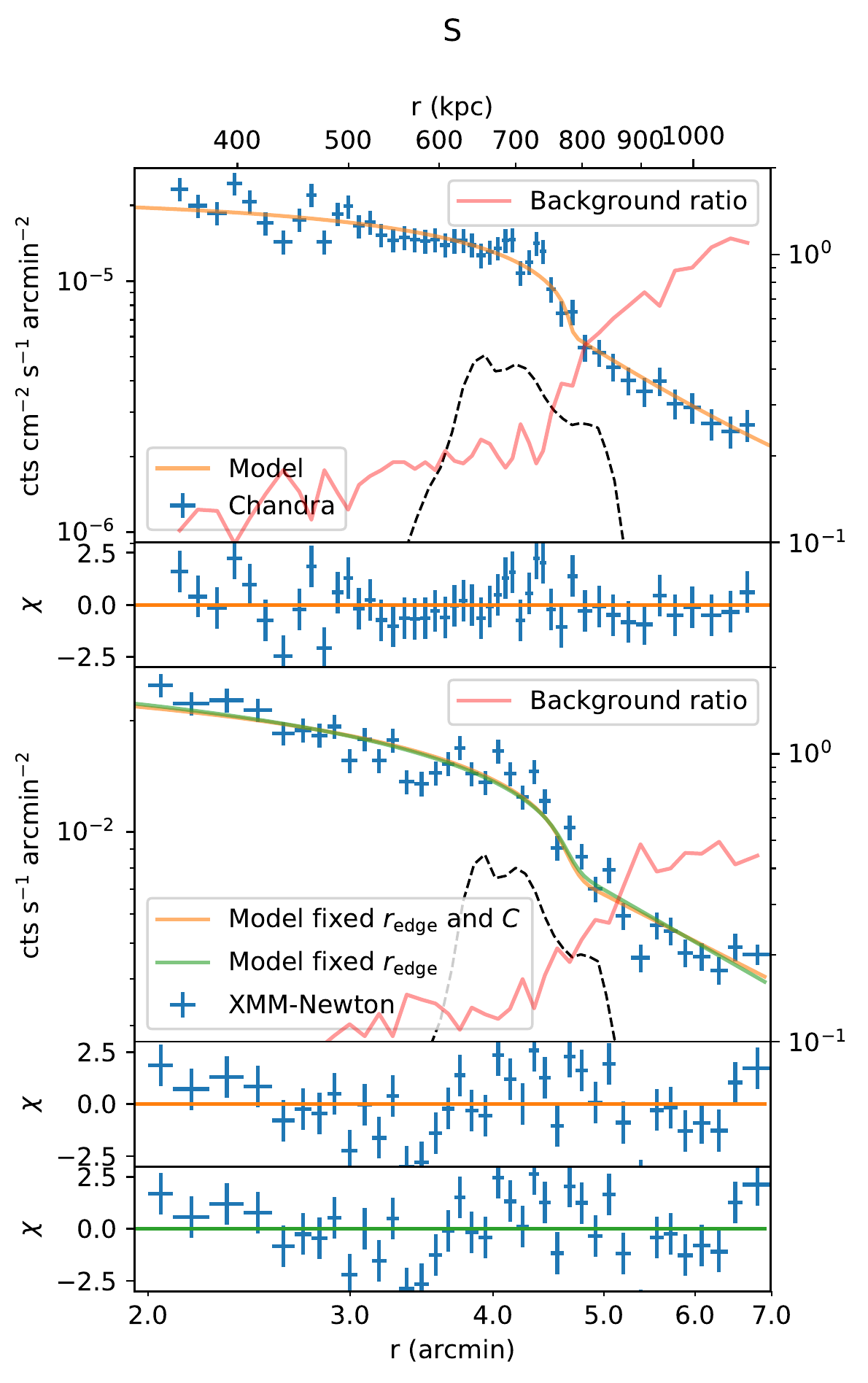}}&
\resizebox{0.32\hsize}{!}{\includegraphics{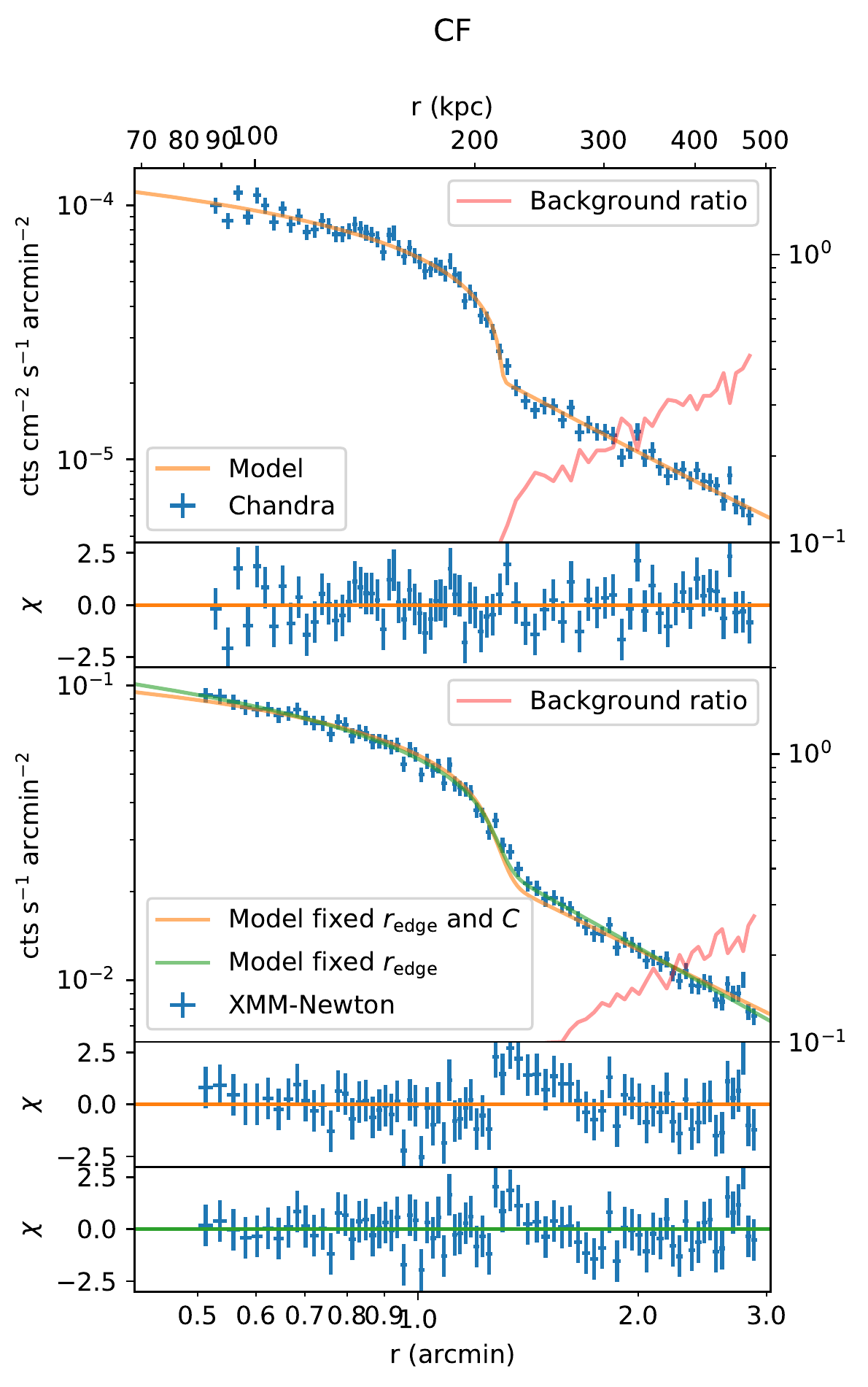}}\\
\end{tabular}
\caption{The surface brightness profile fitting results of south east (SE), south (S) and cold front (CF) regions. The upper panel is the \emph{Chandra} surface brightness. The lower panel is the \emph{XMM-Newton} surface brightness profile fitted by fixed and free $C$ parameters. All \emph{XMM-Newton} models are smoothed by a $\sigma=0.1\arcmin$ gaussian function. Red lines indicate the ratio between the subtracted FWC background counts and the remaining signal, which include the ICM, X-ray background, and soft proton contamination. In regions where this ratio is higher than one, the FWC background dominates. Black dashed lines are the radio surface brightness profiles in an arbitrary unit. }
\label{fig:sbfit}
\end{figure*}

\begin{figure*}
\begin{tabular}{cc}
\resizebox{0.45\hsize}{!}{\includegraphics{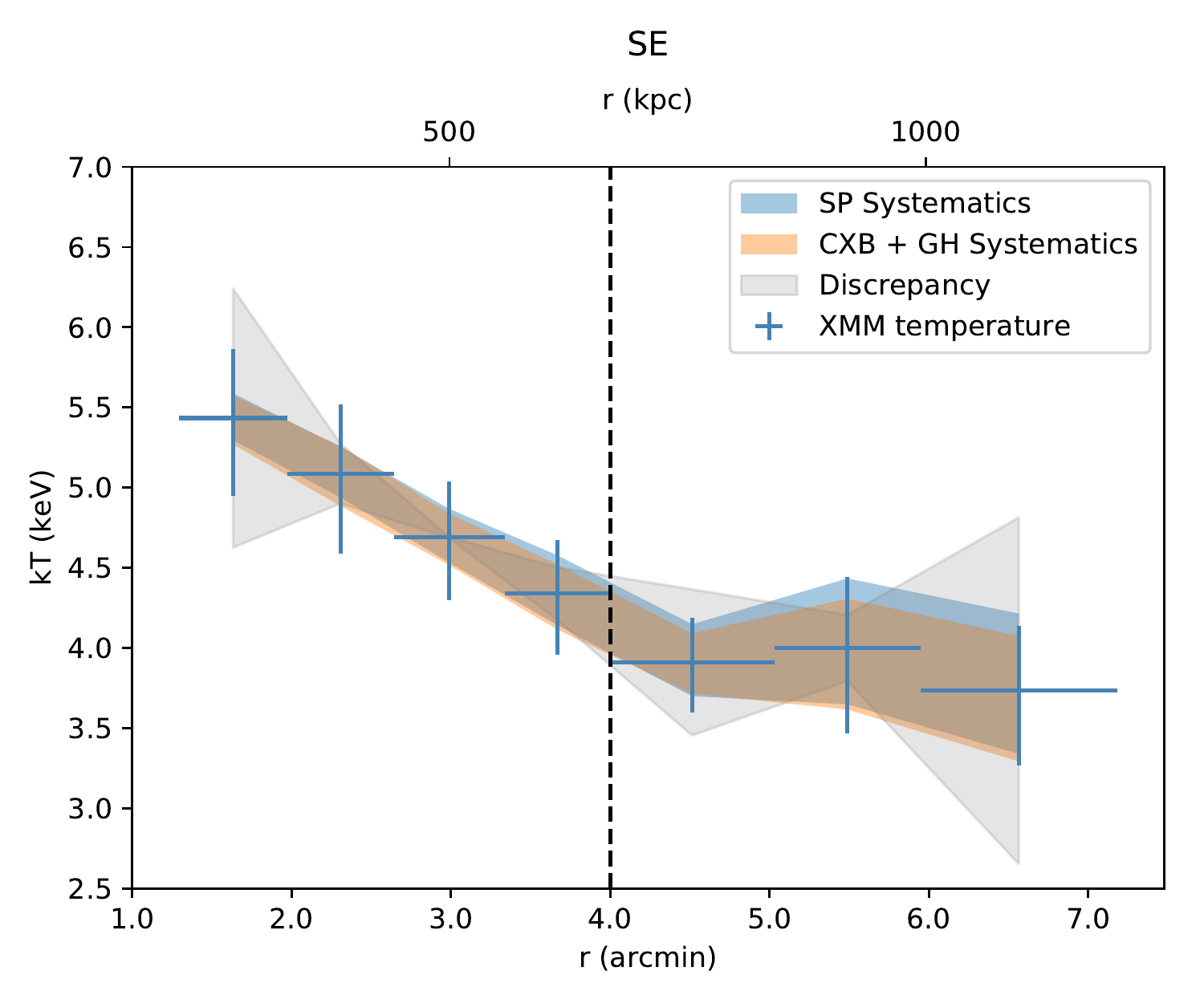}}&
\resizebox{0.45\hsize}{!}{\includegraphics{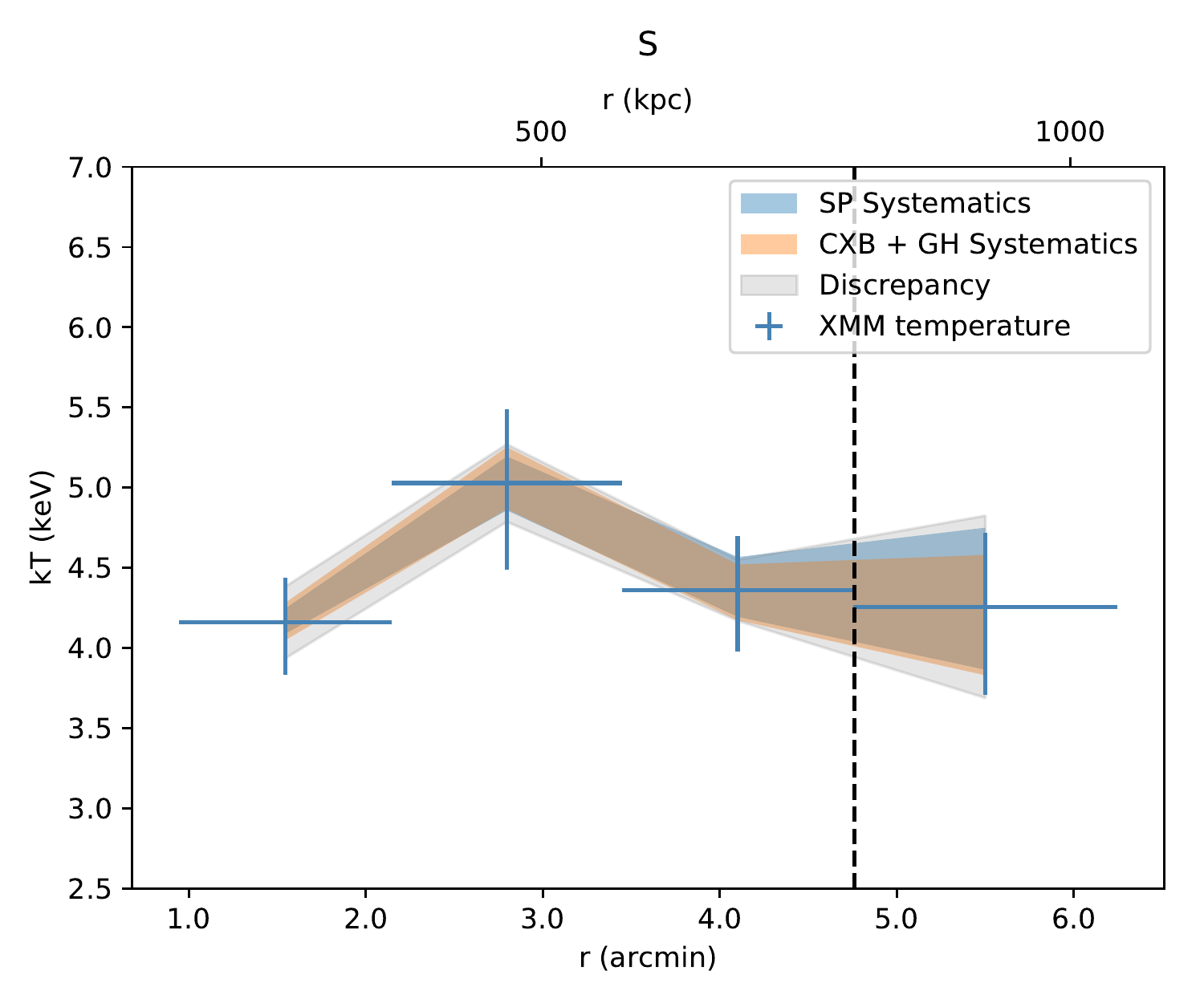}}\\
\multicolumn{2}{c}{\resizebox{0.45\hsize}{!}{\includegraphics{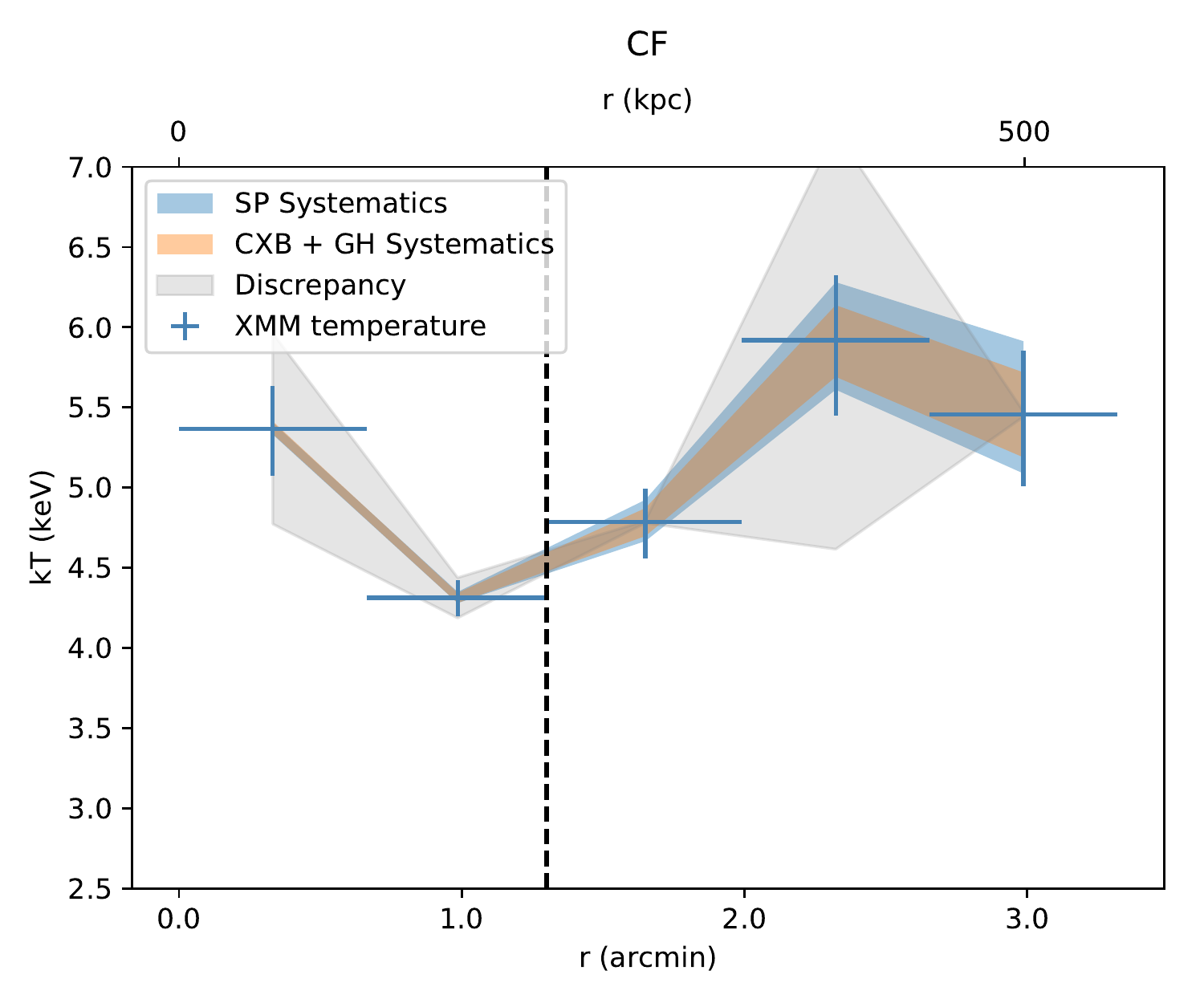}}}\\
\end{tabular}
\caption{The \emph{XMM-Newton} temperature profile of each region. Dashed lines indicate edge locations fitted from surface brightness profiles. Grey bands indicate the temperature discrepancy between MOS and pn.}
\label{fig:tprofile_xmm}
\end{figure*}

Surface brightness profiles and fitted models are plotted in Fig. \ref{fig:sbfit}, fitted parameter values as well as fitting statistics are listed in Table \ref{table:sbfit}. There is a systematic offset between the density jumps measured with \emph{Chandra} and \emph{XMM-Newton}. We use the best-fit $r_\mathrm{edge}$ as the location of the shock/cold front to extract spectra. We also split both the high and low-density sides into several bins when extracting spectra. Temperature profiles are plotted in Fig. \ref{fig:tprofile_xmm}.

\begin{table*}
\caption{Best fit parameters and statistics of surface brightness profiles in Fig. \ref{fig:sbfit}}
\label{table:sbfit}
\centering
\begin{tabular}{@{\extracolsep{4pt}}lccccc}
\hline\hline
&\multicolumn{3}{c}{\emph{Chandra}}&\multicolumn{2}{c}{\emph{XMM-Newton}}\\
\cline{2-4}\cline{5-6}
& $r_\mathrm{edge}$ ($\arcmin$) & $C$ & C-stat / d.o.f. & $C$ &C-stat / d.o.f. \tablefootmark{a}\\
\hline
SE & $4.00\pm0.10$ & $1.33\pm0.13$ & $54.4/54$ & $1.09\pm0.08$ & $70.3/43$\\
S & $4.76\pm0.05$ & $1.74\pm0.15$ & $55.1/44$ & $1.45\pm0.10$ & $86.2/34$\\
CF & $1.30\pm0.01$ & $2.00\pm0.06$ & $75.2/74$ & $1.74\pm0.05$ & $61.3/74$\\
\hline
\end{tabular}
\tablefoot{
\tablefoottext{a} {Fixed $r_\mathrm{edge}$ based on the \emph{Chandra} model.}
}
\end{table*}

\subsubsection{South East}\label{sec:xmm-se}
At the previously reported shock front, the compression factor fitted with our selection region from the \emph{Chandra} profile is identical to \citet{2017NatAs...1E...5V}'s result $C=1.3\pm0.1$, and is slightly higher than \citet{2019ApJ...887...31A}'s result $C=1.19\substack{+0.21\\-0.13}$, but within $1\sigma$ uncertainty. However, it is hard to find this feature in the \emph{XMM-Newton} profile. Fitting with fixed $r_\mathrm{edge}$ and $C$, we obtain C-stat / d.o.f $=76.4/44$. If we free the $C$ parameter, the fitted $C_\mathrm{XMM-Newton}=1.09\pm0.08$, which means the data is consistent with the lack of a density jump, but is consistent with \citet{2019ApJ...887...31A}'s result. The reason that we don't detect an edge in the \emph{XMM-Newton} profile could be the missing pixels around the edge. The radio relic is located very close to bad pixel columns of MOS2, and a CCD gap of pn. The temperature profile (the top left panel in Fig. \ref{fig:tprofile_xmm}) drops linearly and then flattens at larger radii. The temperature at the bright side of the edge is higher than at the other side. Hence, we rule out the possibility of this edge to be a cold front.

\subsubsection{South}
In the south region, a significant surface brightness jump is seen in both \emph{Chandra} ($C_\mathrm{Chandra}=1.74\pm0.15$) and \emph{XMM-Newton} ($C_\mathrm{XMM-Newton}=1.45\pm0.10$) profiles. A simple spherically symmetric double power-law density model cannot fit the \emph{Chandra} profile perfectly. It is a sudden jump with flat or even increasing surface brightness profile on the high-density side. There is an excess above the best-fit model at the edge. The temperature is almost identical across the edge, which is not a typical shock or cold front. We discuss this edge in Sect. \ref{section:discussion_south}.

\subsubsection{Cold front}
The cold front surface brightness profile can be well modelled by the double power-law density model. The density ratio from \emph{Chandra} observation $C_\mathrm{Chandra}=2.00\pm0.06$ is higher than that from the \emph{XMM-Newton}, which is $C_\mathrm{XMM-Newton}=1.74\pm0.05$. Similar to the southern edge, even if we account for the PSF of \emph{XMM-Newton}, the density jump measured by \emph{XMM-Newton} is smaller than that determined using \emph{Chandra}. In addition, the inner power-law component of the density profiles is steeper when measured with XMM-Newton than with Chandra. The energy dependence of the vignetting function and of the effective area (and their uncertainties) can affect this inner slope which, in turn, is correlated with the density jump (a steeper inner power-law leads to a smaller compression factor). This may contribute to the observed differences. The temperature profile confirms that it is a cold front. The temperature reaches the minimum before the cold front and then rises until $r=3\arcmin$. 

\subsection{Global temperature}
We extract spectra from the region with $r_{500}=1.3$ Mpc \citep{2019ApJ...887...31A} to obtain the global temperature. Although we miss MOS1 CCD3 and 6, most of the flux is in the centre CCD, which means our result will not be significantly biased by the missing CCDs. The best-fit temperature is $kT_{500}=4.84\pm0.04\pm0.19$ keV, where the second error item represents the systematics uncertainty. Temperatures of individual detectors are listed in Table \ref{table:t500}. The $kT_\mathrm{500,MOS}$ is about 0.1 keV higher than $kT_\mathrm{500,pn}$. These two measurements agree within their $1\sigma$ uncertainty interval, which is dominated by the systematics from the soft proton model. Compared with \citet{2019ApJ...887...31A}'s result, $kT_{500}=6.5\pm0.1$ keV, the $kT_{500}$ in our work is much lower. 

\begin{figure}
\resizebox{\hsize}{!}{\includegraphics{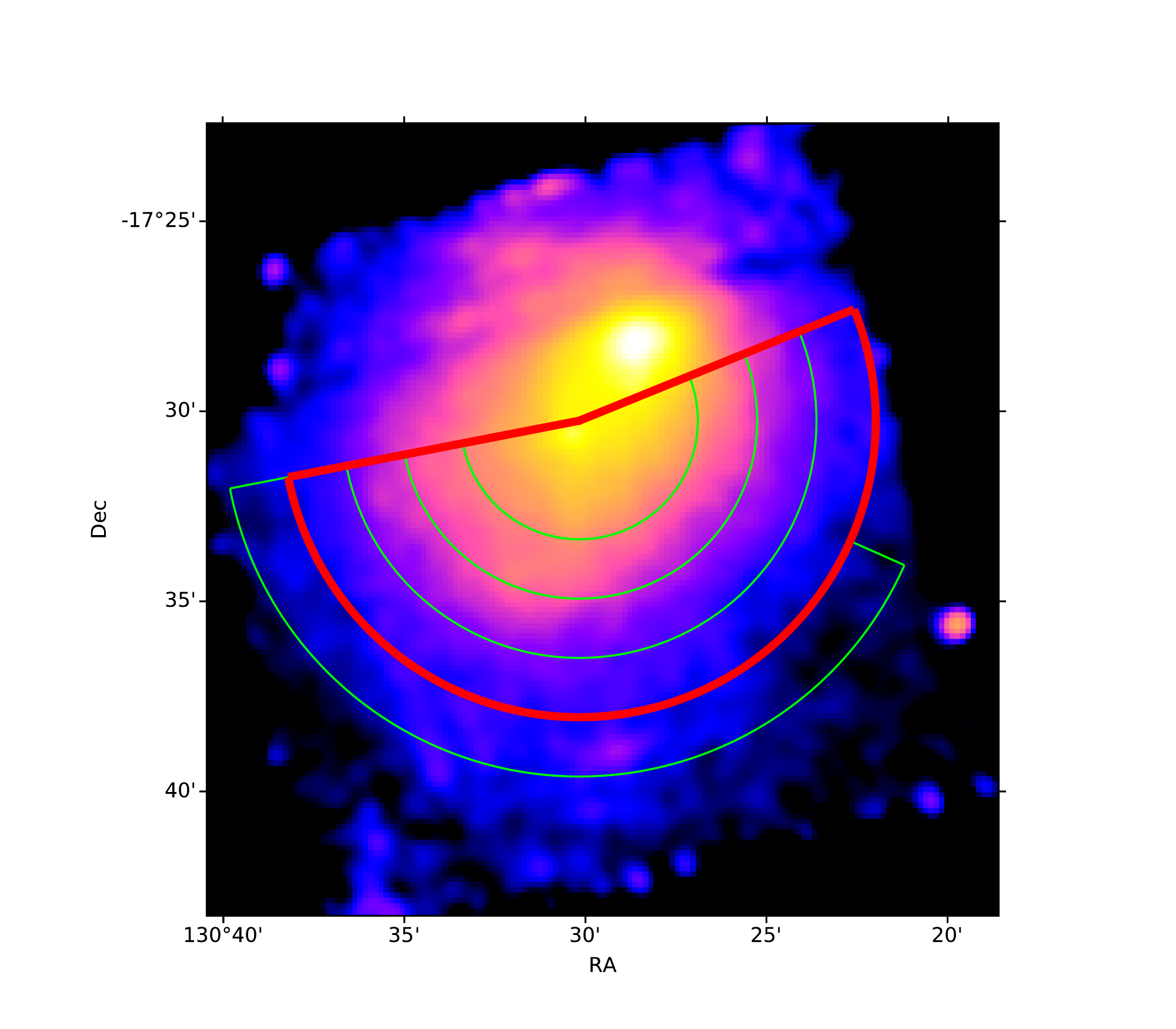}}
\caption{The \emph{Suzaku} semicircle $r_{500}$ selection region (red) and radial bins (green).}
\label{fig:r500_suzaku}
\end{figure}

We have also used \emph{Suzaku} data to check the global temperature. \emph{Suzaku} doesn't suffer from soft proton contamination, and its NXB level is lower than \emph{XMM-Newton}, making it a valuable tool to check our \emph{XMM-Newton} analysis. The \emph{Suzaku} observation doesn't cover the whole $r_{500}$ area. To avoid the missing XIS0 strip, we extract spectra from the south semicircle (see the red region in Fig. \ref{fig:r500_suzaku}). The best-fit results with all detectors as well as with the only front-illuminated (FI, XIS0 + XIS3) and back-illuminated (BI, XIS1) CCDs are listed in Table \ref{table:t500}. The best-fit temperature is $kT_{500}=5.17\pm0.07\pm0.13$ keV, which is slightly higher than the \emph{XMM-Newton}'s result. 

\begin{table}
\caption{$kT_{500}$ from \emph{XMM-Newton} and \emph{Suzaku}.}
\label{table:t500}
\centering
\begin{tabular}{lcc}
\hline
\hline
& $kT_{500}$ (keV) & $\sigma_\mathrm{sys}$\tablefootmark{a} (keV)\\
\hline
\emph{XMM-Newton} & $4.84\pm0.04$ & 0.19\\
MOS & $4.92\pm0.06$ &0.37\\ 
pn & $4.80\pm0.06$ &0.40\\
\hline
\emph{Suzaku} & $5.17\pm0.07$ & 0.13 \\
FI & $5.36\pm0.11$ & 0.13 \\
BI & $4.97\pm0.12$ & 0.13 \\
\hline
\end{tabular}
\tablefoot{\\
\tablefoottext{a} {For \emph{XMM-Newton} spectra, the major systematics is the soft proton component. For \emph{Suzaku} spectra, the systematics is the combined from the CXB and GH component.}
}
\end{table}

\begin{figure*}[t!]
\begin{tabular}{cc}
\resizebox{0.45\hsize}{!}{\includegraphics{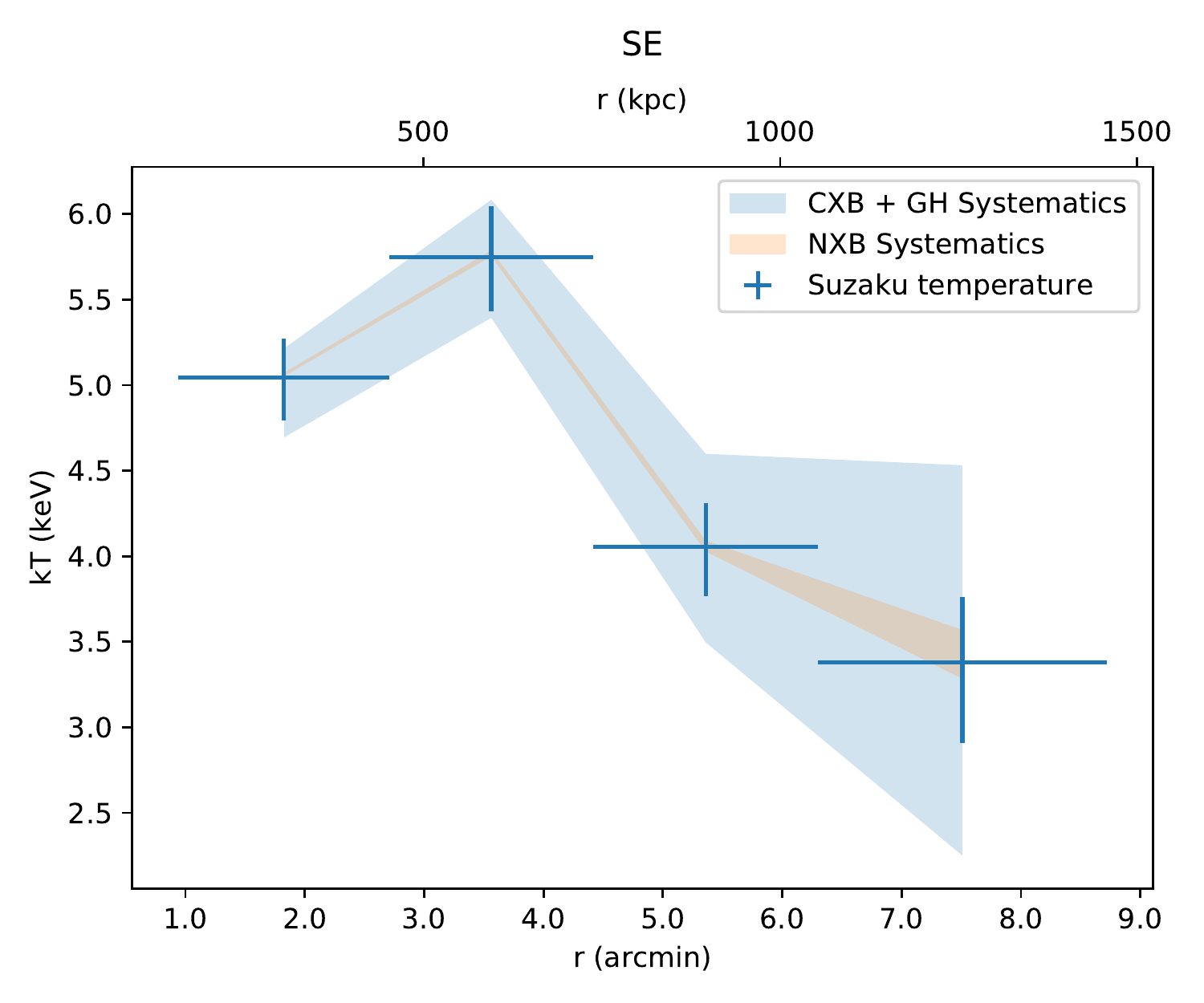}}&
\resizebox{0.45\hsize}{!}{\includegraphics{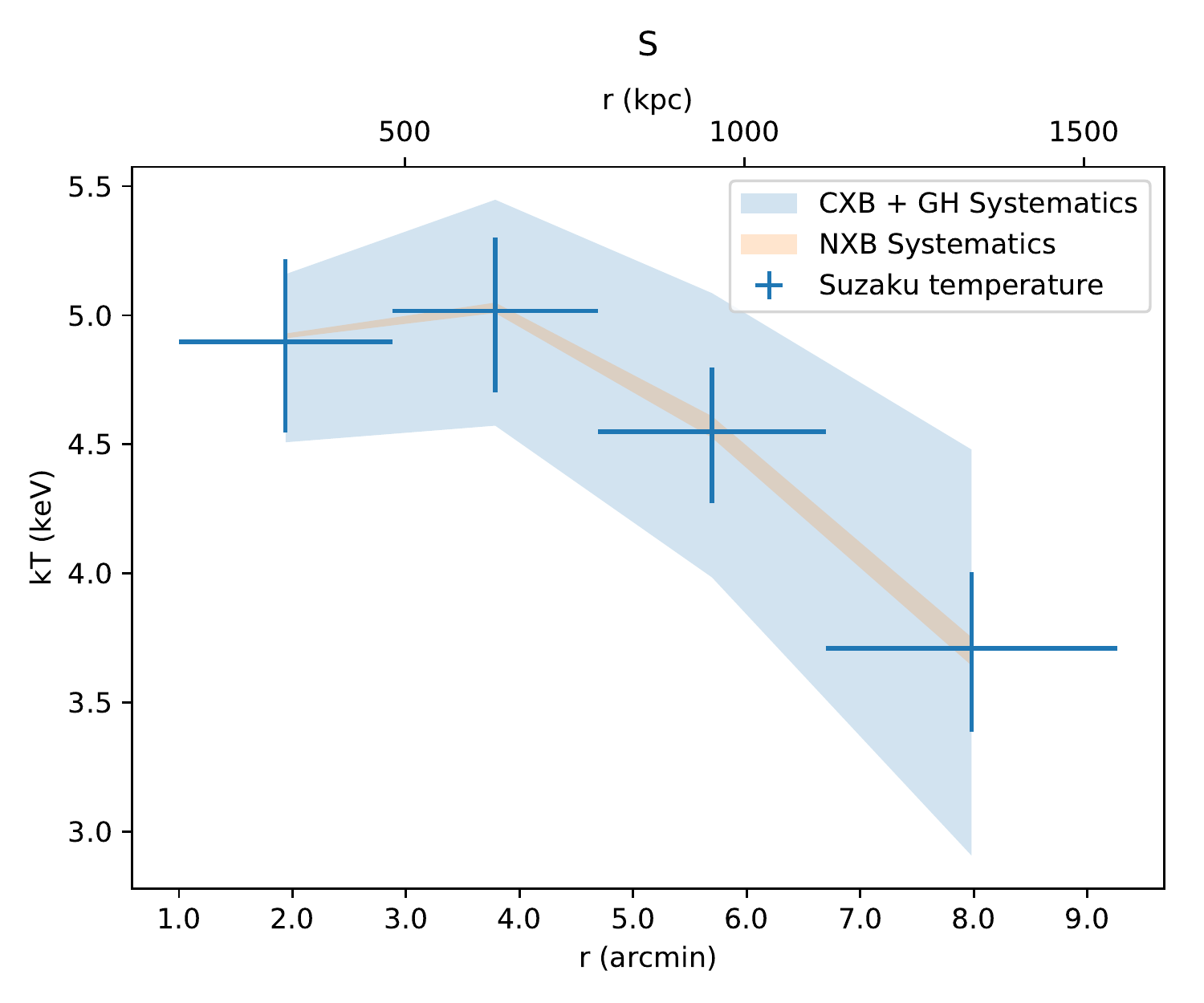}}\\
\resizebox{0.45\hsize}{!}{\includegraphics{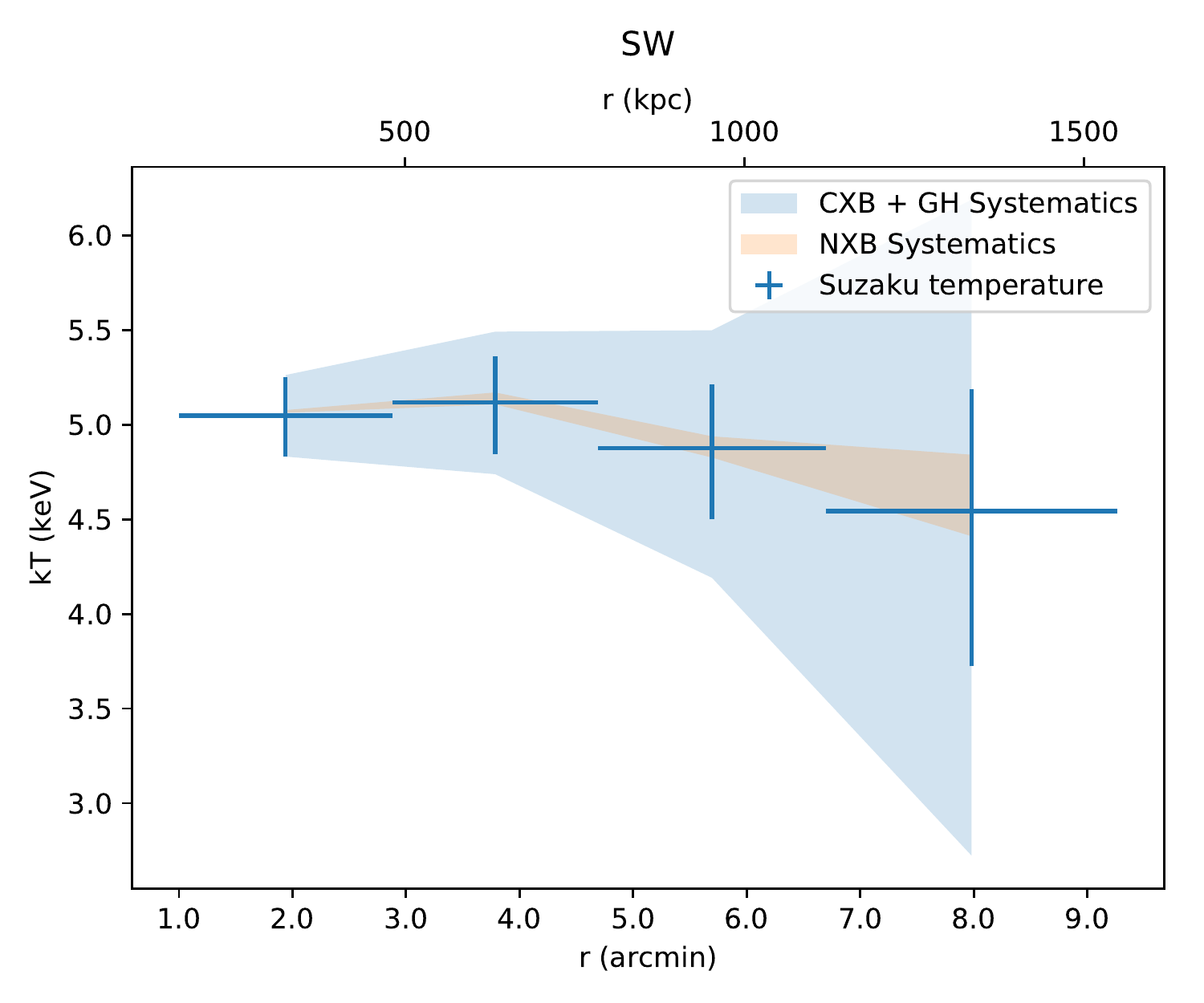}}&
\resizebox{0.45\hsize}{!}{\includegraphics{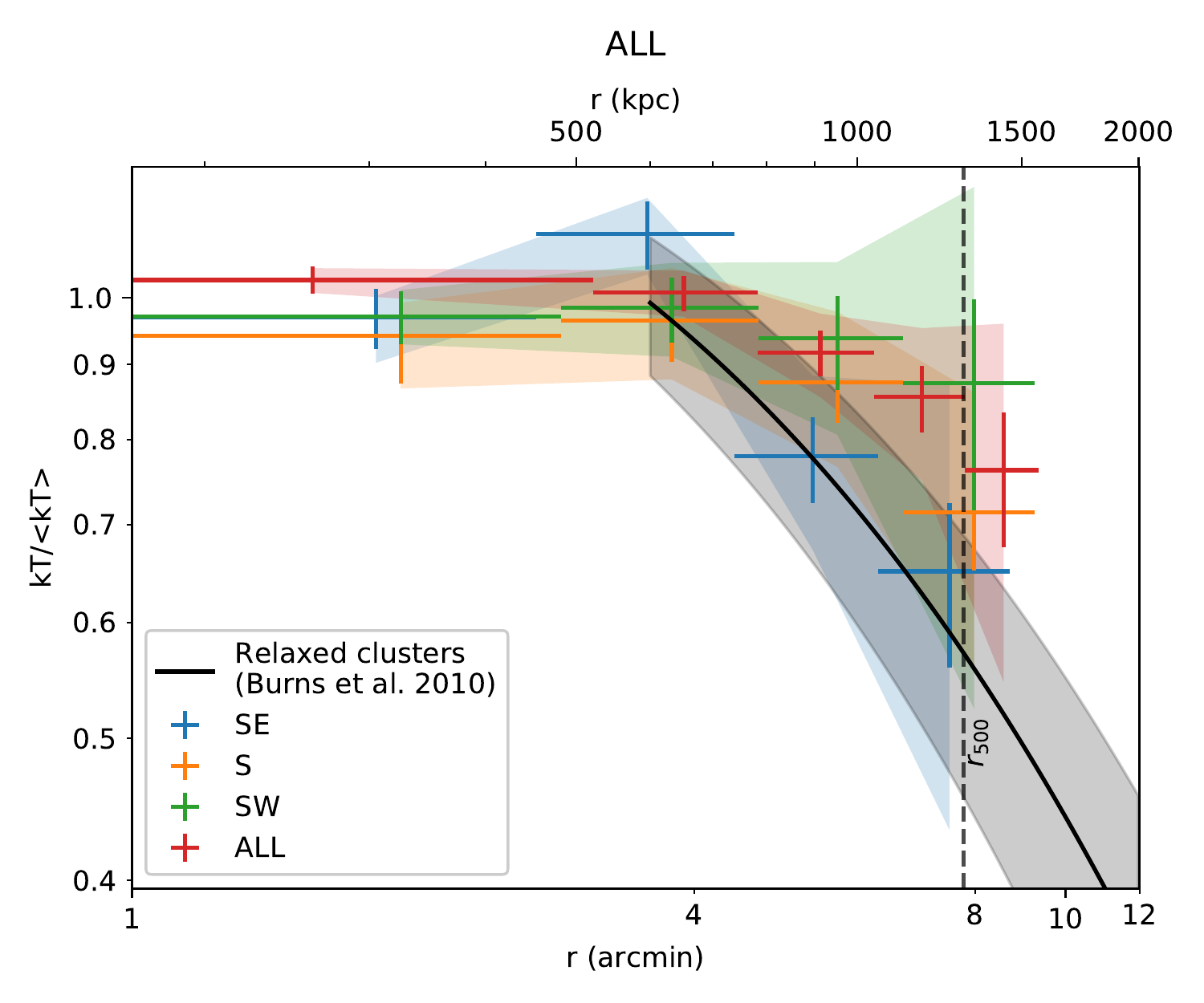}}\\
\end{tabular}
\caption{The \emph{Suzaku} temperature profiles of the SE, S, and SW regions, as well as the comparison to the relaxed temperature profile in the outskirts predicted by numerical simulations.}
\label{fig:tprofile_suzaku}
\end{figure*}

\subsection{Temperature profiles to the outskirts}
Individual \emph{Suzaku} temperature profiles of three sectors are shown in Fig. \ref{fig:tprofile_suzaku}. Apart from three individual directions, we split the \emph{Suzaku} $r_{500}$ region into four annuli (the green regions in Fig. \ref{fig:r500_suzaku}), and define another bin outside of $r_{500}$. We plot all four profiles together with a typical relaxed cluster outskirt temperature profile \citep{2010ApJ...721.1105B}, where we take $\left<kT\right>=5.2$ keV. Burns' curve agrees with \emph{Suzaku} observations of relaxed clusters remarkably \citep{2011PASJ...63S1019A,2013SSRv..177..195R}. For our data, at $r_{500}$, the south temperature profile agrees with the profile of \citet{2010ApJ...721.1105B}. Other three profiles are marginally higher than the typical relaxed cluster temperature profile but within $1\sigma$ systematics. In \citet{2010ApJ...721.1105B}'s work, $\left<kT\right>$ is the averaged temperature between 0.2 to 2.0 $r_{200}$. Because our \emph{Suzaku} observation only covers the $r_{500}$ area, the actual $\left<kT\right>$ can be slightly lower than the value we use. In that case, the south temperature profile can also be marginally higher than the Burns' profile. This cluster is undergoing a major merger, and our results show that the temperature in the outskirts has been disturbed.

\section{Discussion}
\label{sec:discussion}
\subsection{$T_{500}$ discrepancy}
Our measurements of $kT_{500}$ are lower than the result of \emph{Chandra} data. The cross-calibration uncertainties between \emph{XMM-Newton} EPIC and \emph{Chandra} ACIS may be the major reason of this discrepancy. Using the scaling relation of temperatures between EPIC and ACIS $\log kT_\mathrm{EPIC}=0.0889\times\log kT_\mathrm{ACIS}$ \citep{2015A&A...575A..30S},
a 6.5 keV ACIS temperature corresponds to a 5.3 keV EPIC temperature, which is close to our measurement. By contrast, the temperature discrepancy between \emph{XMM-Newton} EPIC and \emph{Suzaku} XIS is relatively small. This discrepancy of $8\%$ is slightly larger than the value from the \emph{Suzaku} XIS and \emph{XMM-Newton} EPIC-pn cross-calibration study  \citep[$5\%$, ][]{2013A&A...552A..47K}.  Because the \emph{Suzaku} extraction region does not cover the cold front, the reported \emph{Suzaku} temperature may be higher than the average value within the entire $r_{500}$ region, explaining this difference.

With our temperature results, we use the $M_{500}-kT_{X}$ relation $h(z)M_{500}=10^{14.58}\times(kT_X/5.0)^{1.71}$ $M_\sun$ \citep{2007A&A...474L..37A} to roughly estimate the mass of the cluster. The $kT_X$ is the temperature from $0.1-0.75r_{500}$. We don't exclude the inner $0.1r_{500}$ part because it is not a relaxed system, and there is no dense cool core in the centre. For $kT_X=5.0$ keV, the $M_{500}-T_{X}$ relation suggests a mass $M_{500}=5.1\times10^{14}$ $M_\sun$. This is less than that from the Planck Sunyaev-Zeldovich catalog, $M_\mathrm{SZ}=(6.6\pm0.3)\times10^{14}$ $M_\sun$ \citep{2016A&A...594A..27P}. However, this underestimation is not surprising. Since the source is undergoing a major merger, the kinetic energy of two sub-halos is still being dissipated into the thermal energy of the ICM. Once the system relaxes, the $kT_X$ would be higher than in the current epoch.

\subsection{Shock properties}
The shock Mach number can be calculated by the Rankine-Hugoniot condition \citep{1959flme.book.....L} either from the density jump or from the temperature jump,
\begin{align}
\mathcal{M}&=\left[\frac{2C}{\gamma+1-C(\gamma-1)}\right]^2\\
\frac{T_1}{T_2}&=\frac{(\gamma+1)/(\gamma-1)-C^{-1}}{(\gamma+1)/(\gamma-1)-C},
\end{align}
where $C$ is the compression factor across the shock, and $\gamma=5/3$ if we assume the ICM is an ideal gas. Because the systematics from the CXB and GH are gaussian, we directly propagate them into the statistical error when estimating the Mach number uncertainty. However, the soft proton systematics is not gaussian, so we use the measured temperature, and the temperature obtained by varying the soft proton component within their $\pm1\sigma$ uncertainties determined in Appendix \ref{appendix:sp} to estimate the \emph{XMM-Newton} Mach number systematics. 

\begin{figure}
\resizebox{\hsize}{!}{\includegraphics{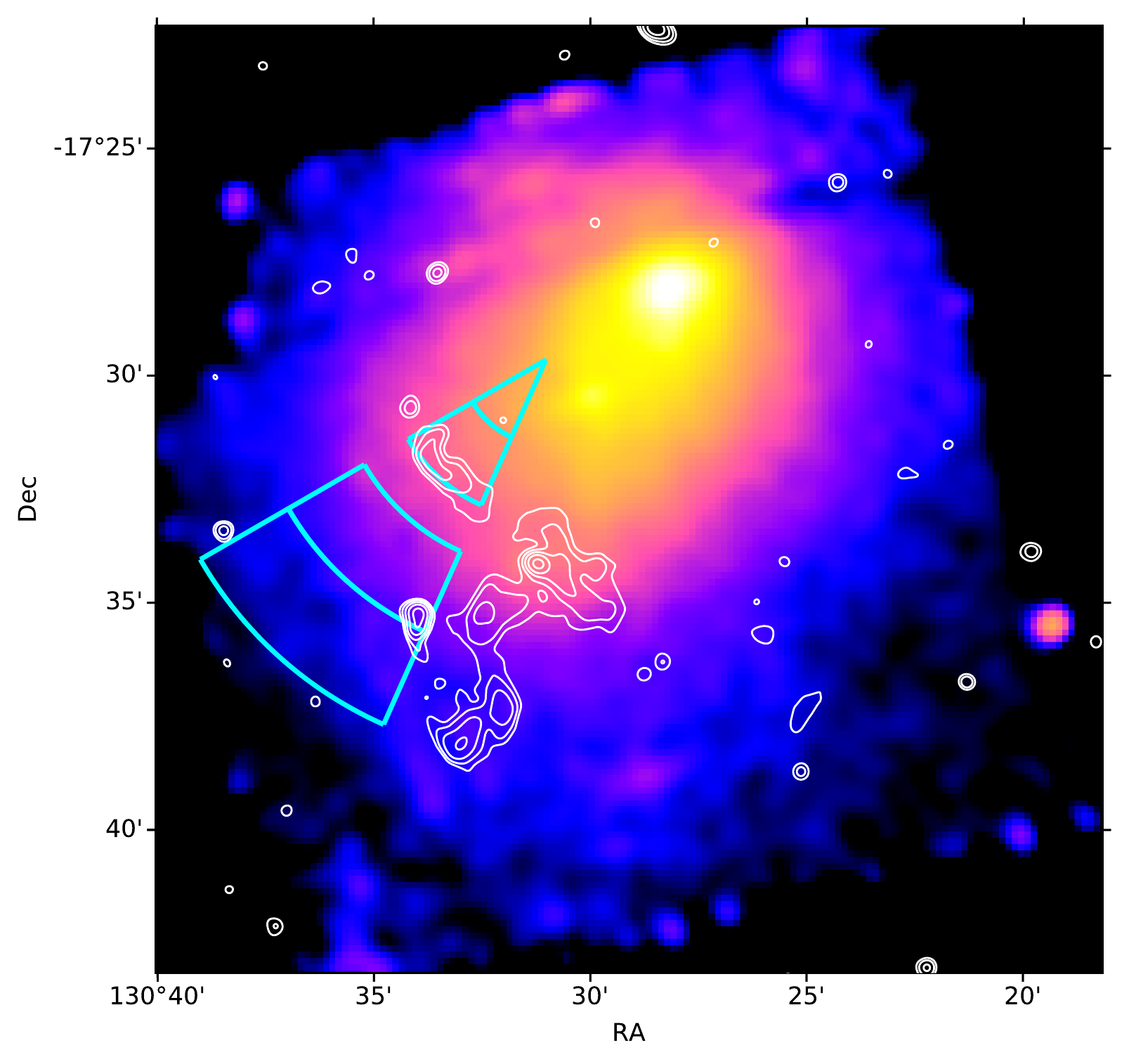}}
\caption{\emph{Suzaku} flux map and the cyan spectral extraction regions are based on radio morphology (white contours).}
\label{fig:suzaku_shock}
\end{figure}

\begin{figure}
\resizebox{\hsize}{!}{\includegraphics{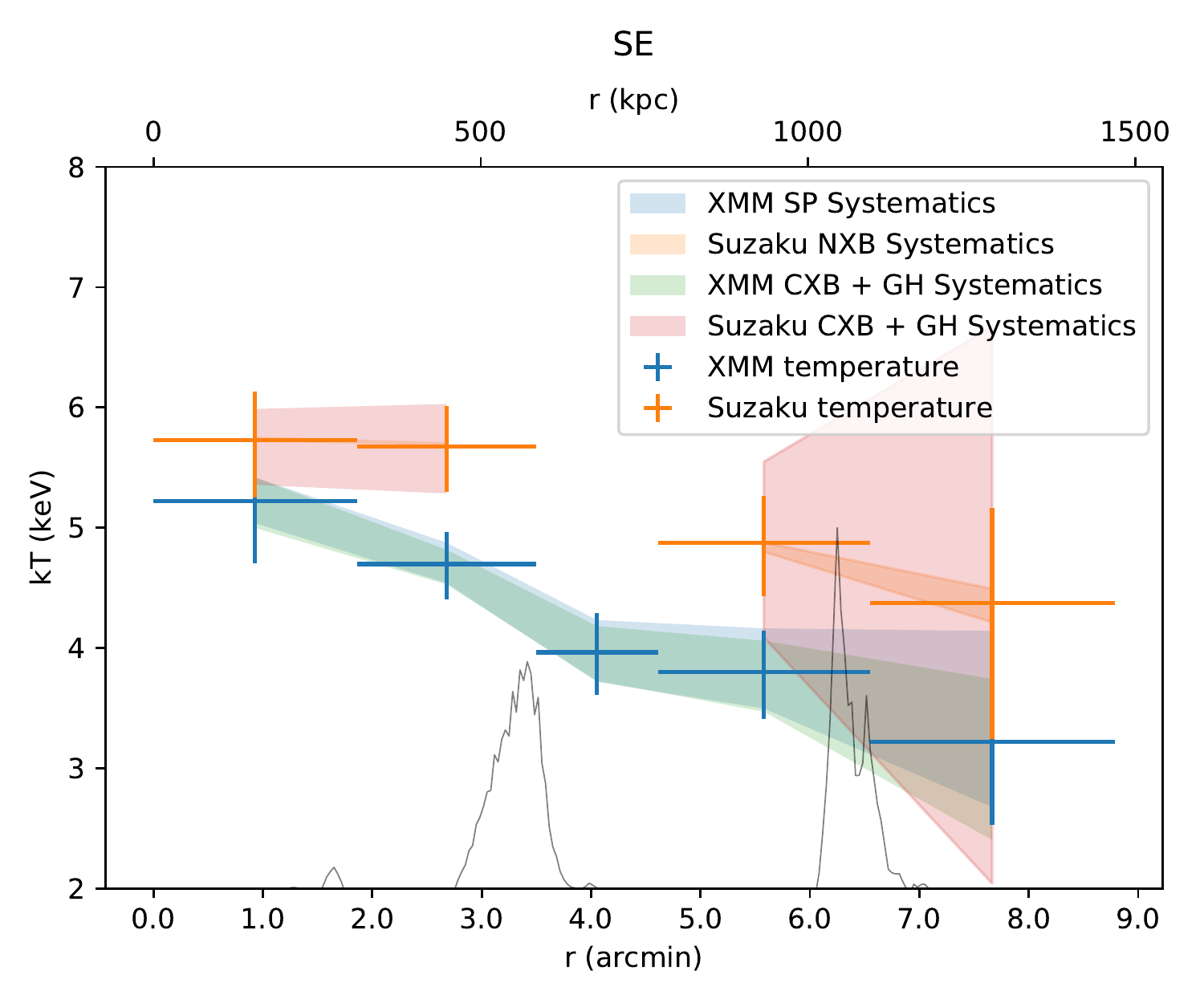}}
\caption{The \emph{Suzaku} and \emph{XMM-Newton} temperature profiles from the radio based selection regions in Fig. \ref{fig:suzaku_shock}. The radio surface brightness profile is plotted in black line.}
\label{fig:suzaku_shock_profile}
\end{figure}

The \emph{Suzaku} south-east sector covers the re-acceleration site, and we see a jump from the second point to the third point in that temperature profile. The \emph{Suzaku} spectral extraction regions are defined unbiasedly. We further inspect the temperature profile based on the radio morphology. The radio-based selection region is shown in Fig. \ref{fig:suzaku_shock}. We intentionally leave a $1.1\arcmin$ gap \citep{2015A&A...582A..87A} between the second and the third bin to avoid photon leakage from the brighter side. We plot both the \emph{XMM-Newton} and \emph{Suzaku} temperature profiles of this sector in Fig. \ref{fig:suzaku_shock_profile}. Because \emph{XMM-Newton} has a much smaller PSF than \emph{Suzaku}, we can use the spectrum from the gap.

\begin{figure}
\resizebox{\hsize}{!}{\includegraphics{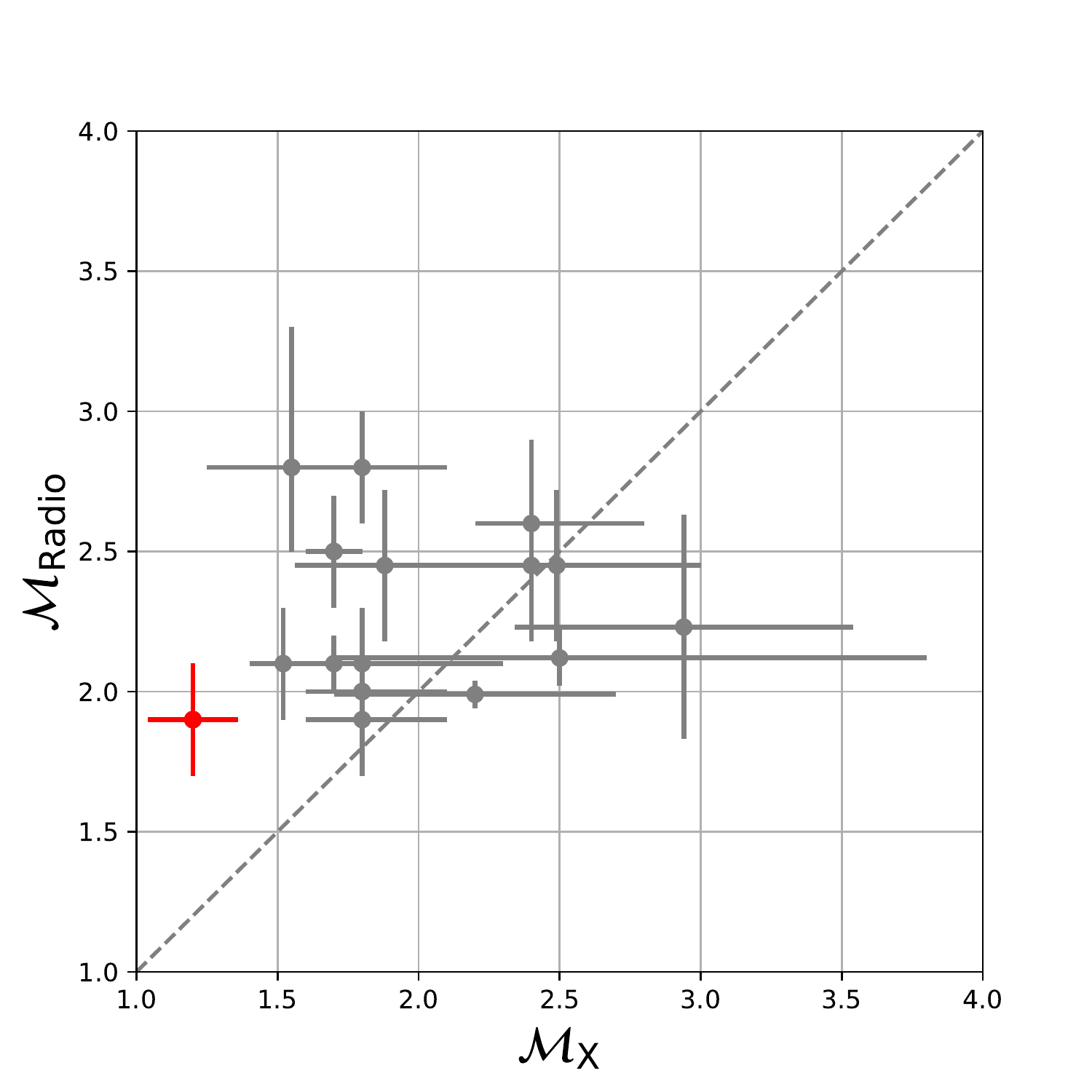}}
\caption{Shock Mach numbers derived from radio spectral index ($\mathcal{M}_\mathrm{Radio}$) against those from the ICM temperature jump ($\mathcal{M}_\mathrm{X}$). The data points of previous studies (grey) are adapted from the Fig. 22 in \citet{2019SSRv..215...16V}. The red point is the south western shock in Abell 3411-3412.}
\label{fig:mach_compare}
\end{figure}

There is a systematic offset between \emph{Suzaku} and \emph{XMM-Newton}. The \emph{Suzaku} temperature is globally higher than the \emph{XMM-Newton} temperature. Both profiles drop from the centre of the cluster to the outskirts. The new \emph{XMM-Newton} temperature profile is similar to the previous one in Sect. \ref{sec:xmm-se}. The temperature decreases from the centre of the cluster and flattens after the radio relic. We use temperatures across the radio relic to obtain the shock Mach number. As a comparison, we calculate the Mach number by the density jump fitted from the \emph{Chandra} surface brightness profile. Results are listed in Table \ref{table:mach}. From our spectral analysis, we confirm the Mach number of this shock is close to the value measured from the surface brightness profile fit. Results from all telescopes point to the value $\mathcal{M}_\mathrm{X}\sim1.2$. This shock is another case where the radio Mach number is higher than the X-ray Mach number (see Fig. \ref{fig:mach_compare}). Such a low Mach number supports the re-acceleration scenario. Note that our calculation does not account for the presence of a ``relaxed'' temperature gradient in the absence of a shock. This could further reduce the Mach number, but the conclusion that the re-acceleration mechanism is needed would remain robust.

\begin{table*}
\caption{The comparison of the south-east shock Mach number obtained from different instruments and methods. The second error in the \emph{XMM-Newton} measurement is from the soft proton systematics.}
\centering
\begin{tabular}{lll}
\hline\hline
Instrument&$\mathcal{M}_T$&$\mathcal{M}_\mathrm{SB}$\\
\hline
\emph{XMM-Newton} EPIC & $1.19\pm0.15\pm0.03$ & \\
\emph{Suzaku} XIS & $1.17\pm0.23$ & \\
\emph{Chandra} ACIS & & $1.20\pm0.07$\\
&& $1.13\substack{+0.14\\-0.08}$ \citep{2019ApJ...887...31A}\\
\hline
\end{tabular}
\label{table:mach}
\end{table*}

\subsection{The mystery of the southern edge}\label{section:discussion_south}
The density jump at the southern edge is strong, and \citet{2019ApJ...887...31A} claim it is a cold front from the sub-cluster Abell 3412. From our spectral fitting, the temperature inside and outside of the southern edge is $kT=4.36\pm0.34$ keV, and $kT=4.26\pm0.46$ keV, respectively. The projected temperature jump is $1.02\pm0.14$. From the surface brightness fitting, the de-projected density jump is $C=1.7\pm0.2$. This value corresponds to a de-projected temperature jump of $1.48\pm0.25$ under the assumption of Rankine-Hugoniot shock conditions, and a temperature jump $0.59\pm0.07$ under the assumption that it is a cold front in pressure equilibrium. Neither the shock scenario nor the cold front scenario matches the measured lack of temperature jump.

%Projection effects could underestimate the real temperature jump. 
To obtain the de-projected temperature jump, we simply assume that the spectrum from the high-density side is a double-temperature spectrum. The temperature of one of the components is the same as that from the low-density side. We assume that the discontinuity structure is spherically symmetric, and calculate the volume ratio between the intrinsic and projected components in the high-density side. We fit spectra from both sides simultaneously. For the high-density side spectrum, we couple one CIE temperature to that of the low-density spectrum. We also couple the normalisation of that component to that of the low-density spectrum with a factor of the volume ratio. We leave the other two temperature and normalisation parameters free. The de-projected temperature ratio is then $1.08\pm0.17$ with a systematic uncertainty 0.10. 
This value is $\sim1.3\sigma$ offset from the shock scenario but is $\sim2.6\sigma$ offset from the cold front scenario. Therefore, the temperature jump we measured prefers the shock scenario. Also, the pressure across the edge is out of equilibrium. The pressure jump implies the supersonic motion of the gas.

The presence of a huge density jump but a marginal temperature jump suggests an excess of surface brightness on the bright side of the edge. Also, the \emph{Chandra} surface brightness profile shows a tip beyond the best-fit double power law density model. Because the BCG of Abell 3412 (see Fig. \ref{fig:abell3411}) is located only $1\arcmin$ away from the southern edge, the surface brightness excess may be due to the remnant core of the sub-cluster Abell 3412. We are therefore looking at a more complex superposition of a core and a shock. The second possibility is that the excess emission may be associated with one galaxy in the cluster, which contains highly ionised gas. The gas is being stripped from the galaxy while it moves in the cluster (e.g. ESO 137-001 \citealt{2006ApJ...637L..81S}). The third possibility is the excess emission could be inverse Compton (IC) radiation from the radio jet tail on top of the X-ray edge. We estimate the upper limit of IC emission based on the equation from \citet{2014IJMPD..2330007B}
\begin{align}
F_\mathrm{IC}(\nu_\mathrm{X})=&1.38\times10^{-34}\left(\frac{F_\mathrm{Syn}(\nu_\mathrm{R})}{\mathrm{Jy}}\right)\left(\frac{\nu_\mathrm{X}/\mathrm{keV}}{\nu_\mathrm{R}/\mathrm{GHz}}\right)^{-\alpha}\nonumber\\
&\times\frac{(1+z)^{\alpha+3}}{\left<B_{\mu\mathrm{G}}^{1+\alpha}\right>}\ell(\alpha),
\end{align}
where $\left<B_{\mu\mathrm{G}}^{1+\alpha}\right>$ is the emission weighted magnetic field strength and $\ell(\alpha)$ is a dimensionless function. In Abell 3411, the radio spectral index at the southern edge is $\alpha\sim1$ \citep{2017NatAs...1E...5V}, at which $\ell=3.16\times10^{3}$. In the third southern spectral extraction region, the averaged radio flux at 325 MHz is $1.2\times10^{-3}$ Jy arcmin$^{-2}$. Usually, in the ICM, the magnetic field value $B\sim1$ to few $\mu$G. If we use $\left<B\right>=1$ $\mu$G to estimate the upper limit of the X-ray IC flux, the corresponding flux density is $3.24\times10^{-24} $ erg s$^{-1}$ Hz$^{-1}$ cm$^{-2}$ arcmin$^{-2}$. The converted photon density is $7.8\times10^{-9}$ ph s$^{-1}$ keV$^{-1}$ cm$^{-2}$ arcmin$^{-2}$ at 1 keV. In the 1.2 -- 4.0 keV band, the contribution of the IC emission is $2.8\times10^ {-8}$ ph s$^{-1}$ cm$^{-2}$ arcmin$^{-2}$, which is about two orders of magnitude lower than the total source flux. This possibility is therefore ruled out.

\subsection{The location of the bow shock}
\begin{figure}[]
\resizebox{\hsize}{!}{\includegraphics{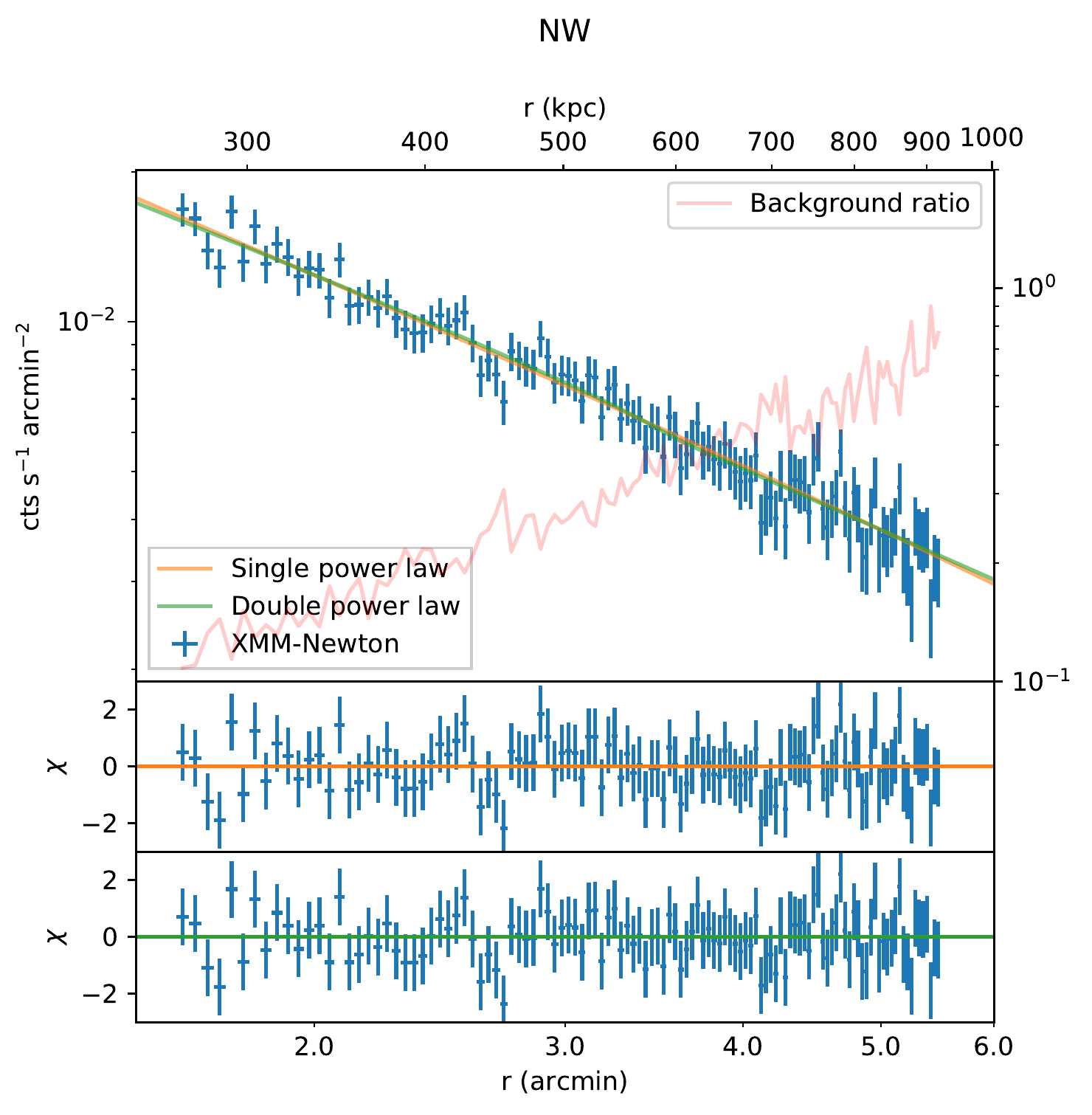}}
\caption{The \emph{XMM-Newton} surface brightness profile of the north west region. No density jump is found.}
\label{fig:sb_nw}
\end{figure}

In front of the ``bullet'', \citet{2019ApJ...887...31A} claim the detection of a bow shock with $\mathcal{M}=1.15\substack{+0.14\\-0.09}$ at $r=3.48\substack{+0.61\\-0.71}$ arcmin. The significance of the density jump is low, and the uncertainty of the location is large. To confirm this jump, we extract the \emph{XMM-Newton} surface brightness profile in front of the ``'bullet'' using the same region definition as \citet{2019ApJ...887...31A}  (see Fig. \ref{fig:sb_nw}). We fit the profile using both single power law and double power law models. The double power law model returns a statistics C-stat$/\mathrm{d.o.f.}=98.4/115$ with density jump $C=1.056\pm0.061$. As a comparison, the single power law model returns a statistics C-stat$/\mathrm{d.o.f.}=99.8/118$. A single power law model can fit this profile well. 

So far, radio observation cannot pinpoint the bow shock because this cluster has neither a radio relic nor a radio halo edge in the northern outskirts. One other method to predict the bow shock location is to use the relation between the bow shock stand-off distance and the Mach number \citep{2002ASSL..272....1S,schreier1982compressible}. However, \citet{2016MNRAS.458..681D} found that most of the bow shocks in galaxy clusters have longer stand-off distance than the expected value. For an extreme case Abell 2146 \citep{2010MNRAS.406.1721R}, the difference can reach a factor of 10 \citep{2016MNRAS.458..681D}. Recently, from simulations, \citet{2019MNRAS.482...20Z} found the unexpected large stand-off distance can be due to the de-acceleration of the cold front speed after the core passage, while the shock front can move faster. 

The offset between the projected BCG (see Fig. \ref{fig:abell3411}) and the X-ray peak positions imply the merging phase. For the sub-cluster Abell 3411, the BCG lags behind the X-ray peak by $\sim17\arcsec$. Without a weak-lensing observation, we consider the position of the BCG as the bottom of the gravitational potential well of the dark matter halo. When two sub-clusters undergo the first core passage, the position of the dark matter halo will usually be in front of the gas density peaks (e.g. the Bullet cluster, \citealt{2006ApJ...648L.109C}) because dark matter is collision-less, but the ICM is collisional. When the dark matter halo reaches the apocentre, the ambient gas pressure drops quickly so the gas could catch up and overtake the mass peak (e.g. Abell 168, \citealt{2004ApJ...610L..81H}). Hence, the location of the Abell 3411's BCG indicates the dark matter halo has almost reached its apocentre. The dynamic analysis also suggests the two sub-clusters are near their apocentres \citep{2017NatAs...1E...5V}. Thus, the stand-off distance could be much larger than the expected value. The stand-off distance calculated from the bow shock location reported by \citet{2019ApJ...887...31A} almost matches the Mach number $\mathcal{M}\sim1.2$. We speculate that the real bow shock location could be far ahead of the reported location. Unfortunately, in the north-western outskirts, the \emph{XMM-Newton} counts are dominated by the background, and the \emph{Suzaku} observation doesn't cover that region. We are unable to probe the bow shock by thermodynamic analysis.

\section{Conclusion}
\label{sec:conclusion}
We analyse the \emph{XMM-Newton} and \emph{Suzaku} data to study the thermodynamic properties of the merging system Abell 3411-3412. We calibrate the \emph{XMM-Newton} soft proton background properties based on one Lockman hole observation and apply the model to fit the Abell 3411 spectra (Appendix \ref{appendix:sp}). Our work updates the knowledge of this merging system. We summarise our results as follows:
\begin{enumerate}
\item We measure $T_{500}=4.84\pm0.04\pm0.19$ with \emph{XMM-Newton} and $T_{500}=5.17\pm0.07\pm0.13$ in the southern semicircle with \emph{Suzaku}. The corresponding mass from the $M_{500}-T_\mathrm{X}$ relation is $M_{500}=5.1\times10^{14}$ $M_\sun$.
\item The \emph{Chandra} northern bullet-like sub-cluster and southern edges are detected by \emph{XMM-Newton} as well, while the south-eastern edge shows no significant density jump in the \emph{XMM-Newton} surface brightness profile.
\item The southern edge was claimed as a cold front previously \citep{2019ApJ...887...31A}. With our \emph{XMM-Newton} analysis, the temperature jump prefers a shock front scenario. There is a clear pressure jump indicating supersonic motions, although the geometry seems to be more complicated, with a possible superposition of a shock and additional stripped material from the Abell 3412 sub-cluster.
\item Both \emph{Suzaku} and \emph{XMM-Newton} results confirm the south-eastern edge is a $\mathcal{M}\sim1.2$ shock front, which agrees with the previous result from \emph{Chandra} surface brightness fit \citep{2017NatAs...1E...5V,2019ApJ...887...31A}. Such a low Mach number supports the particle re-acceleration scenario at the shock front.
\end{enumerate}

\begin{acknowledgement}
We thank the anonymous referee for constructive suggestions that improved this paper.
X.Z. is supported by the the China Scholarship Council (CSC). R.J.vW. acknowledges support from the ERC Starting Grant ClusterWeb 804208. SRON is supported financially by NWO, The Netherlands Organization for Scientific Research. This research made use of Astropy,\footnote{\url{http://www.astropy.org}} a community-developed core Python package for Astronomy \citep{2013A&A...558A..33A, 2018AJ....156..123A}. This research is based on observations obtained with \emph{XMM-Newton}, an ESA science mission with instruments and contributions directly funded by ESA Member States and NASA. This research has made use of data obtained from the \emph{Suzaku} satellite, a collaborative mission between the space agencies of Japan (JAXA) and the USA (NASA). This research has made use of data obtained from the \emph{Chandra} Data Archive and the \emph{Chandra} Source Catalog, and software provided by the \emph{Chandra} X-ray Center (CXC) in the application package CIAO. 

\end{acknowledgement}

\bibpunct{(}{)}{;}{a}{}{,}
\bibliographystyle{aa}
\bibliography{abell3411}

\begin{appendix}
\section{Light curves of EPIC CCDs}\label{appendix:oofov}
\begin{figure}
\resizebox{\hsize}{!}{\includegraphics{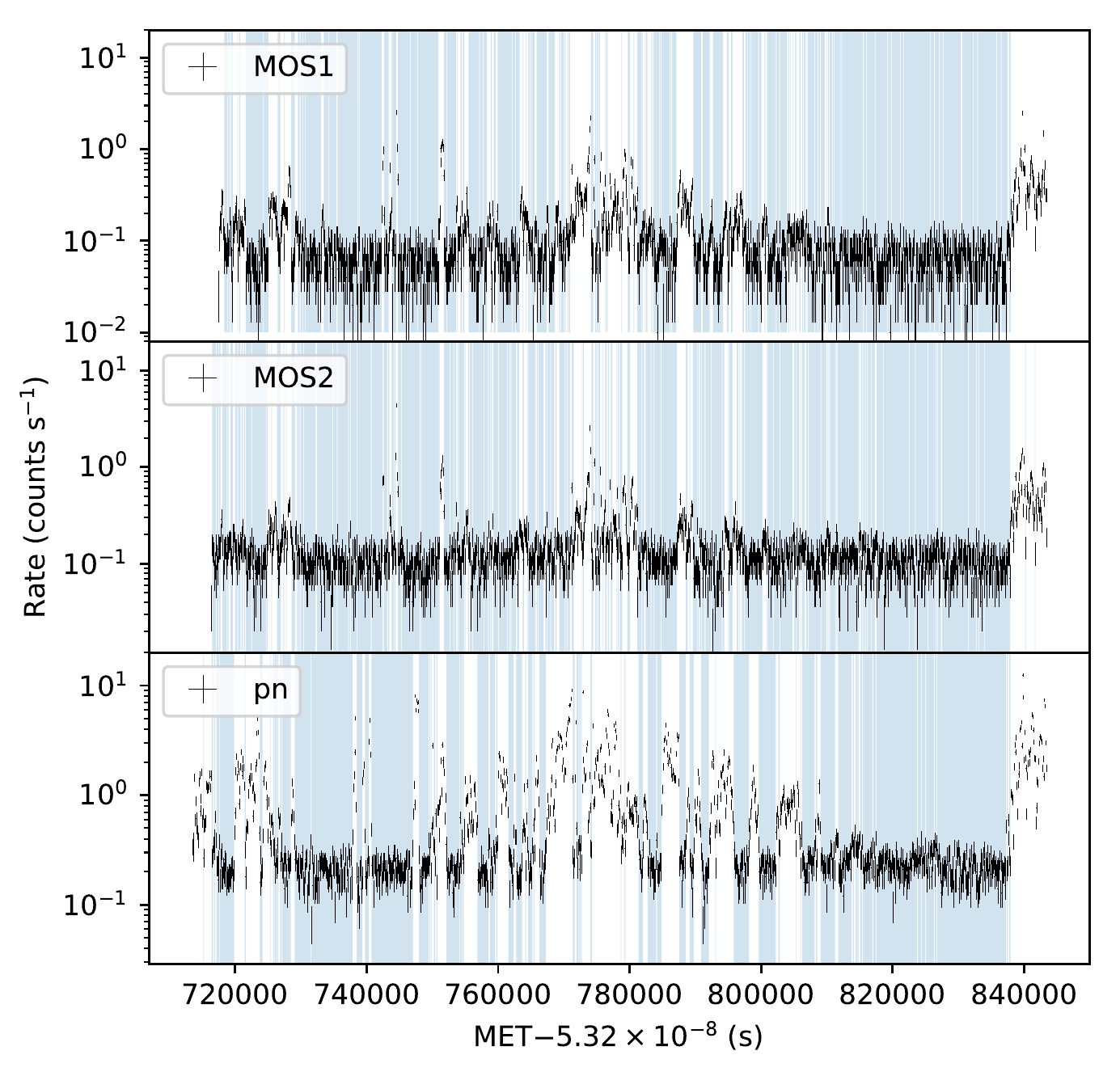}}
\caption{One-hundred second binned 10 --12 keV light curves of the three EPIC detectors. The filtered GTIs are shown as blue shadows.}
\label{fig:fov_lc}
\end{figure}

In Table \ref{table:obs}, the MOS2 GTI is about 8 ks larger than MOS1. We plot the light curve and filtered GTIs of each EPIC detector in Fig. \ref{fig:fov_lc}. Although the light curves of two MOS detectors have a similar trend, some flares are only significant in MOS1. This explains why we obtain less GTI for MOS1.

\begin{figure}
\resizebox{\hsize}{!}{\includegraphics{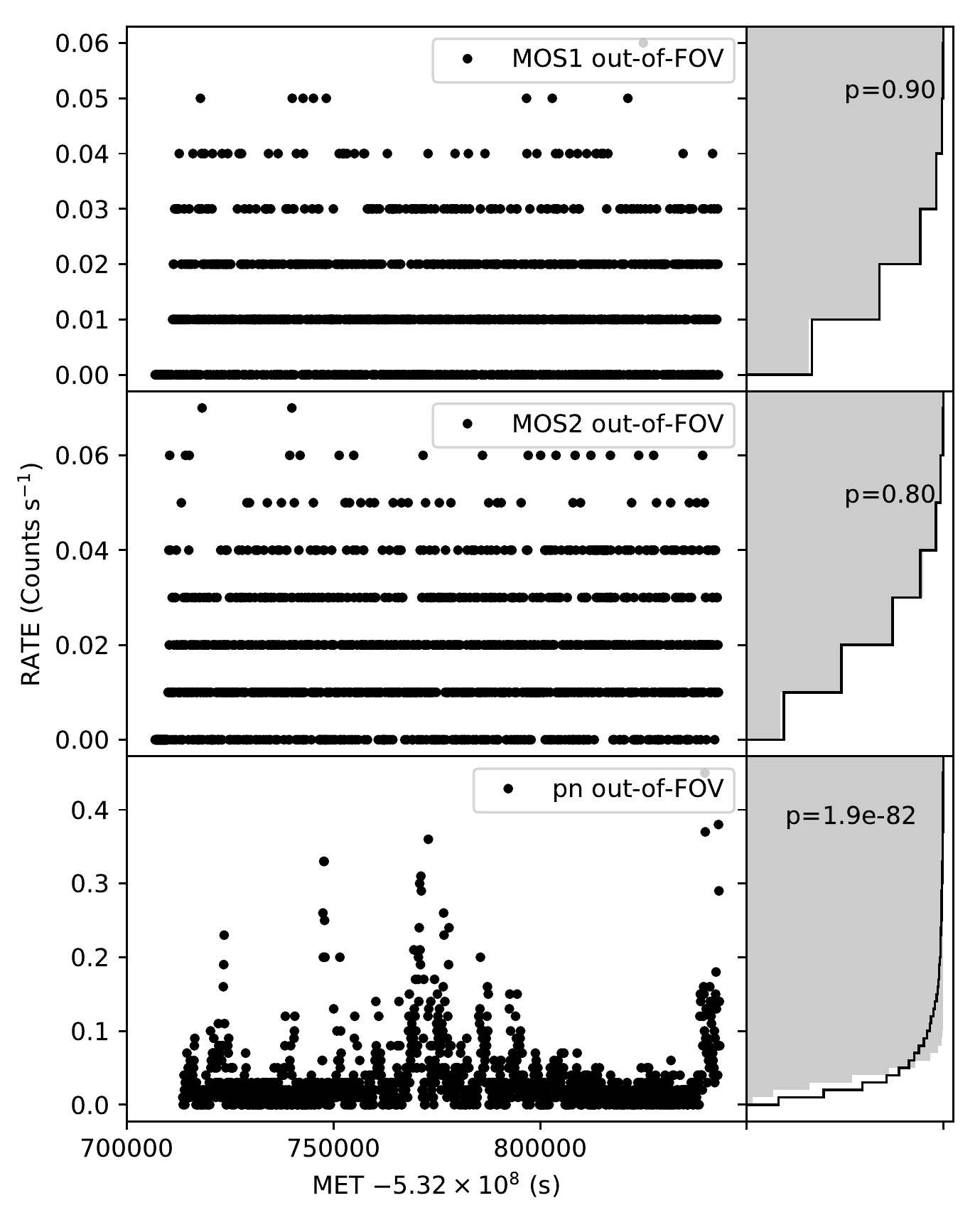}}
\caption{Out-of-FOV 10 --12 keV light curves of three EPIC detectors. Right panels are cumulative density functions (c.d.f.) of count rates. For each light curve, the c.d.f. of the Poisson distribution with $\mu$ from the data is plotted as grey area. The p-values to reject the null hypothesis that the count rate distribution is Poissonian are labelled.}
\label{fig:corner_lc}
\end{figure}

Out-of-FOV detector pixels are usually used for particle background level estimation. However, EPIC-pn CCD's out-of-FOV corners suffer from soft proton flares as well. We select 100 s binned pn out-of-FOV light curves with selection criteria \texttt{FLAG==65536 \&\& PATTERN==0 \&\& (PI IN [10000:12000])}. To make a comparison, we extract light curves of the two MOS CCDs with the same out-of-FOV region expressions as in Sect. \ref{sec:data-reduction}, which are from \citet{2008A&A...478..575K}. Light curves are plotted in Fig. \ref{fig:corner_lc}. We use a two-sample Kolmogorov-Smirnov (KS) test to check whether the cumulative density function (c.d.f.) of the count rate matches a Poisson distribution. For the pn CCD, the p-value is less than 0.05. Therefore, the null hypothesis that the count rate follows a Poisson distribution is rejected. The KS test suggests that the out-of-FOV region of the pn detector is significantly contaminated by soft proton flares, while MOSs' out-of-FOV area is clean enough to be used as reference for the particle background level estimation. 

\section{Soft proton modelling} \label{appendix:sp}
Our observation suffered from significant soft proton contamination. Although we adopt strict flare filtering criteria, the contamination in the quiescent state is not negligible. Inappropriate estimations of the soft proton flux, as well as its spectral shape, would introduce considerable systematics to fit the results. The integrated flare state soft proton spectra from MOS are studied by \citet{2008A&A...478..575K}. They are smooth and featureless, with the shape of an exponential cut off power law. The spectrum is harder when flares are stronger. 

The soft proton spectra of pn during flares have not been studied yet. To investigate the soft proton background properties, including spectral parameters, vignetting functions, etc., we analyse one observation of the Lockman Hole (ObsID: 0147511201), which is also heavily contaminated by soft proton flares. That observation was also performed with the medium filter. Flare state time intervals are defined by $\mu+2\sigma$ filtering criteria on 100 s binned light curves in the 10 -- 12 keV energy intervals. The flare state proportion of pn is $\sim87\%$. The pure flare state spectra are simply calculated by subtracting quiescent state spectra from flare state spectra. 

\subsection{Spectral analysis}
\begin{figure}
\resizebox{\hsize}{!}{\includegraphics{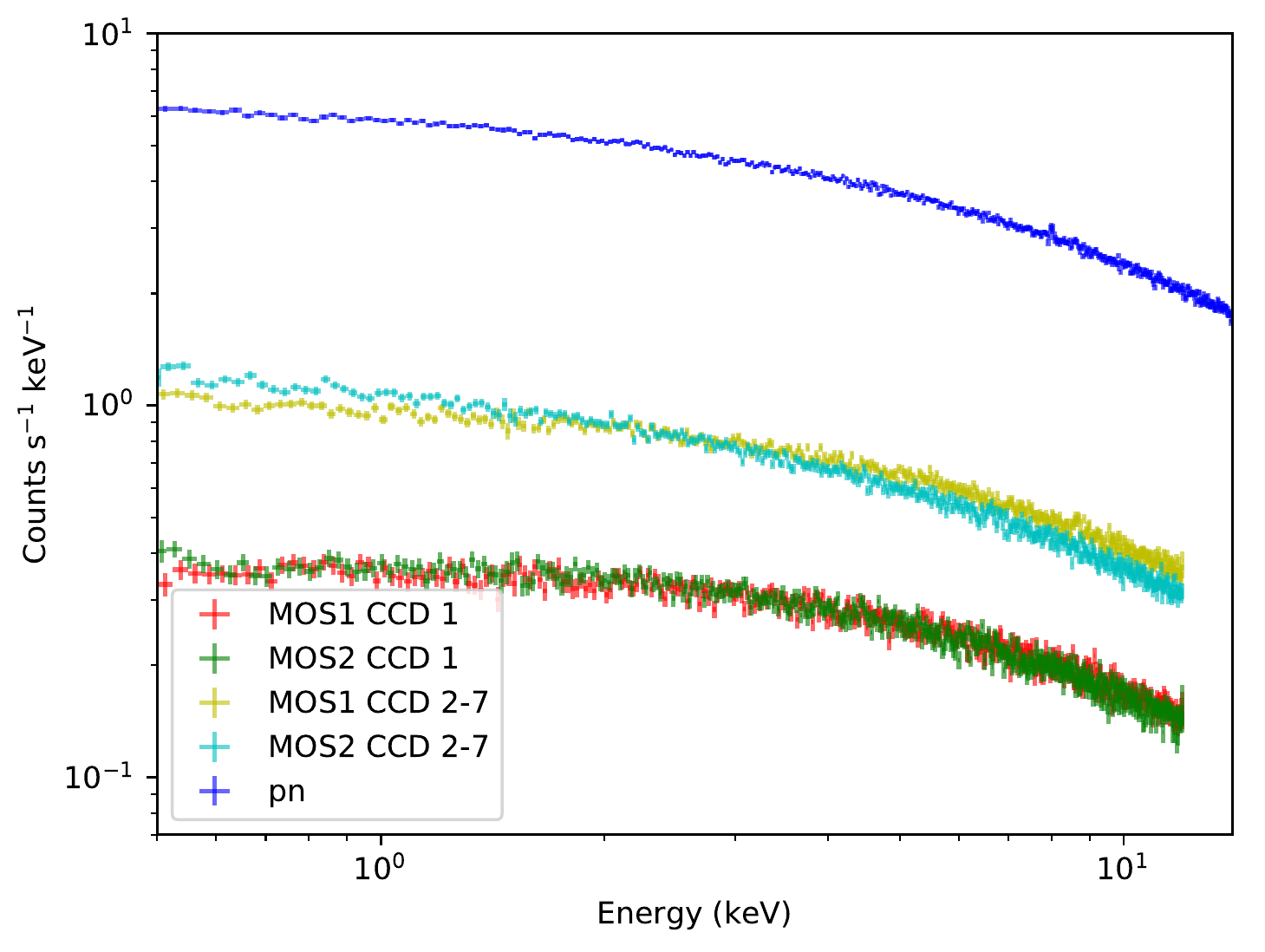}}\\
\caption{Flare state soft proton spectra in $12\arcmin$ radius. }
\label{fig:sp_spectrum}
\end{figure}

Flare state soft proton spectra in the 0.5 -- 14.0 keV band within a $12\arcmin$ radius are plotted in Fig. \ref{fig:sp_spectrum}. The spectra of the centre and outer MOS CCDs are plotted individually. The shape of the pn spectrum is basically coincident with that of MOS. They are all smooth and featureless and can be described as a cut off power law. The spectra from central MOS CCDs are coincident with each other. However, the spectra from outer CCDs are slightly different. 

RMFs generated by \texttt{rmfgen} are calibrated on photons and include the photon redistribution jump at the Si K edge. However, we don't see any feature there in the soft proton spectra. As a result, we use \texttt{genrsp} in FTOOL to generate a dummy RMF for fitting. A SPEX built-in generalised power law model is used to model the soft proton spectra. The generalised power law can be expressed as 
\begin{equation}
F(E)=AE^{-\Gamma}e^{\eta(E)},
\end{equation}
where $A$ is the flux density at 1 keV, $\Gamma$ is the photon index, and $\eta(E)$ is given by
\begin{equation}
\eta(E)=\frac{r\xi+\sqrt{r^2\xi^2+b^2(1-r^2)}}{1-r^2},
\end{equation}
with $\xi=\ln(E/E_0)$ and 
$r=(\sqrt{1+(\Delta\Gamma)^2}-1)/|\Delta\Gamma|$, where $E_0$ is the break energy, $\Delta\Gamma$ is photon index difference after the break energy, and $b$ is the break strength. Instead of using flux density $A$, we have adjusted the model implementation to use 2 -- 10 keV integrated luminosity $L$ as a normalisation factor.

\begin{figure*}
\begin{tabular}{ccc}
\resizebox{0.31\hsize}{!}{\includegraphics{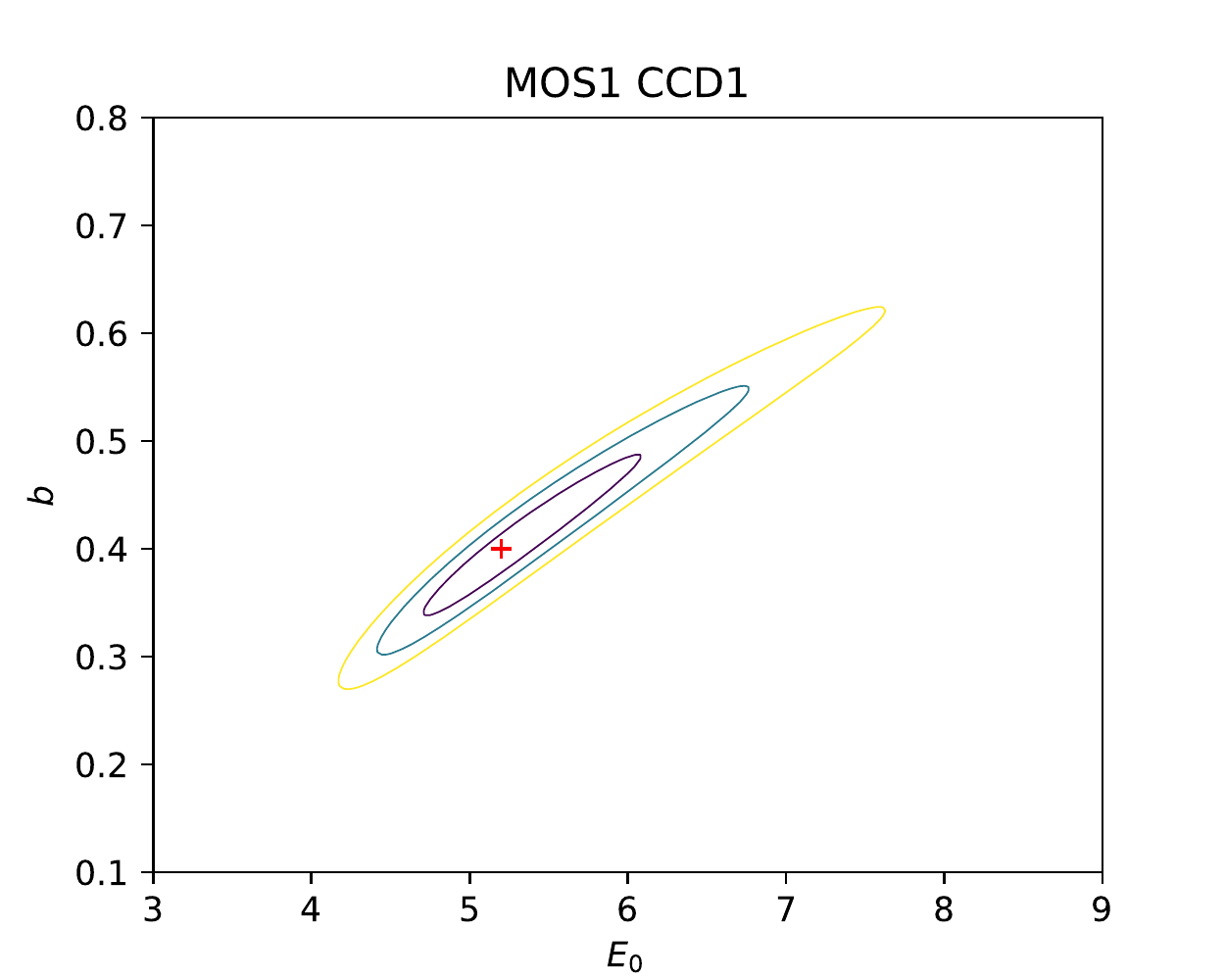}}&
\resizebox{0.31\hsize}{!}{\includegraphics{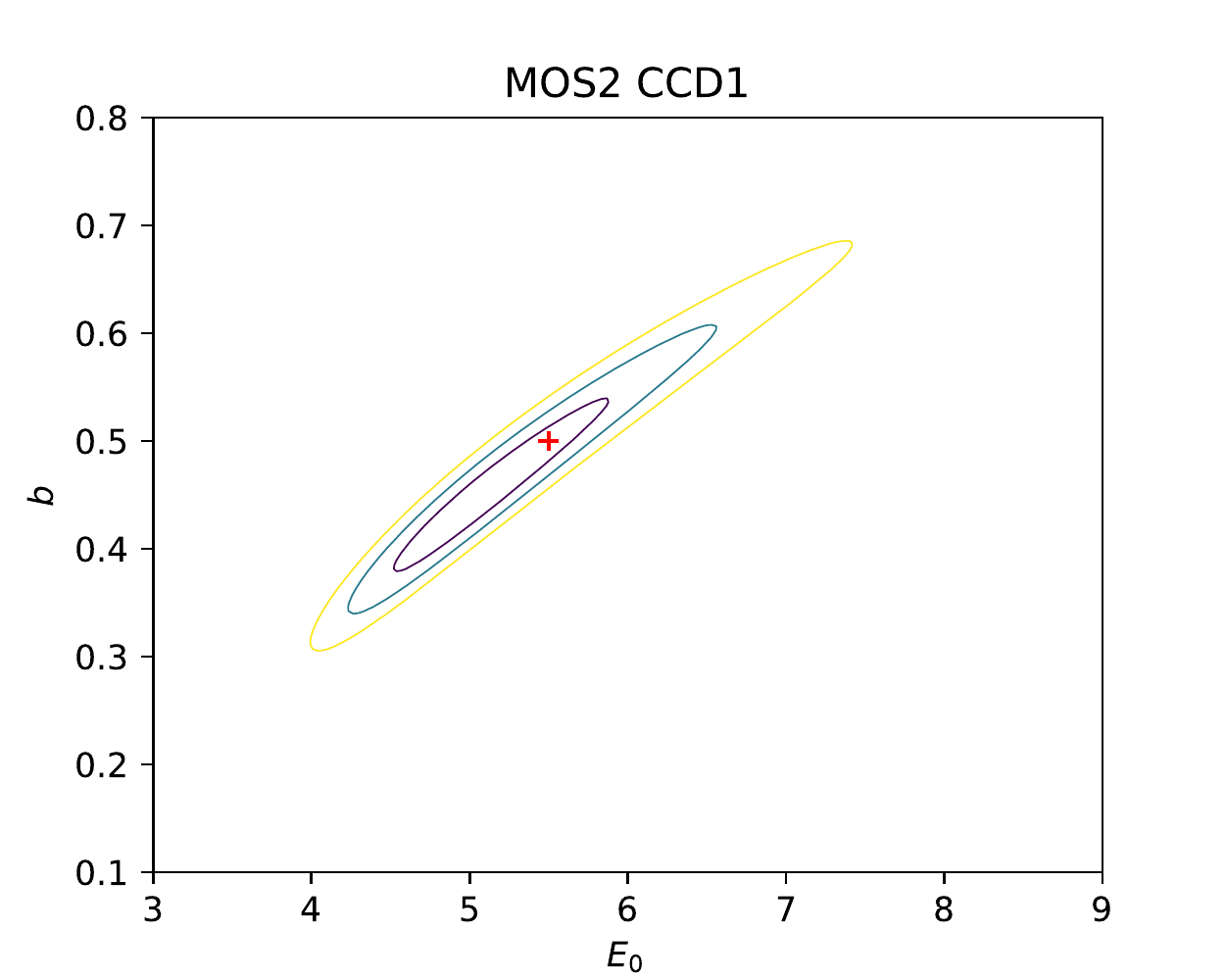}}&
\resizebox{0.31\hsize}{!}{\includegraphics{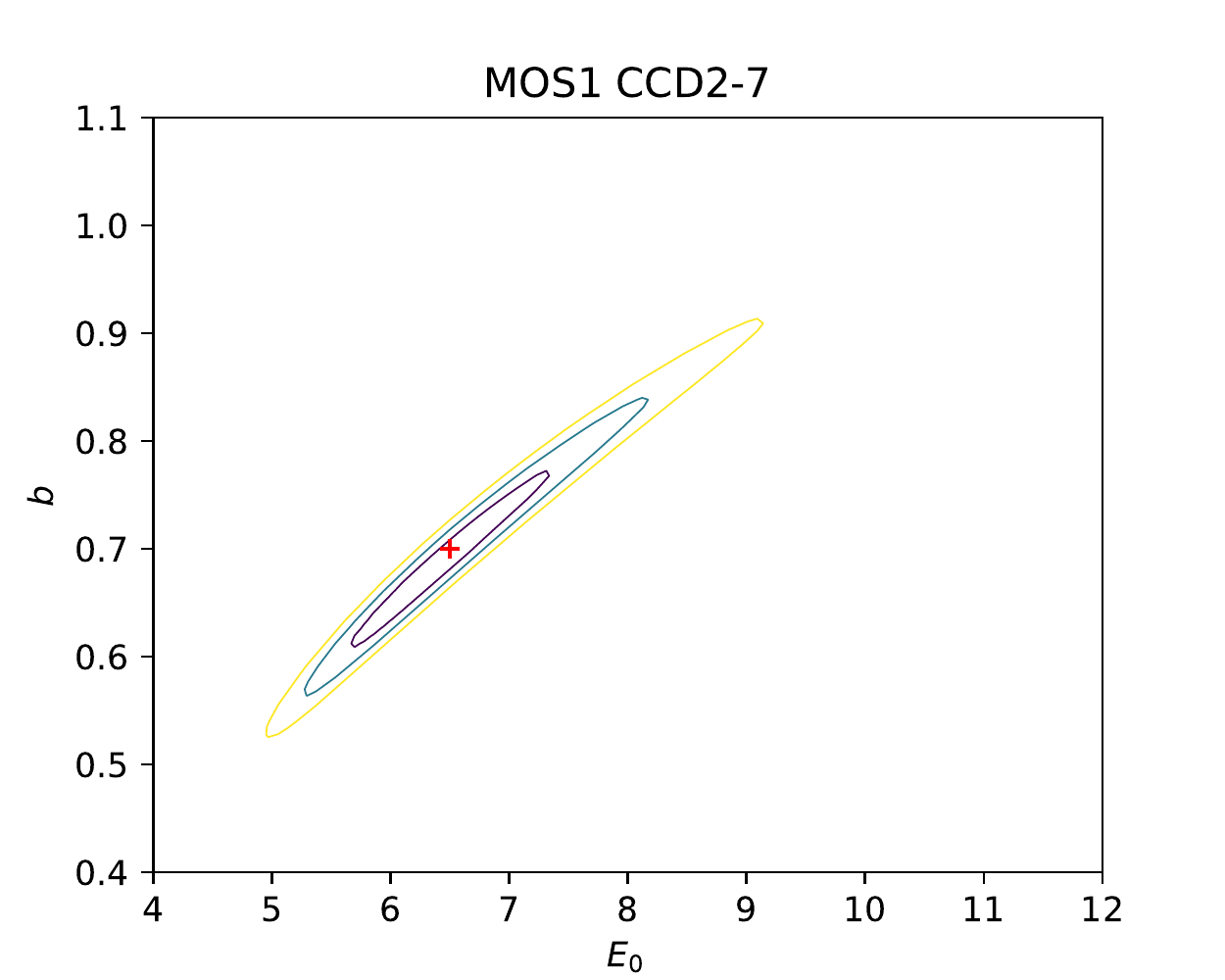}}\\
\resizebox{0.31\hsize}{!}{\includegraphics{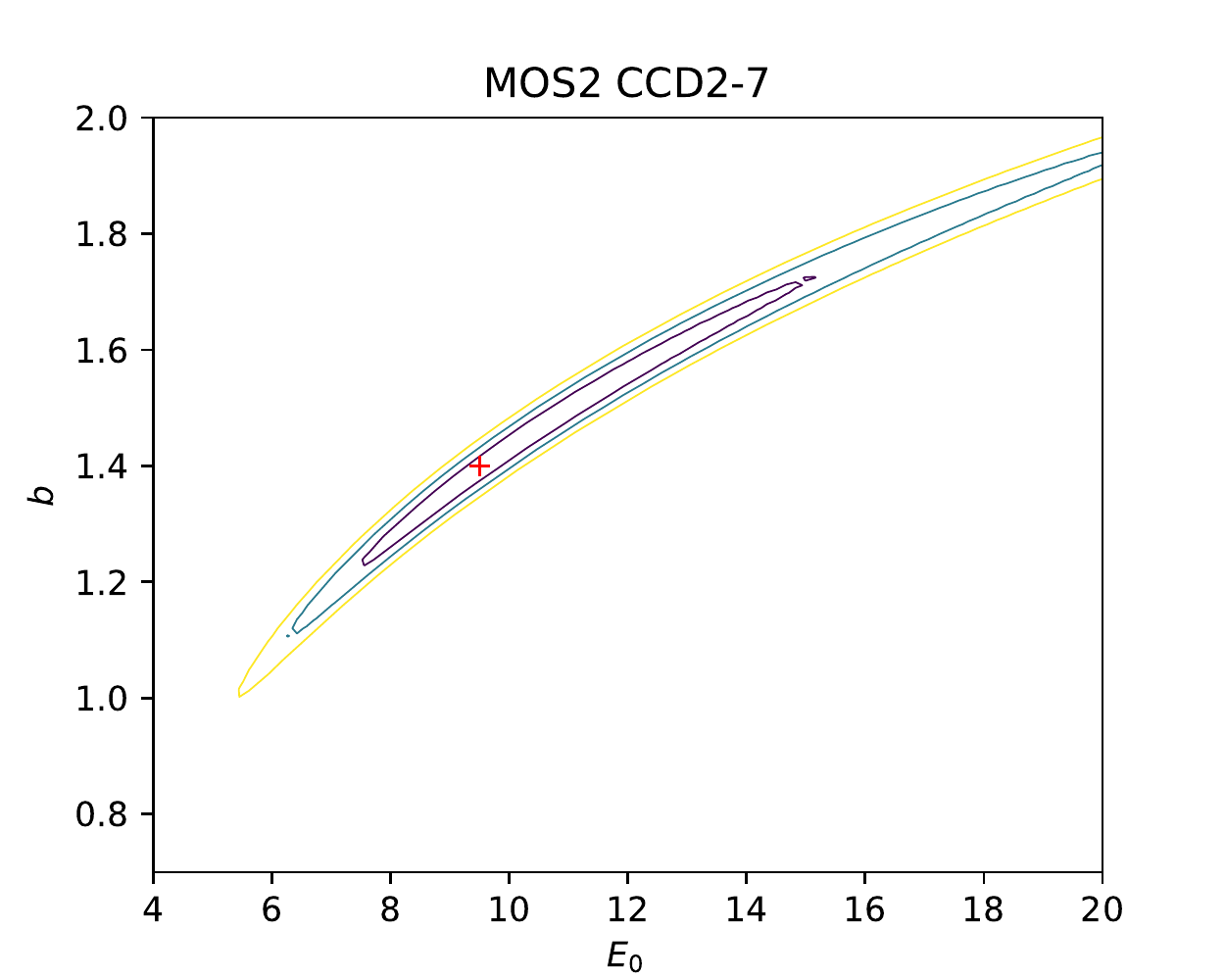}}&
\resizebox{0.31\hsize}{!}{\includegraphics{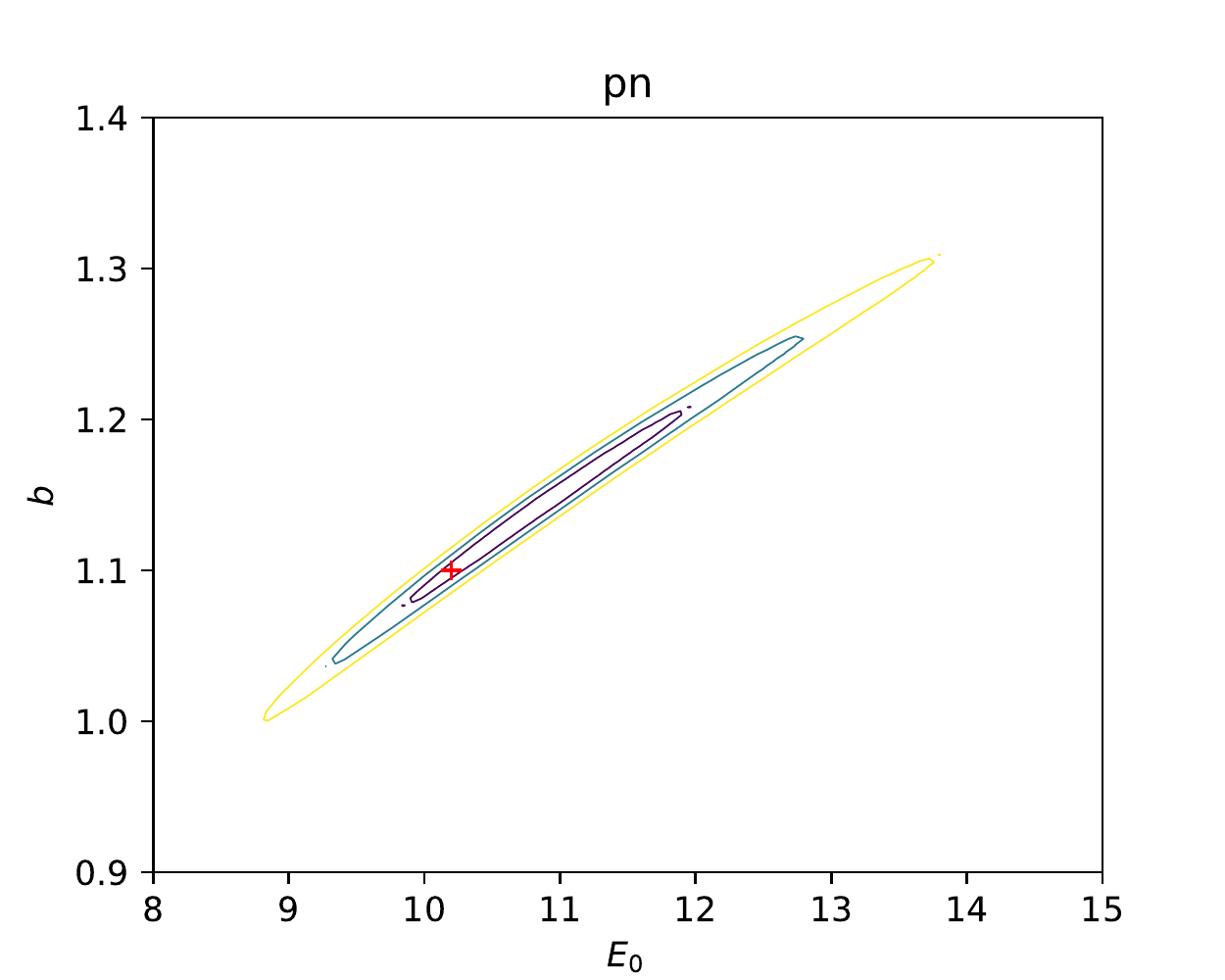}}&
\end{tabular}
\caption{Best-fit statistics for the Lockman hole soft proton flare state spectra in $E_0$ vs $b$ space. The chosen parameters are plotted as red crosses. Contours from inner to outer are $1\sigma$, $2\sigma$, and $3\sigma$ confidence levels.}
\label{fig:sp_param}
\end{figure*}

We fit the integrated pn, MOS centre CCD and MOS outer CCDs spectra with the model described above. To solve the degeneracy of parameters, we fix all $\Gamma$ to 0, hence the photon index after the break energy $\Gamma_2=\Delta\Gamma$. 
We manually choose sets of $E_0$ and $b$ values inside the $1\sigma$ contours from the parameter diagrams (see Fig. \ref{fig:sp_param}). All fixed and fitted parameters, as well as the fit statistics, are listed in Table \ref{sp_par}.

\begin{table*}
\caption{Best fit parameters of flare state soft proton spectra in the Lockman Hole observation.}
\label{sp_par}
\centering
\begin{tabular}{lcccc}
\hline\hline
& $\Gamma_2$ & $E_0$ (keV) & $b$ & C-stat/d.o.f.\\
\hline
MOS1 CCD 1 &$0.941\pm0.008$&5.2&0.4&622/890\\
MOS2 CCD 1 &$1.060\pm0.008$&5.5&0.5&658/890\\
MOS1 CCD 2-7 &$1.206\pm0.006$&6.5&0.7&807/457\\
MOS2 CCD 2-7 &$1.629\pm0.008$&9.5&1.4&728/457\\
pn &$1.632\pm0.004$&10.2&1.1&840/539\\
\hline

\end{tabular}
\end{table*}

\subsection{Vignetting function}\label{appendix:sp-vig}
The soft proton vignetting function is different from that of X-rays. To determine the spatial distribution of soft proton counts, we study the vignetting behaviour of different CCDs from the Lockman hole observation. We calculate surface brightness profiles with our surface brightness profile analysis tool. We take total count maps as the source images and quiescent state count maps as backgrounds. A uniform dummy exposure map is applied. The surface brightness profile of the residual soft protons reflects the vignetting behaviour.  

\begin{figure}
\resizebox{\hsize}{!}{\includegraphics{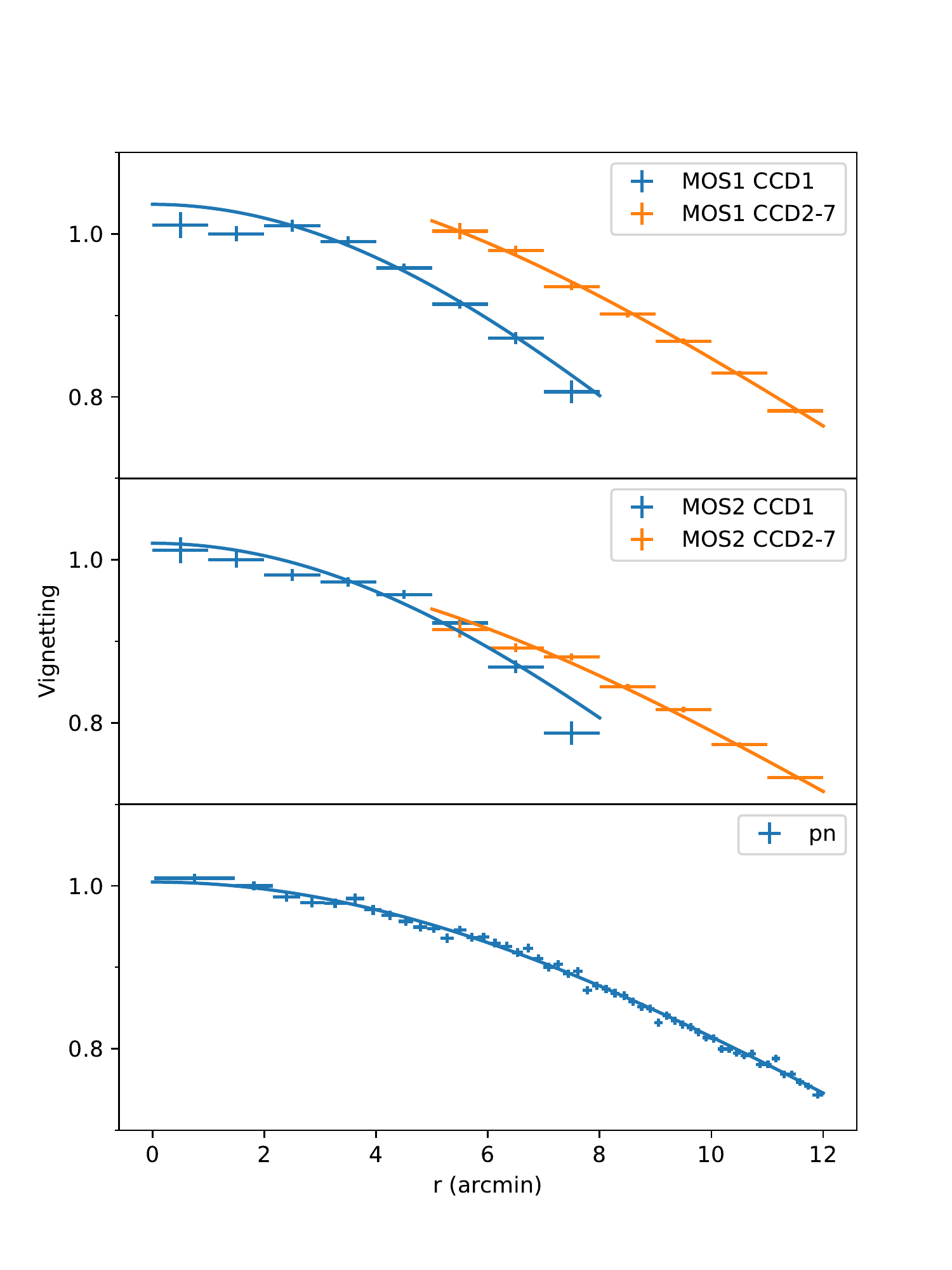}}
\caption{Vignetting functions of flare state soft protons detected by the EPIC CCDs in the 2 -- 10 keV band determined from the Lockman hole observation. Each set of measurements is normalised to the second data point. Solid lines are best-fit $\beta$ models.}
\label{fig:sp_vig}
\end{figure}

The count weighted vignetting functions in the 2 -- 10 keV band are shown in Fig. \ref{fig:sp_vig}. Because the MOS outer chips are closer to the mirror, there is a gap the vignetting functions of the central and outer CCDs. The vignetting behaviours of MOS1 and MOS2 centre CCDs are similar, but different in the outer CCDs. We fit vignetting functions with $\beta$ profiles \citep{1976A&A....49..137C}
\begin{equation}
\label{eq:beta}
S(r)=S_0\left[1+\left(\frac{r}{r_0}\right)^2\right]^{0.5-3\beta},
\end{equation}
where $r_0$ is fixed to $40\arcmin$. Best fit parameters are listed in Table \ref{sp_vig}.

\begin{table}
\caption{The best fit parameters of 2 -- 10 keV soft proton vignetting functions. Parameter $r_0$ is fixed to $40\arcmin$.}
\label{sp_vig}
\centering
\begin{tabular}{lcc}

\hline\hline
& $S_0$ (arbitrary unit)& $\beta$\\
\hline
MOS1 CCD1&$1.824\pm0.008$&$2.35\pm0.10$ \\
MOS1 CCD 2-7&$1.906\pm0.009$&$1.51\pm0.03$\\
MOS2 CCD1&$1.778\pm0.008$&$2.17\pm0.10$ \\
MOS2 CCD 2-7&$1.743\pm0.008$&$1.45\pm0.03$\\
pn &$8.316\pm0.011$&$1.325\pm0.010$\\
\hline
\end{tabular}
\end{table}

\subsection{Self-calibration}\label{appendix:sp-calibration}

\begin{figure}
\resizebox{\hsize}{!}{\includegraphics{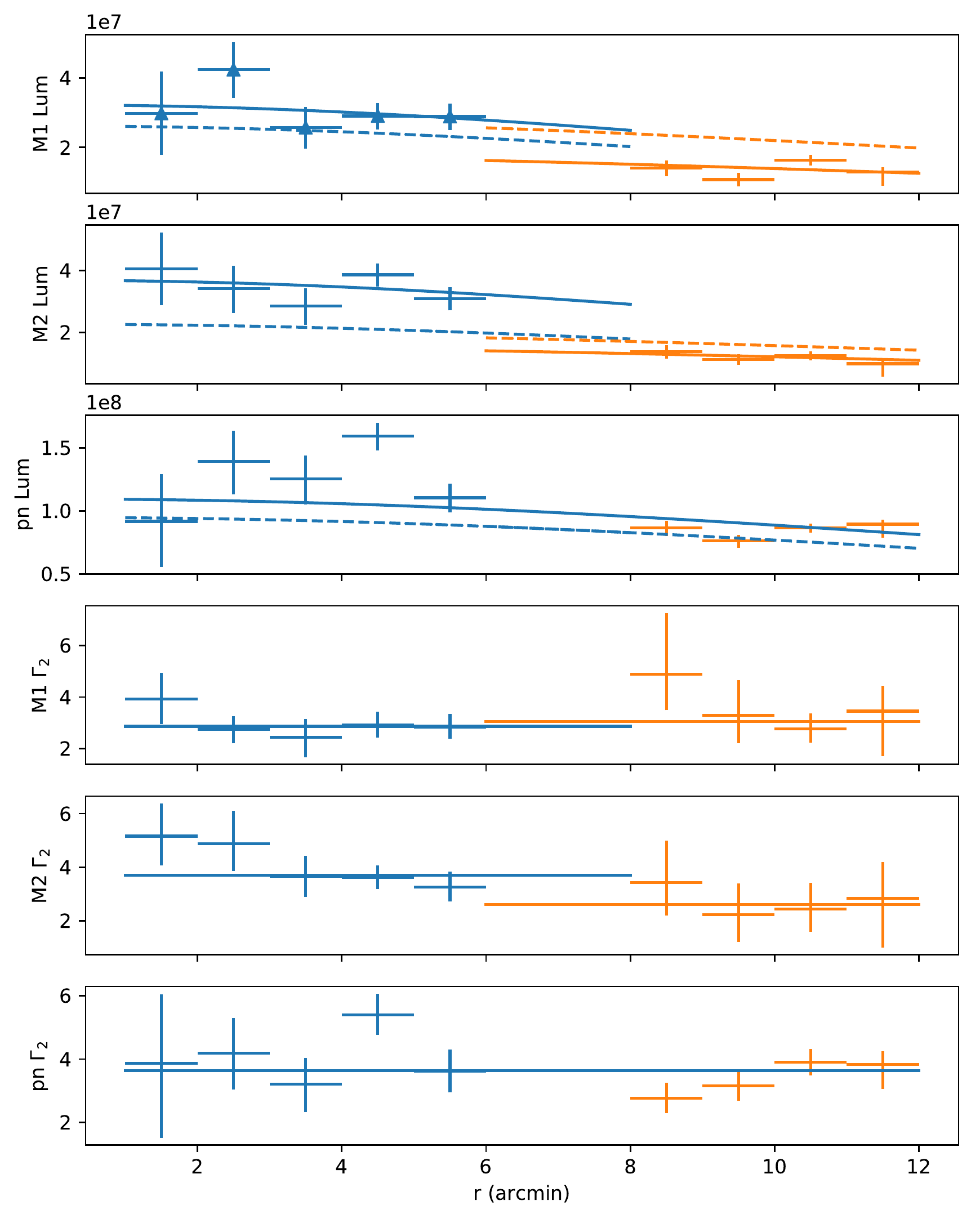}}
\caption{Radial profiles of $L$ and $\Gamma_2$ from the Abell 3411 observation. Blue points are from the centre MOS CCD region, orange points are from outer MOS CCDs region. The best fit radial models are plotted with lines.}
\label{fig:sp_radial}
\end{figure}

We extract MOS and pn spectra from the Abell 3411 observation, separating the MOS centre and outer CCD region. We exclude the union of the MOS and pn bad pixel regions using additional region selection expressions. The selected regions are annuli centred at the pn focal point from $1\arcmin$ to $12\arcmin$ with width $1\arcmin$. From the central MOS CCD region, we extract spectra up to $r=6\arcmin$. From the outer MOS CCD region, we extract spectra from $r=8\arcmin$. There are $3\times5$ centre region spectra and $3\times4$ outer region spectra in total. The energy range 0.5 -- 14.0 keV is used for spectral fitting. Spectral components are the same as described in Table \ref{table:components}. We first fit FWC spectra with an exponential cut off power law and delta lines. We freeze the FWC continuum with the fitted cut-off power-law parameters and fit the soft proton and ICM components. We add two delta lines at 0.56 and 0.65 keV to fit the SWCX radiation. The other free parameters are $\Gamma_2$ and $L$ of soft proton components, $norm$, $T$ and $Z$ of the ICM. The best-fit values of a subset of the most relevant parameters are plotted in Fig. \ref{fig:sp_radial}. 

We use constant models to fit five $\Gamma_2$ profiles, and use the vignetting models from Appendix \ref{appendix:sp-vig} to fit five luminosity profiles individually. We fix each $\beta$ parameter but thaw the normalisation. Best-fit soft proton $L_0$s and $\Gamma_2$s are listed in the second and the third column of Table \ref{table:sp_calibrated}. The best-fit model profiles are plotted with solid lines in Fig. \ref{fig:sp_radial}. If we assume the detector responses to SP are identical over time and in both flare and quiescent states, we can estimate the luminosity profiles in a second way. For this, we need to calculate the ratio of normalisations among different detectors. From Appendix \ref{appendix:sp-vig} we have surface brightness radial profiles at the flare state
\begin{equation}
S_\mathrm{Det}^\mathrm{Flare}(r) = \int_2^{10}\mathrm{Vig}_\mathrm{Det}(r)F_\mathrm{Det}^\mathrm{Flare}(E,r) \mathrm{d}E,
\end{equation}
where $F_\mathrm{Det}^\mathrm{Flare}$ are the flare state spectrum models of different CCDs at radius $r$. Vig is the vignetting function. Det can be MOS1 center, MOS1 outer, MOS2 center, MOS2 outer and pn. We take pn as a reference. The count rate ratio between other detectors and pn is 
\begin{equation}
\xi_\mathrm{CR,Det/pn}^\mathrm{Flare}(r)=\frac{\mathrm{Vig}_\mathrm{Det}(r)}{\mathrm{Vig}_\mathrm{pn}(r)}\frac{\int_2^{10}F_\mathrm{Det}^\mathrm{Flare}(E,r) \mathrm{d}E}{\int_2^{10}F_\mathrm{pn}^\mathrm{Flare}(E,r) \mathrm{d}E}.
\end{equation}
The energy flux ratio between other detectors and pn at the quiescent state can be easily calculated.
\begin{align}
\xi_\mathrm{E,Det/pn}^\mathrm{Quiescent}(r)=&\frac{\mathrm{Vig}_\mathrm{Det}(r)}{\mathrm{Vig}_\mathrm{pn}(r)}\frac{\int_2^{10}EF_\mathrm{Det}^\mathrm{Quiescent}(E,r) \mathrm{d}E}{\int_2^{10}EF_\mathrm{pn}^\mathrm{Quiescent}(E,r)\mathrm{d}E}\nonumber\\
=&\xi_\mathrm{CR,Det/pn}^\mathrm{Flare}(r)\frac{\int_2^{10}F_\mathrm{pn}^\mathrm{Flare}(E,r) \mathrm{d}E}{\int_2^{10}F_\mathrm{Det}^\mathrm{Flare}(E,r) \mathrm{d}E}\times\nonumber\\
&\frac{\int_2^{10}EF_\mathrm{Det}^\mathrm{Quiescent}(E,r) \mathrm{d}E}{\int_2^{10}EF_\mathrm{pn}^\mathrm{Quiescent}(E,r)\mathrm{d}E}.
\end{align}
With the best-fit quiescent state $\Gamma_2$, we obtain $\xi_\mathrm{E}^\mathrm{Quiescent}$s and list them in the fourth column of Table \ref{table:sp_calibrated}. We couple the MOS $L$ parameter to that of pn with the scale factor $\xi_\mathrm{E}^\mathrm{Quiescent}$ to fit the $L$ profiles simultaneously. The best-fit $L_0$ of pn is $(9.5\pm0.2)\times10^{37}W$. We plot this set of luminosity models with dashed lines in Fig. \ref{fig:sp_radial}.

The systematics of radial luminosity models include two parts: one is the offset between the measured model (solid lines) and the empirical model on the basis of the Lockman Hole observation (dashed lines); another one is from the intrinsic scatter that makes the $\chi^2/\mathrm{d.o.f.}$ of each profile in Fig. \ref{fig:sp_radial} larger than 1. 

The offset systematics $\eta_\mathrm{off}$ are calculated by the formula $\eta_\mathrm{off}=|L_0^\mathrm{dashed}-L_0^\mathrm{solid}|/L_0^\mathrm{solid}$. The intrinsic systematics $\eta_\mathrm{in}$ are calculated such that 
\begin{equation}
\sum_i\frac{\left(L_i-\hat{L}_i\right)^2}{\sigma_i^2+\eta_\mathrm{in}^2L_i^2}=\mathrm{d.o.f.},
\end{equation}
where $\hat{L}_i$ is the model luminosity at the $i$th point. The total systematics are then $\eta_\mathrm{total}^2=\eta_\mathrm{off}^2+\eta_\mathrm{in}^2$. We list the offset, intrinsic, and total systematics in the fifth to seventh column of Table \ref{table:sp_calibrated}.

\begin{table*}
\caption{The best fit parameters and systematics of quiescent state soft proton components.}
\label{table:sp_calibrated}
\centering
\begin{tabular}{lcccccc}
\hline\hline
& $L_0$ ($10^{37}$W) & $\Gamma_2$ & $\xi^\mathrm{Quiescent}_\mathrm{E}$ & $\eta_\mathrm{off}$& $\eta_\mathrm{in}$&$\eta_\mathrm{total}$\\
\hline
MOS1 CCD1& $3.22\pm0.25$ & $2.86\pm0.26$ & 0.29 & 0.19 & 0 &0.19\\
MOS2 CCD1& $3.68\pm0.24$ & $3.70\pm0.29$ & 0.34 & 0.38 & 0 &0.38\\
MOS1 CCD 2-7& $1.78\pm0.12$ & $3.05\pm0.46$ & 0.16 & 0.58 & 0.15 &0.60\\
MOS2 CCD 2-7& $1.54\pm0.12$ &$2.60\pm0.58$& 0.14 & 0.30 & 0 &0.30\\
pn &$10.94\pm0.28$&$3.64\pm0.20$ & 1.00 & 0.13 & 0.18 &0.22\\
\hline
\end{tabular}
\end{table*}

We apply the self-calibrated soft proton model to spectra from regions of interest. Parameters $E_0$ and $b$ are fixed based on the values in Table \ref{sp_par}. $\Gamma_2$ is fixed given in Table \ref{table:sp_calibrated}. $L$ is fixed to the value calculated from the vignetting function \ref{eq:beta} with $\beta$ values from Table \ref{sp_vig} and normalisation values from the column $L_0$ in \ref{table:sp_calibrated}. 

\section{Cosmic X-ray Background}\label{appendix:cxb}
Thanks to the \emph{Chandra} observation, we are able to study the $\log N-\log S$ relationship of point sources in the Abell 3411-3412 field. The result can help us to constrain the \emph{XMM-Newton} point source detection limit as well as to optimise point source exclusion in the \emph{Suzaku} analysis.

\subsection{Point source flux and the $\log N-\log S $ relation}
We use \texttt{wavdetect} to detect point sources in this field. An exposure weighted PSF map is provided for source detection. The wavelet size is set as 1.0, 2.0, and 4.0. The task returns 147 sources in total. After visual inspection, 113 sources are left. We use \texttt{roi} to extract source and background regions for each point source. The background regions are set as elliptical annuli from 1.5 to 2.0 times the source radius. We extract spectra for each source from each observation using the task \texttt{specextract}. A point source aperture correction is applied. For each point source, all source and background spectra, as well as response files from each observation, are combined by \texttt{combine\_spectra}. We model each point source spectrum with an absorbed power-law model. The energy range 0.5 -- 7.0 keV is used for spectral fitting. We fit spectra with both a fixed $\Gamma=1.41$ and a free $\Gamma$. If the relative error of $\Gamma$ is less than $10\%$, we adopt the best-fit $\Gamma$ and the corresponding flux. We exclude sources with zero fitted flux or $\Gamma>5$ and compile the rest 101 sources into a catalogue.  

\begin{figure}
\resizebox{\hsize}{!}{\includegraphics{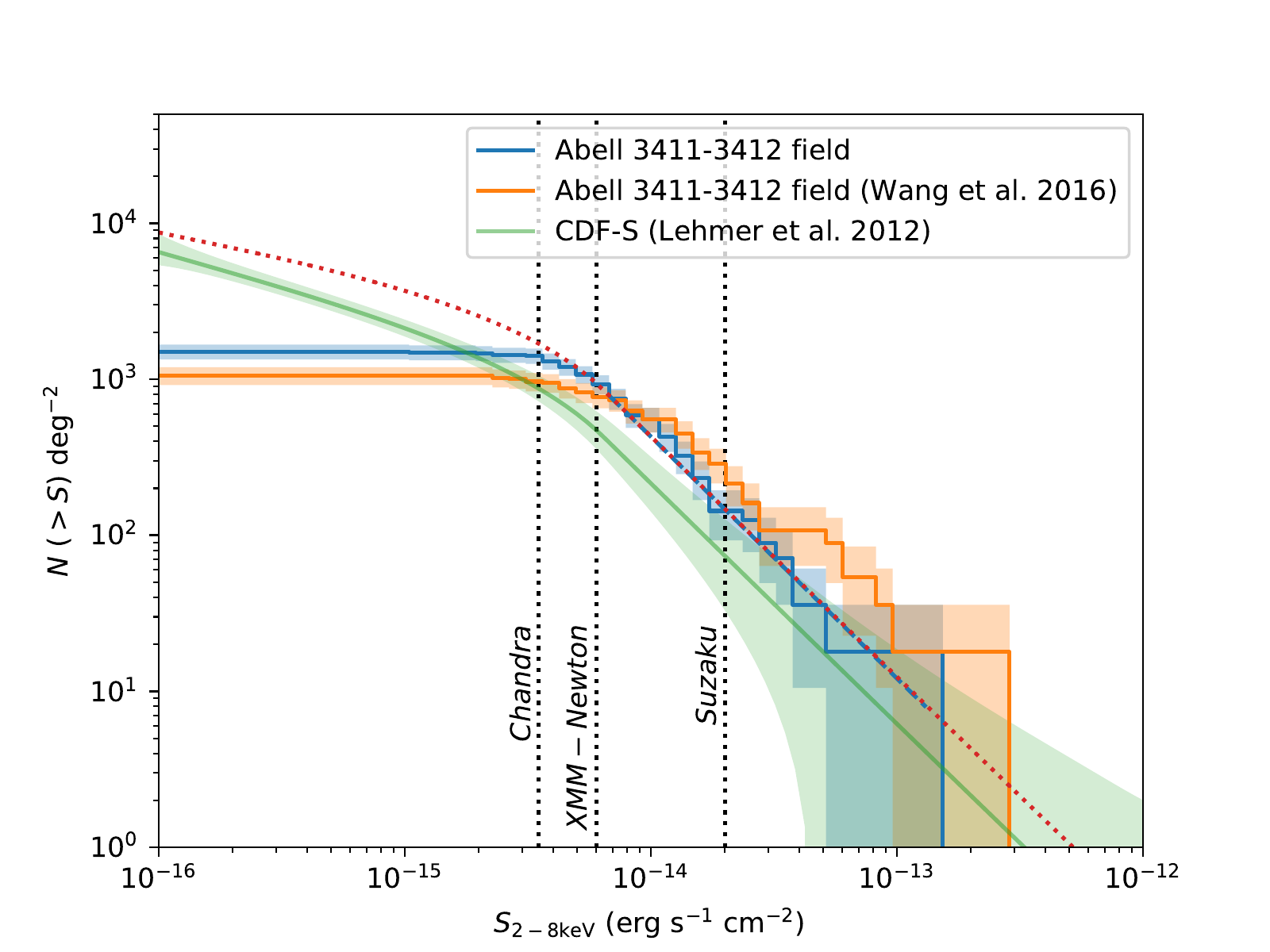}}
\caption{$\log{N}-\log{S}$ curve from \emph{Chandra} observations. The blue steps are from our analysis of the Abell 3411-3412 field. The best-fit power law is plotted as a blue dashed line. The model we assume to calculate the unresolved CXB in this field is shown as the red dotted line. The orange steps are from \citet{2016ApJS..224...40W}'s catalogue of this field. As a comparison, the curve from CDF-S is plotted as a green line. The error of the CDF-S curve includes both Poisson error and best-fit parameters' uncertainties. The detection limits of \emph{Chandra} and \emph{XMM-Newton} as well as the \emph{Suzaku} point source exclusion limit in this paper are marked as vertical dotted lines.}
\label{fig:lognlogs}
\end{figure}

To check the $\log{N}-\log{S}$ relation of our sample, we plot the cumulative source number curve in Fig. \ref{fig:lognlogs}. The $\log{N}-\log{S}$ relationship from the Chandra Deep Field South (CDF-S) has been well studied by \citet{2012ApJ...752...46L}. X-ray point sources in the range $10^{-15}<S<10^{-13}$ (erg s$^{-1}$ cm$^{-2}$) are dominated by AGNs (including X-ray binaries), and their distribution can be expressed by a broken power law,
\begin{equation}
\frac{\mathrm{d}N}{\mathrm{d}S} = \begin{dcases*}
K(S/S_\mathrm{ref})^{-\beta_1} & $(S\le f_b)$\\
K(f_b/S_\mathrm{ref})^{\beta_2-\beta_1}(S/S_\mathrm{ref})^{-\beta_2} &$(S>f_b)$
\end{dcases*},
\end{equation}
where $S_\mathrm{ref}=10^{-14}$ erg s$^{-1}$ cm$^{-2}$, and $f_b=6.4\pm1.0\times10^{-15}$ erg s$^{-1}$ cm$^{-2}$ is the power law break flux. The power law index after the break flux is $\beta_2=2.55\pm0.17$. We plot the total CDF-S cumulative $ \log{N}-\log{S}$ curve in Fig. \ref{fig:lognlogs} as well. The normalisation of the Abell 3411-3412 field is higher than that of the CDF-S. To cross check our point source flux analysis, we look up the catalogue compiled by \citet{2016ApJS..224...40W}, which covers the Abell 3411-3412 field. In their work, for each source, the 0.3 -- 8.0 keV flux is calculated with a fixed $\Gamma=1.7$ and a free $n_\mathrm{H}$. We assume $n\mathrm{H}=4.8\times10^{20}$ cm$^{-2}$ and convert the 0.3 -- 8.0 keV flux to a 2 -- 8 keV flux. The cumulative curve from that catalogue is also plotted. Though the methods of point source flux calculation are different from our work, the $\log{N}-\log{S}$ curve is consistent with ours at the faint end. At the bright end, the discrepancy is due to the assumptions for flux calculation. The consistency of results from two independent analyses proves that in this field, the number of point sources is much higher than the average value in the CDF-S. We fit our cumulative curve from $6\times10^{-15}$ to $1\times10^{-13}$ erg s$^{-1}$ cm$^{-2}$ using a single power law model with a fixed cumulative index $\alpha=\beta_2+1=1.55$. The ratio between our normalisation to that from CDF-S is $K_\mathrm{A3411}/K_\mathrm{CDF-S}=2.03\pm0.03$. 

\subsection{Detection limit and CXB flux}
From the cumulative $\log N-\log S$ curve, the \emph{Chandra} detection limit in this field is $\sim3.5\times10^{-15}$ erg s$^{-1}$ cm$^{-2}$.
The detection limit of \emph{XMM-Newton} is $6\times10^{-15}$ erg s$^{-1}$ cm$^{-2}$, see Sect. \ref{section:spectrum-xmm} for details. The \emph{Suzaku} detection limit is much higher because of the large PSF radius of $r_\mathrm{HEW}=1\arcmin$. Excluding more point sources would make the spectrum fitting less biased by unresolved sources but would decrease the signal statistics at the same time. We exclude from the \emph{Suzaku} analysis only point sources detected by \emph{Chandra} with a flux above $2\times10^{-14}$ s$^{-1}$ cm$^{-2}$. Point source coordinates are from the compiled \emph{Chandra} catalogue. We inspect the chosen sources on the flux maps and additionally include two sources. One is a super soft source (130.527\degr, -17.569\degr), whose photon index $\Gamma=3.8$ makes the 2 -- 8 keV flux $F=7.2\times10^{-15}$ erg s$^{-1}$ cm$^{-2}$ to be below our exclusion limit. Another one is (130.561\degr, -17.657\degr), which is just at the edge of the \emph{Chandra} field, but the source is bright in the \emph{XMM-Newton} flux map. We shift all source coordinates to match the \emph{Suzaku} astrometry.

The unresolved point source flux from CXBTools \footnote{\url{http://doi.org/10.5281/zenodo.2575495}} is based on the $\log{N}-\log{S}$ relation from \citet{2012ApJ...752...46L}'s work. Since we have found the point source density in our field is twice higher than that of CDF-S, we need to take that into account to estimate the unresolved CXB level of \emph{XMM-Newton} and \emph{Suzaku} spectral components properly. The number density of point sources below the \emph{Chandra} detection limit is unknown, but we speculate the $\mathrm{d}N/\mathrm{d}S$ curve of our field will converge into the curve of CDF-S when $S$ is small. We assume the convergent point is $S_\mathrm{cov}=1.4\times10^{-16}$ erg s$^{-1}$ cm$^{-2}$ and at the breaking point of \citet{2012ApJ...752...46L}'s curve $f_\mathrm{b}$, the differential curve normalisation is twice as the curve from CDF-S. We can express this relationship with the equation as below:
\begin{equation}
\left(\frac{\mathrm{d}N}{\mathrm{d}S}\right)_\mathrm{CDF-S}\left(S_\mathrm{cov}\right)=2\times\left(\frac{\mathrm{d}N}{\mathrm{d}S}\right)_\mathrm{CDF-S}\left(S_\mathrm{cov}\right) \times\left(\frac{S_\mathrm{cov}}{f_\mathrm{b}}\right)^{\alpha}
\end{equation}
The solution of the function is $\alpha=0.18$. Hence, in our field, the differential $\log{N}-\log{S}$ relation is 
\begin{equation}
\left(\frac{\mathrm{d}N}{\mathrm{d}S}\right)_\mathrm{A3411}=2\times\left(\frac{\mathrm{d}N}{\mathrm{d}S}\right)_\mathrm{CDF-S}\times\begin{dcases*}
\left(\frac{S}{f_\mathrm{b}}\right)^{0.18} & ($S_\mathrm{cov}\le S<f_\mathrm{b}$) \\
1 & ($S\ge f_\mathrm{b}$) \\
\end{dcases*},
\end{equation}
and the unit of source flux is erg s$^{-1}$ cm$^{-2}$. The cumulative curve of this modified $\log N-\log S$ model is plotted as the red dotted line in Fig. \ref{fig:lognlogs}.

\begin{figure}[t!]
\resizebox{\hsize}{!}{\includegraphics{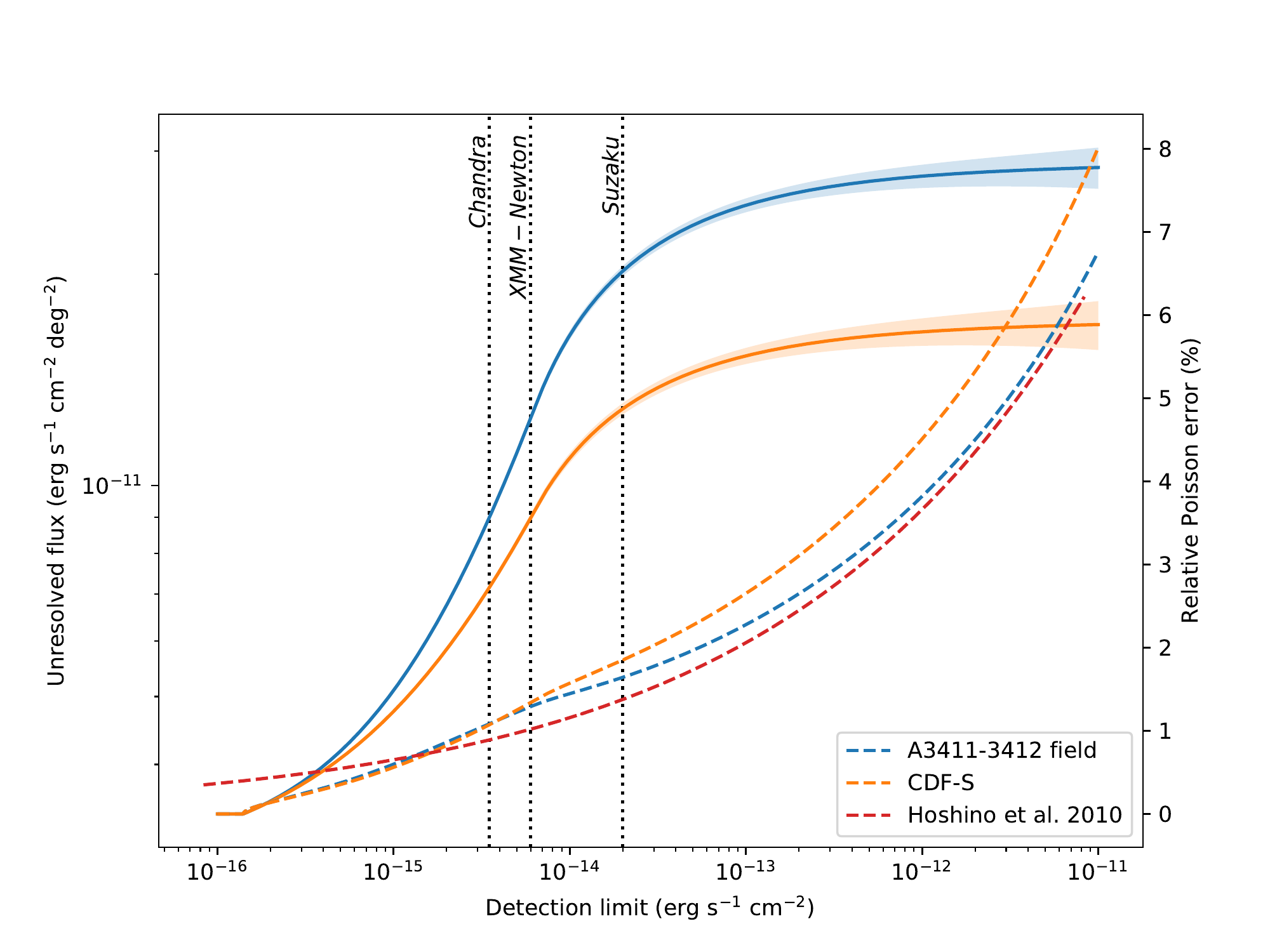}}
\caption{The unresolved CXB flux in the 2 -- 8 keV band as a function of point source detection limit. The curves of CDF-S and our field are plotted with orange and blue lines, respectively. Dashed lines indicate the relative poisson error for 1 deg$^2$ sky area.}
\label{fig:cxbflux}
\end{figure}

We apply the differential $\log{N}-\log{S}$ relation to estimate unresolved CXB flux in this paper. The unresolved CXB flux and its Poisson uncertainty can be expressed as
\begin{align}
F(S<S_\mathrm{lim})&=A\left[F(S<S_\mathrm{cov})+\int_{S_\mathrm{cov}}^{S_\mathrm{lim}}S\left(\frac{\mathrm{d}N}{\mathrm{d}S}\right)_\mathrm{A3411}\mathrm{d}S\right] \\
\sigma_F^2 &= A\int_{S_\mathrm{cov}}^{S_\mathrm{lim}}S^2\left(\frac{\mathrm{d}N}{\mathrm{d}S}\right)_\mathrm{A3411}\mathrm{d}S \label{equation:cxberror},
\end{align}
where the unresolved flux below $1.4\times10^{-16}$ erg s$^{-1}$ cm$^{-2}$ is $3.4\times10^{-12}$ erg s$^{-1}$ cm$^{-2}$ deg$^{-2}$ \citep{2006ApJ...645...95H}, and $A$ is the sky area of the selection region. The unresolved flux as a function of the detection limit is plotted in Fig. \ref{fig:cxbflux} together with the relative error for a 1 deg$^2$ sky area. For comparison, we over-plot the empirical relative error curve from \citet{2010PASJ...62..371H}.

We note that for \emph{XMM-Newton} data analysis, the actual CXB residual luminosity is not uniform due to the ICM emission. The detection limit extends to fainter point sources further out in radius as the ICM emission decreases. The point source sensitivity (in cgs units) can be expressed as 
\begin{equation}
F=1.609\times10^{-9}\bar{E}\frac{S^2}{2At}\left(1+\sqrt{1+\frac{4BP}{S^2}}\right),
\end{equation}
where $\bar{E}$ is the averaged photon energy, $S$ the signal-to-noise ratio, $A$ the effective area, $t$ the exposure time, $B$ the background counts per PSF beam, and $P$ the PSF size. We use a quadratic function to model the radial increase of the PSF size and use a linear function to model the vignetting effect. The background $B$ in the formula is composed of the cluster emission and other background components. The cluster emission is modelled by a $\beta$ model (see Eq.\ref{eq:beta}). The estimated 2D standard deviation of the residual CXB flux inside $10\arcmin$ is $\sim12\%$. As a comparison, based on the uncertainty curve in Fig. \ref{fig:cxbflux}, the CXB uncertainty contributed by the cosmic variance in a 3 arcmin$^{2}$ selection region is $\sim45\%$ and this value will be larger in a smaller selection region. Therefore we only take the cosmic variance into account. For \emph{Suzaku} data analysis, we adopt the point source exclusion limit based on the Chandra point source catalogue. Therefore we can assume a uniform residual CXB flux.

\end{appendix}

\end{document}